\documentclass[12pt]{article}
\usepackage{a4wide}
\usepackage[centertags]{amsmath}
\usepackage{amssymb}
\usepackage[sort&compress,numbers]{natbib}
\usepackage{ifpdf}
\usepackage{setspace}
\usepackage{amsfonts}
\usepackage{color}
\usepackage{threeparttable}
\usepackage{cancel}
\usepackage{tikz}
\usetikzlibrary{decorations.pathmorphing}

\usepackage{yfonts}
\usepackage{ulem}

\usetikzlibrary{decorations.markings}
\ifpdf
\usepackage[colorlinks=true,
linkcolor=black,
citecolor=black,
urlcolor=blue,
filecolor=blue,
pdfstartview=FitV,
pdftitle={},
pdfauthor={},
pdfsubject={Hydrodynamics},
pdfkeywords={hydrodynamics, AdS/CFT},
pdfpagemode=None,
bookmarksopen=true
]{hyperref}
\else
\usepackage[hypertex]{hyperref}
\fi
\usepackage{rotating}

\usepackage{authblk}

\makeatletter
\@addtoreset{equation}{section}

\newcommand{\qn}{{\mathfrak{q}}}
\newcommand{\wn}{{\mathfrak{w}}}

\makeatletter
\renewcommand\section{\@startsection {section}{1}{\z@}%
	{-3.5ex \@plus -1ex \@minus -.2ex}
	{2.3ex \@plus.2ex}%
	{\normalfont\large\bfseries}}
\renewcommand\subsection{\@startsection{subsection}{2}{\z@}%
	{-3.25ex\@plus -1ex \@minus -.2ex}%
	{1.5ex \@plus .2ex}%
	{\normalfont\bfseries}}

\def\sec#1{section~\ref{#1}}

\title{{
Characteristic momentum of Hydro+ and a bound on the speed of sound near the QCD critical point
}}
\author[1]{Navid Abbasi\footnote{abbasi@lzu.edu.cn}}  
\author[2]{Matthias Kaminski\footnote{mski@ua.edu}}
\affil[1]{\small{\it School of Nuclear Science and Technology, Lanzhou University, 222 South Tianshui Road, Lanzhou
		730000, China, }}
\affil[2]{\small{\it Department of Physics and Astronomy, 
		University of Alabama, 514 University Boulevard, Tuscaloosa, AL 35487, USA.}}

\begin{document}

\setlength{\baselineskip}{16pt}
\begin{titlepage}
\maketitle

\vspace{-36pt}

\begin{abstract}
%
Near the critical point in the QCD phase diagram, hydrodynamics breaks down  at a momentum 
where the frequency of the fastest hydrodynamic mode becomes comparable with the decay rate of the slowest non-hydrodynamic mode. 
Hydro+ was developed as a framework which extends the range of validity of  hydrodynamics beyond that momentum value. This was achieved through coupling the hydrodynamic modes to the slowest non-hydrodynamic mode. 
In this work, analyzing the spectrum of linear perturbations in Hydro+, we find that a slow mode 
falls out of equilibrium if its momentum is greater than a characteristic momentum value. 
That characteristic momentum turns out to be set by the branch points of the dispersion relations.  
These branch points occur at the critical momenta of so-called spectral curves and are related to the radius of convergence of the derivative expansion. 
The existence of such a characteristic momentum scale suggests that a particular class of slow modes has no remarkable effect on the flow of the plasma. 
Based on these results and previously derived relations to the stiffness of the equation of state, we find a temperature-dependent upper bound for the speed of sound near the critical point in the QCD phase diagram.
  \end{abstract}
\thispagestyle{empty}
\setcounter{page}{0}
\end{titlepage}

\renewcommand{\baselinestretch}{1}  
\tableofcontents
\renewcommand{\baselinestretch}{1.2}  
\section{Introduction}
\label{intro}
In the previous 15 years, most of the heavy-ion-collision experiments have been set for high collision energies at RHIC and LHC~\cite{Bzdak:2019pkr,Rajagopal:2019xwg}.  
In these experiments, the non-vanishing net baryon charge coming from the incident nuclei mostly ends up at high rapidity.  
Meanwhile, the Quark-Gluon-Plasma (QGP) forming at mid-rapidity carries nearly no baryon charge, i.e. $\mu_B = 0$.  
In order to study the QGP at nonzero $\mu_B$, one requires 
heavy ion collisions at lower collision energies. 
This is the goal of the ongoing Beam Energy Scan (BES) program at RHIC~\cite{Stephanov:1998dy,Stephanov:2004wx}	(see also \cite{Bzdak:2019pkr} for a comprehensive review).\footnote{High baryon densities are also a focus of study at the 
	Compressed Baryonic Matter (CBM) experiment at the FAIR facility at GSI, the Multi-Purpose-Detector (MPD) at the NICA accelerator in Dubna, and the CSR-External target Experiment (CEE) at the HIAF facility in China.} 
The BES program provides us with the opportunity to detect signatures of the QCD critical point, if such a point exists within the region of the phase diagram which is accessible to experiment~\cite{Stephanov:2006zvm,Yin:2018ejt}.

In order to maximize the discovery potential of the experimental efforts, it is desirable to identify 
signatures originating from the critical fluctuations appearing at all length scales near the critical point, a behavior that is familiar from any second order phase transition. 
In thermal equilibrium, such fluctuations have been well-understood for a long time~\cite{Halperin}. 
However, during a heavy ion collision, critical fluctuations can not possibly stay in thermal equilibrium~\cite{Berdnikov:1999ph}.

If there is a critical point in the equilibrium phase diagram of QCD, the phase space trajectory of the droplet of QGP formed in a heavy ion collision may 
pass near the critical point as it expands and cools down. 
Near the critical point, the equilibration time scale diverges. In other words, the critical fluctuations fall out-of-equilibrium \cite{Kawasaki}. 
One naturally expects these out-of-equilibrium fluctuations to modify the equation of state (EoS), which eventually affects 
the hydrodynamic evolution. 
The recently developed Hydro+ framework~\cite{Stephanov:2017ghc} is an approach to self-consistently study the effect of critical fluctuations on the evolution of hydrodynamic variables such as the fluid velocity and baryon chemical potential near the critical point.

Coming closer to the critical point, there exists an increasing number of long-lived non-hydrodynamic modes, which we will refer to as {\it slow modes}. This effect is known as {\it critical slowing down}.  
If the decay rate of these modes is of order of the frequency of the hydrodynamic modes, i.e. $\Gamma_{\boldsymbol{Q}}\sim\omega$, the standard hydrodynamic approximation fails to work. 
Heuristically, Hydro+~\cite{Stephanov:2017ghc} couples these slow non-hydrodynamic modes, $\phi_{\boldsymbol{Q}}$, to the hydrodynamic modes. Hence, Hydro+ has a larger regime of validity than hydrodynamics. In this way, Hydro+ is capable of describing fluctuations closer to the critical point than hydrodynamics is.

The spectrum of linear perturbations around thermal equilibrium in (neutral) {\it single-mode} Hydro+ contains three modes with momentum vector $\boldsymbol{q}$ obeying distinct dispersion relations $\omega(\boldsymbol{q})$: two sound modes together with a {\it single} non-hydrodynamic mode, $\phi_{\boldsymbol{Q}_{0}}$. 
The effect of this slow mode is more important at large values of $q=|\boldsymbol{q}|$, i.e. $q \gtrsim\Gamma_{\boldsymbol{Q}_0}/c_s$; the equation of state 
becomes stiffer\footnote{This terminology refers to the increasing rigidity of the QGP near the critical point. It behaves more like an incompressible fluid closer to the critical point.} 
and consequently, the sound velocity increases near the critical point at such large momenta. 

In this work, for the first time, we compute the branch point singularities of the Hydro+ dispersion relations. These occur at critical values of the momentum, $q_c$, and they can be related to the radius of convergence of dispersion relations~\cite{Withers:2018srf,Grozdanov:2019kge,Grozdanov:2019uhi,Heller:2020hnq}. We further consider the effect of these singularities on the observable $\Delta c_s^2$, the shift in the speed of sound in QCD plasma due to the fluctuations near the critical point. 
For this purpose,
we explicitly compute the spectrum of linear perturbations, $\omega(q)$, in a single-mode Hydro+\footnote{By a single-mode Hydro+ we mean a toy model consisting of the coupling between hydrodynamics and $\phi_{\boldsymbol{Q}_0}$ at a fixed $\boldsymbol{Q}_0$. }, and thereby we determine the critical momentum $q_c$.

We find that at $q<q_c$, the $\phi_{\boldsymbol{Q}_{0}}$-mode is still a fast-decaying mode, that is to say that standard hydrodynamics continues to hold in this range.  However, at $q\gtrsim q_c$, the critical slowing down phenomenon is inevitable. 
Our work is guided by a 
recent line of research studying the radius of convergence of hydrodynamics in holographic models~\cite{Withers:2018srf,Grozdanov:2019kge,Grozdanov:2019uhi,Heller:2020hnq,Abbasi:2020ykq,Abbasi:2020xli,Asadi:2021hds,Jeong:2021zhz,Grozdanov:2021gzh,Jeong:2021zsv,Wu:2021mkk,Baggioli:2020loj,Arean:2020eus,Heller:2020uuy,Baggioli:2021ujk,Jansen:2020hfd,Heller:2021oxl,Huh:2021ppg,Liu:2021qmt}. In the present work, however, we work entirely in field theory while holography makes its one and only appearance in \sec{gravity}. 

In the second part of this work, we apply our single-mode Hydro+ results to QCD plasma near the critical point in the QCD phase diagram. We use a set of standard assumptions from the literature (see~\cite{Rajagopal:2019xwg} and references therein), and extract the characteristic momentum $q_c$ near the critical point within the kinetic framework~\cite{Stephanov:2017ghc}. 
Let us emphasize that our goal is not to provide phenomenological values in this work. Instead, we study qualitatively how the existence of the characteristic momentum scale, the critical momentum $q_c$, affects physical predictions, especially the effective speed of sound near the critical point of the QCD phase diagram.

This paper is structured as follows. First, in \sec{single_section}, we introduce Hydro+ as well as the analysis of spectral curves leading to the computation of the radius of convergence of the hydrodynamic expansion when linearized in the hydrodynamic field variables. We close that section by elaborating on the spectrum of linear perturbations in single-mode Hydro+. 
In \sec{QCD}, we first review the effects of fluctuations near the critical point. Then by use of the results obtained in \sec{single_section}, the radius of convergence as a function of the temperature $T$ is computed. Next, \sec{stiffness} is devoted to investigating the impact of convergence on the stiffness of the equation of state near the critical point, and consequently its effect on the speed of sound.   
In \sec{range_of_applicabaility} we discuss the limitations of our study. 
We discuss a possible gravity dual of Hydro+ in \sec{gravity}. Finally, in \sec{conclusion}, we end with the review of our results and mention some possible follow-up directions.

\section{Convergence radius of single-mode Hydro$+$}
\label{single_section}
%
In this section, we first present a brief overview of the relevant aspects of {\it Hydro+} and subsequently a brief overview of {\it spectral curves} and how they determine the radius of convergence of hydrodynamics. Combining these two concepts, we then compute the radius of convergence of single-mode Hydro+.  

\subsection{A brief overview of Hydro+}
In a system with partially equilibrated states, the evolution is locally described by hydrodynamics in terms of conserved densities which are referred to as {\it hydrodynamic fields}, energy density $\epsilon(t,\boldsymbol{x})$, charge density $n(t,\boldsymbol{x})$ and momentum density $w(t,\boldsymbol{x})\,u^{\mu}$ \cite{Kovtun:2012rj}. Usually, all other microscopic (non-hydrodynamic) modes, corresponding to non-conserved quantities, decouple from these densities. However, nearing a critical point in the phase diagram, an increasing number of non-hydrodynamic (gapped) modes is known to become long-lived, as their decay rate $\Gamma$ decreases and the correlation length increases.  As a result the standard hydrodynamics fails to work near the critical point. Hydro+ is a framework that systematically combines the dynamics of these long-lived modes with that of the conserved densities. A detailed review of Hydro+ can be found in \cite{Stephanov:2017ghc}.

In order to introduce the key concepts behind Hydro+ which are of relevance to our analysis, let us consider a general out-of-equilibrium two-point function, $G({\boldsymbol{x}_1, \boldsymbol{x}_2})$, of a conserved operator, for example the energy-momentum tensor. 
If the scale at which the midpoint $(\boldsymbol{x}_1+\boldsymbol{x}_2)/2$  
varies is much larger than the scale of the $|\boldsymbol{x}_1-\boldsymbol{x}_2|$-dependence, then $G$  can be replaced with a continuous set of local modes, $G_{\boldsymbol{Q}}(\boldsymbol{x})$. Here $\boldsymbol{Q}$ is a continuous index which denotes the momentum associated with the critical fluctuation. The  separation of scales  mentioned above can also be written as $q\ll Q$ where $q$ is the momentum of the hydrodynamic perturbation. 

Among the critical fluctuations, the two-point function of $m=s/n$, with the entropy density $s$, is referred to as $G_{m}(\boldsymbol{x}_1,\boldsymbol{x}_2)$, and it is special. 
The corresponding local modes $\phi_{\boldsymbol{Q}}$ are the slowest-varying non-conserved modes in the system. 
The presence of such long-lived non-hydrodynamic modes near the critical point is what is referred to as the \textit{critical slowing down}. If the decay rate of these modes is of order of the frequency of the conserved modes, i.e. $\Gamma_{\boldsymbol{Q}}\sim\omega$, the standard hydrodynamic description fails to work. 
Then Hydro+ comes into play. It actually extends the regime of validity of hydrodynamics near the critical point~\cite{Stephanov:2017ghc} through coupling the dynamics of $\phi_{\boldsymbol{Q}}$ with that of the hydrodynamic modes.

\subsection{A brief overview of spectral curves \& hydrodynamic convergence}
Recently, methods from complex analysis have been developed in order to compute the radius of convergence of the hydrodynamic derivative expansion from the so-called {\it spectral curve} of a given theory~\cite{Grozdanov:2019kge,Grozdanov:2019uhi,Withers:2018srf}. In the present paper we will extend these methods to Hydro+, but before that we start with a short review.  

In hydrodynamics, the spectral curve  arises from the determinant of a system of hydrodynamic perturbation equations that encodes the hydrodynamic dispersion relations. 
It is an implicit function of frequency $\omega$ and momentum $\boldsymbol{q}$ taking the form $F(\omega, \boldsymbol{q}) = 0$.  
Here we consider rotation-invariant theories and states, hence  the dependence on $\boldsymbol{q}$ is through $|\boldsymbol{q}|^2=q^2$.
As an example, consider  the spectral curve $F(\omega, q^2) = c_s^2 \omega^2-q^2=0$\footnote{This is actually the general structure of the spectral curve in gapless theories, expanded to leading order about $\omega(0)=0$.} at small momentum and frequency, corresponding to the hydrodynamic description in the absence of any derivative (viscous) correction; it encodes the sound dispersion relation $\omega=\pm c_s q$, with the speed of sound, $c_s$. 
At arbitrary values of the momentum and frequency, the analytic structure of spectral curves can be very complicated. Let us limit our discussion to a description whose spectral curve is an analytic function, which is the case for Hydro+ truncated at leading order\footnote{Considering Hydro+ to subleading order can be viewed similar to Mueller-Israel-Stewart theory (MIS). Similar to the dissipative shear tensor in MIS, we here may interpret the slow mode as a resummed version of contributions of all non-hydrodynamic modes to all orders in derivatives.}\footnote{See \cite{Heller:2020hnq} for theories with non-analytic spectral curve.}. 

Given an analytic spectral curve, there is a relation between the regime of validity of linear hydrodynamics and {\it critical points of the spectral curve}.\footnote{These critical points of spectral curves have nothing to do with the critical point which occurs in the QCD phase diagram. The identical naming is a coincidence.}  To elaborate on this relation, let us recall that
the critical points of spectral curves can  be computed as those points $(\omega^*, q^*)$, which satisfy 
\begin{equation}\label{eq:criticalPointConditions}
F(\omega, q)|_{(\omega^*, q^*)} = 0 \,,  \quad   
\partial_\omega (\omega, q)|_{(\omega^*, q^*)} = 0\,.
\end{equation}
Note that, in general, $q^*$ and $\omega^*$ are complex-valued. 
In general, solutions $\omega(q)$ of eq.~\eqref{eq:criticalPointConditions} are referred to as dispersion relations; one example, the sound dispersion relation was mentioned above. A subset of all critical points are also branch points in the sense of complex analysis, singularities of the dispersion relations. Therefore, it is no surprise that those critical points limit the radius of convergence of the hydrodynamic derivative expansion in momentum space and also in position space~\cite{Heller:2020uuy}.

In a gapless theory, the spectral curve encodes the dispersion relations $\omega(q)$ that pass through $(\omega=0, q=0)$. These are the so-called hydrodynamic dispersion relations. If $F$ is analytic, such $\omega(q)$ can be found as a Puiseux series about $(\omega=0, q=0)$ \cite{Grozdanov:2019kge,Grozdanov:2019uhi}. This series then may have a finite radius of convergence, $q_c$, in the complex $q-$plane. 
This radius of convergence is set by one of the critical points of $F$, namely the one closest to the origin; i.e. it is that $q^*$ which has the minimal magnitude $q_c=\min \{ |q^*| \}$. This $q_c$ is what is referred to as the radius of convergence of the derivative expansion of $\omega(q)$.  In other words, $\omega(q)$ is an analytic function within the disc $q<q_c$, and develops a non-analyticity at $q=q_c$.

In summary, in order to determine the radius of convergence of linear hydrodynamics from the spectral curve,  one has to first find all branches of Puiseux series passing through $(\omega=0, q=0)$  \cite{Abbasi:2020ykq}. Each of these branches corresponds to one of the hydrodynamic dispersion relations $\omega(q)$. Then for a particular $\omega(q)$ of interest, e.g. the sound dispersion $\omega(q)$, the convergence radius of the derivative expansion is the distance from the nearest singularity of $\omega(q)$ to the origin. That singularity may be located at a complex-valued momentum.

\subsection{Single mode Hydro$+$}
\label{single_subsection}
As mentioned above, there are situations where one (or a set of) non-hydrodynamic modes decays so slowly that it couples to the hydrodynamic densities, for example, near the critical point in the QCD phase diagram. Let us focus on one single mode and call it $\phi$. Then the partial-equilibrium states are those not only satisfying the constraint related to the conserved densities, but also satisfying the constraint that the expectation value of the slow mode takes on a particular value $\phi$. Accordingly, the partial equilibrium entropy density $s(\epsilon,n)$ should be modified to include the $\phi$ mode as $s_{(+)}(\epsilon,n,\phi)$~\cite{Stephanov:2017ghc}. Then one finds  
\begin{equation}\label{}
ds_{+}=\,\beta_{(+)}d\epsilon -\alpha_{(+)}dn-\pi d \phi \, ,
\end{equation}
with $\alpha_{(+)}$ and $\beta_{(+)}$ 
generalizing $\alpha=\mu/T$ and $\beta=1/T$, respectively, to the partial-equilibrium states now including the slow mode $\phi$. 
Here, $\pi(\epsilon, n, \phi)$ is the thermodynamic  force returning $\phi$ to its equilibrium value.
In a complete equilibrium, where $\phi$ reaches its equilibrium value  $\bar{\phi}(\epsilon,n)$,
\begin{equation}\label{}
\pi(\epsilon, n,\bar{\phi}(\epsilon,n))=0 \, .
\end{equation}
The hydrodynamic equations coupled to the relaxation equation of $\phi$ are then given by
\begin{eqnarray}\label{Depsilon}
D\epsilon&=&-w_{(+)}\theta -\partial_{(\mu}u_{\nu)}\, \Pi^{\mu\nu},\\\label{Dn}
Dn&=&- n \theta - \partial \cdot \Delta J,\\\label{Du}
w_{(+)}D u^{\nu}&=&-\partial^{\nu}_{\perp}p-\delta_{\perp\lambda}^{\nu}\, \partial_{\mu}\Pi^{\mu\lambda},\\\label{Dphi}
D\phi&=&-\gamma_{\pi}\pi-A_{\phi}\theta+\cdots\label{phi_EoM}\,,
\end{eqnarray}
with $D=u\cdot \partial,\, \theta=\partial\cdot u$, the fluid velocity $u^\mu$, the viscous stress tensor $\Pi^{\mu\nu}$, the partial equilibrium enthalpy $w_{(+)}=\epsilon+p_{(+)}$, 
$A_\phi$ is the compression/expansion susceptibility of $\phi$, $\gamma_\pi$ parametrizes the strength of the {\it returning force} driving $\phi$ towards its equilibrium value, and 
dots stand for the higher order derivative corrections. Imposing the constraints induced by the second law of thermodynamics and substituting $\phi(\epsilon, n, \pi)$ into eq.\eqref{phi_EoM}, this equation takes the form
\begin{equation}\label{Dpi}
D\pi=-\Gamma_{\pi}\pi-\frac{\beta p_{\pi}}{\phi_{\pi}}\theta+\cdots\, ,
\end{equation}
where the relaxation rate is defined as $\Gamma_\pi \equiv \gamma_\pi/\phi_\pi$, with $\phi_\pi=\left( \frac{\partial \phi}{\partial \pi}\right )_{\epsilon, n}$, and $p_\pi=\left(\frac{\partial p}{\partial \pi}\right)_{\epsilon,n}$.  

It is easy to show that from the linearized hydrodynamic equations the dispersion relations of the two sound modes together with the slow mode are given as the three roots of the following equation \cite{Stephanov:2017ghc}
\begin{equation}\label{Hydro+}
F(\omega,q^2)=\,\omega^2-q^2\left(c_s^2+\frac{\omega}{\omega+i \Gamma_{\pi}} \,\frac{\beta p_{\pi}}{\phi_{\pi}w}\right)=\,0\,.
\end{equation}
$F$ is actually the spectral curve of Hydro+ leading order in derivative corrections.
The real part of the expression inside the parentheses is the square of the effective velocity of sound.
Thus the presence of the slow mode leads to an enhancement of the value of $c_s^2$ given by~\cite{Stephanov:2017ghc}  
\begin{equation}\label{Delta_cs_2_single}
\Delta c_s^2=\frac{\omega^2}{\omega^2+ \Gamma_{\pi}^2} \,\Delta c_s^2(\infty),\,\,\,\,\,\,\,\,\,\,\Delta c_s^2(\infty)=\,\frac{\beta p_{\pi}}{\phi_{\pi}w}\,.
\end{equation}
We now define the dimensionless quantities 
\begin{equation}\label{dimensionless}
\wn=\frac{\omega}{\Gamma_{\pi}},\,\,\,\,\,\,\,\,\qn=\frac{c_s\,q}{\Gamma_{\pi}},\,\,\,\,\,\,\,\,\alpha=\frac{\Delta c_s^2(\infty)}{c_s^2} \, ,
\end{equation}
through which the spectral curve~\eqref{Hydro+} simplifies to
\begin{equation}\label{spectral}
F(\wn,\qn^2)=\,\wn^2-\qn^2\,\frac{i+ \wn+\,\alpha\,\wn}{i+\wn}=\,0 \, .
\end{equation}
To our knowledge, the spectral curve has not been written in the form~\eqref{spectral} before. We will see that this form has several advantages. 
This spectral curve is a polynomial of order three. Thus we find here, for the first time, the dispersion relations which are given by the three algebraic roots $\wn(\qn)$ of~\eqref{spectral}, namely  
\begin{equation}\label{omega_s}
\begin{split}
\wn_{1}(\qn)=&\,-\frac{i}{12}\left(4+\frac{2^{7/3}(-1+3(1+\alpha)\qn^2)}{3\,\mathcal{D}(\qn)}-\frac{2^{2/3}}{3}\mathcal{D(\qn)}\right)\, ,\\
\wn_{2}(\qn)=&\,-\frac{i}{12}\left(4+\frac{2^{4/3}(-i+\sqrt{3})(-1+3(1+\alpha)\qn^2)}{\mathcal{D}(\qn)}-2^{2/3}(i+\sqrt{3})\mathcal{D(\qn)}\right)\, , \\
\wn_{3}(\qn)=&\,-\frac{i}{12}\left(4+\frac{2^{4/3}(-i-\sqrt{3})(-1+3(1+\alpha)\qn^2)}{\mathcal{D}(\qn)}-2^{2/3}(1+i\sqrt{3})\mathcal{D(\qn)}\right) \, ,
\end{split}
\end{equation}
where 
\begin{equation}
\mathcal{D}(\qn)=\,\left(2i+9i(2-\alpha)\qn^2+\,3\sqrt{3}\sqrt{-4-4\qn^4(1+\alpha^3)+\qn^2(-8+20\alpha+\alpha^2)}\right)^{1	/3}.
\end{equation}
Continuing these dispersion relations to complex momenta, there are four square-root singularities, branch points, in the dispersion relation of each mode. These stem from the polynomial of order four under the square-root in $\mathcal{D}(\qn)$. These branch points are located at the roots of
\begin{equation}\label{singularities}
\begin{split}
(\qn_1^*)^2=&\frac{\alpha^2+20\alpha-8+\sqrt{\alpha-8}\,(\alpha^{3/2}-8\alpha^{1/2})}{8(1+\alpha)^3}\,,\\\,\,\,\,\,\,\,\,(\qn_2^*)^2=&\frac{\alpha^2+20\alpha-8-\sqrt{\alpha-8}\,(\alpha^{3/2}-8\alpha^{1/2})}{8(1+\alpha)^3}\,.
\end{split}
\end{equation}
Depending on whether $\alpha<8$ or $\alpha>8$,  the branch points occur at complex or real momenta, respectively, distinguishing the following cases:
\begin{itemize}
	\item[(i)] When $\alpha<8$, $(\qn^*_1)^2$ and $(\qn^*_2)^2$ are complex, however $|\qn^*_1|^2=|\qn^*_2|^2$.	
	\item[(ii)] At $\alpha=8$, $(\qn^*_1)^2=\,(\qn^*_2)^2$ and both become real.	
	\item[(iii)] When $\alpha>8$, $(\qn^*_1)^2$ and $(\qn^*_2)^2$ are both real and $(\qn^*_1)^2<(\qn^*_2)^2$.	
\end{itemize}
In the next subsections, we discuss the spectrum for the above three cases in detail.

Obviously, the existence of  branch points \eqref{singularities} limits the range of analyticity of the dispersion relations~\eqref{omega_s}. About the point $(\wn, \qn)=(0,0)$, each of the dispersion relations \eqref{omega_s} is analytic
 within the disc $|\qn|\le \min\{|\qn^*_1|, |\qn^*_2|\}$ in the complex $\qn$-plane. Then, according to the Puiseux theorem \cite{Grozdanov:2019kge}, $\wn(\qn)$ can be computed as a convergent series about $\qn=0$; the radius of convergence of this series is set by
\begin{equation}\label{q_c_0} 
\qn_c=\min\{|\qn^*_1|, |\qn^*_2|\}\,.
\end{equation}
We see that $\qn_c$ is a characteristic momentum scale in the theory which depends only on $\alpha$. 
\subsection*{Asymptotics of the spectrum}
Consider the asymptotic behavior of the dispersion relations~\eqref{omega_s} for the two cases $\qn\ll1$ and $\qn\gg1$:
\begin{itemize}
\item For $\, \qn\ll1$,  Hydro+ dispersion relations reduce to the standard hydrodynamic ones, given by
\begin{equation}
\begin{split}
\wn_{1,2}=&\,\pm\qn-\frac{1}{2}\,i\,\alpha\,\qn^2+O(\qn^3)\, ,\\
\wn_3=&-i+\,i\,\alpha\,\qn^2+O(\qn^4) \, .
\end{split}
\end{equation}
The first two modes are nothing but the ordinary sound modes; $\omega=\pm c_s q$ (see \eqref{dimensionless}). The third mode is a fast decaying non-hydrodynamic mode whose dynamics decouples from the conserved quantities.
\item
For $	
 \,\qn\gg1$, hydrodynamics seizes to work. The slow mode comes to couple with the sound modes, yielding
 \begin{equation}\label{asymptote}
 \begin{split}
 \wn_{1,2}=&\,\pm\sqrt{1+\alpha}\,\qn-\frac{i \,\alpha}{2(1+\alpha)}+O(\frac{1}{\qn})\,,\\
 \wn_3=&-\frac{i }{1+\alpha}+O(\frac{1}{\qn^2})\,.
 \end{split}
 \end{equation}
 The first two modes are the sound modes which do now propagate faster than the ordinary sound modes. 
 This is equivalent to saying that the equation of state (EoS) has become stiffer. The third mode is a  non-hydrodynamic (gapped) mode. It is important to note that:
 \begin{enumerate}
 	\item For $\alpha>2$, the gapped slow mode is the longest-lived mode. This implies that Hydro+ is necessary in this range. 
 	\item For $0<\alpha<2$, the dissipative sound modes are the longest-lived modes. Nevertheless, the decay rate of the slow mode is comparable with that of sound modes. Therefore, although the slow mode decays faster than the sound modes in this range, Hydro+ should still be applied.
	\item For $\alpha=2$, the sound modes and the slow modes decay at the same rate for $\qn\gg 1$, see figure~\ref{spectrum_before}. 
 	\end{enumerate}
 \end{itemize}
\subsection*{Spectrum at $\alpha<8$}
\begin{figure}[tb]
	\centering
	\includegraphics[width=0.55\textwidth]{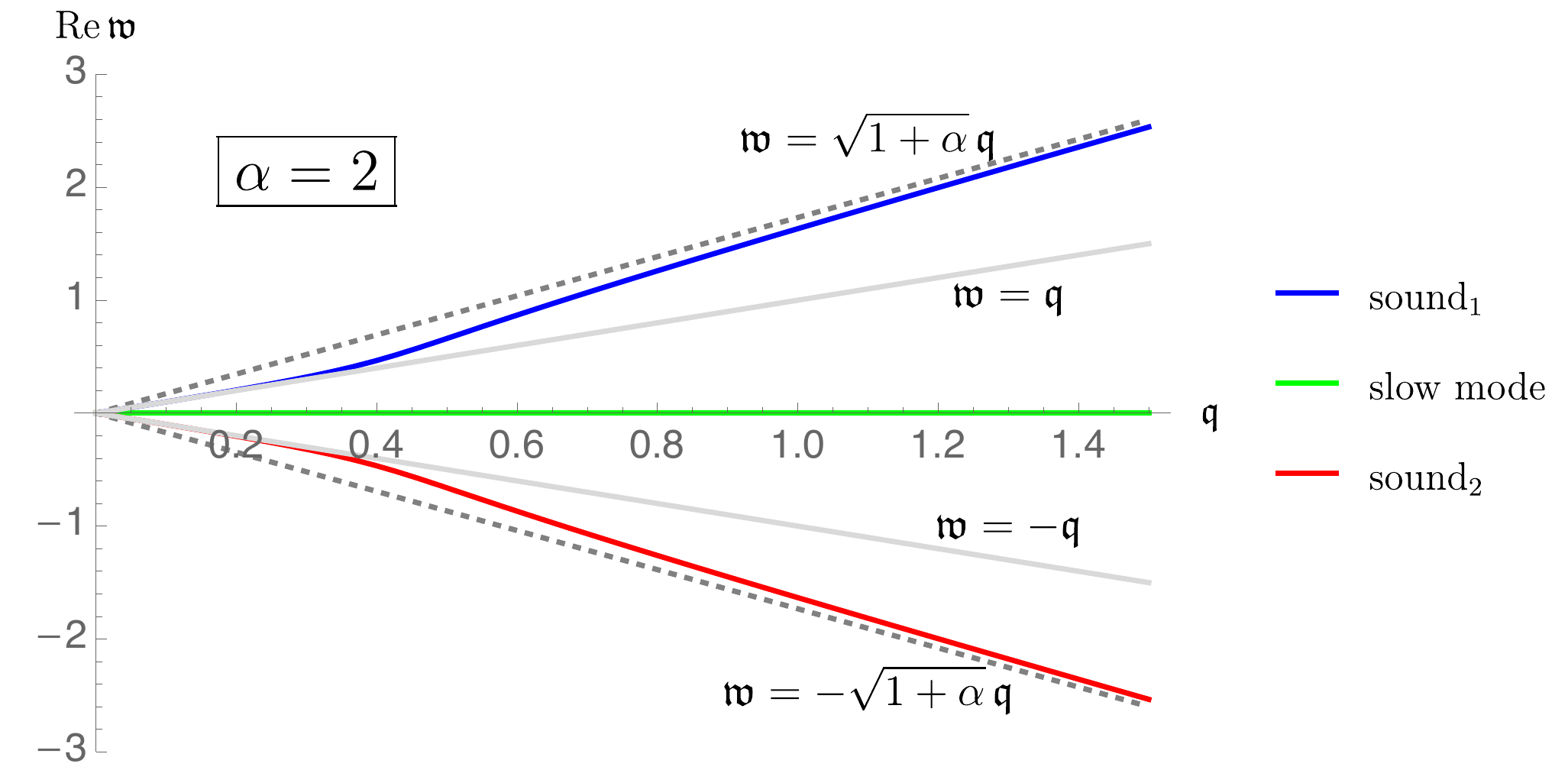}\includegraphics[width=0.55\textwidth]{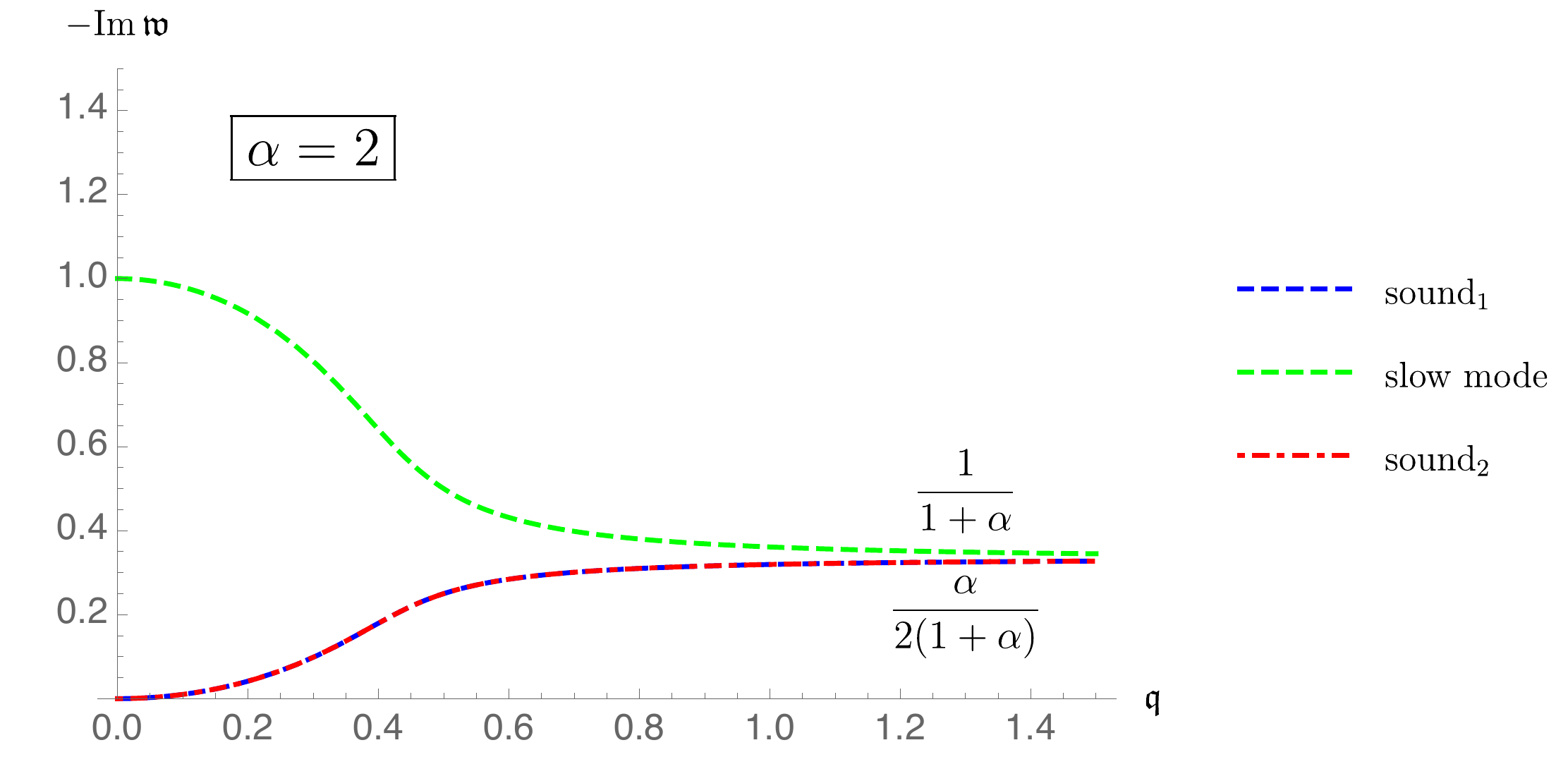}
	\caption{{\it Dispersion relations for $\alpha < 8 $ (displayed is the example $\alpha=2$). } The slow mode (green curves) is purely imaginary. 
		Sound modes (blue and red curves) have complex frequencies, whose velocities are $c_s$ at $\qn\ll1$ while they  tend to $\sqrt{1+\alpha}\,c_s$ at $\qn\gg1$. This is due to the backreaction of the slow mode on the hydrodynamic evolution. 
		The sound modes are the longest-lived modes for all values of momentum $\qn$. Note that away from this very special case, for any $2<\alpha<8$ the slow mode will be longer lived than the two sound modes, i.e. the green curve will be closer to the $\qn$-axis than the blue and red curves.  
		Compare this with figure 2 in~\cite{Stephanov:2017ghc}.}
	\label{spectrum_before}
\end{figure}

As it was sown earlier, for $\alpha<8$, each of the three dispersion relations has two complex-valued singularities. For concreteness, let us consider the case $\alpha=2$.\footnote{This special case has been studied in \cite{Stephanov:2017ghc}. It is special in the sense that in the large-$\qn$ limit all modes have the same imaginary part for $\alpha=2$, as can be seen in figure~\ref{spectrum_before}.} The location of branch points is given by
\begin{equation}\label{branch_alpha_l_8}
\mathcal{Q}_{1,2}\equiv\qn^{*2}=\,0.166667\pm\,0.096225\,i\,.
\end{equation}
Note that $|\mathcal{Q}_1|=|\mathcal{Q}_2|$. The corresponding spectrum of modes has been illustrated in figure~\ref{spectrum_before}.  As it can be seen in the right panel, in this special case, $\phi$ mode decays at the same rate as the sound modes decay at large momenta.
\par\bigskip 
\noindent

\subsection*{Spectrum at $\alpha>8$}
\begin{figure}[tb]
	\centering
	\includegraphics[width=0.55\textwidth]{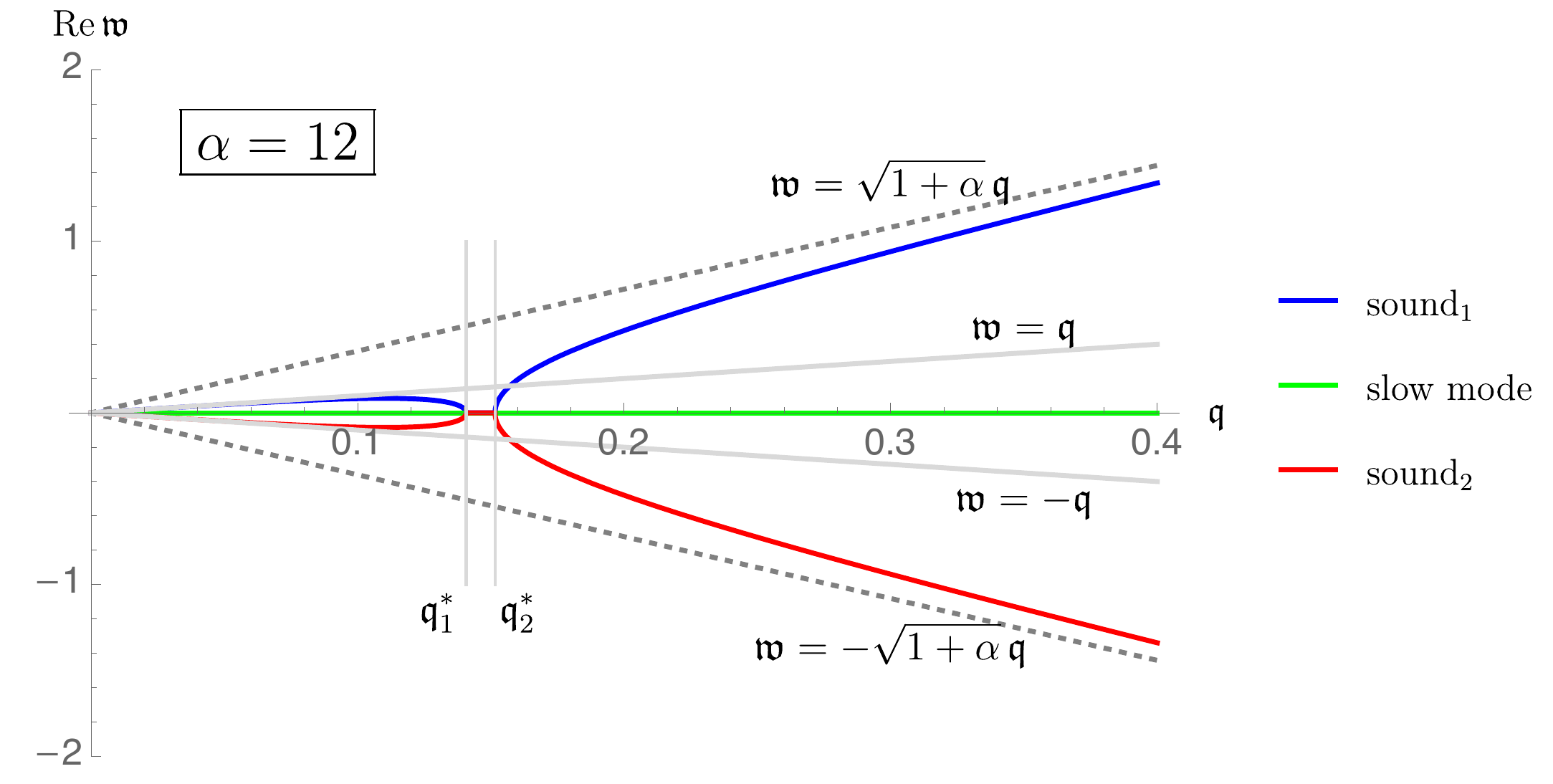}\includegraphics[width=0.55\textwidth]{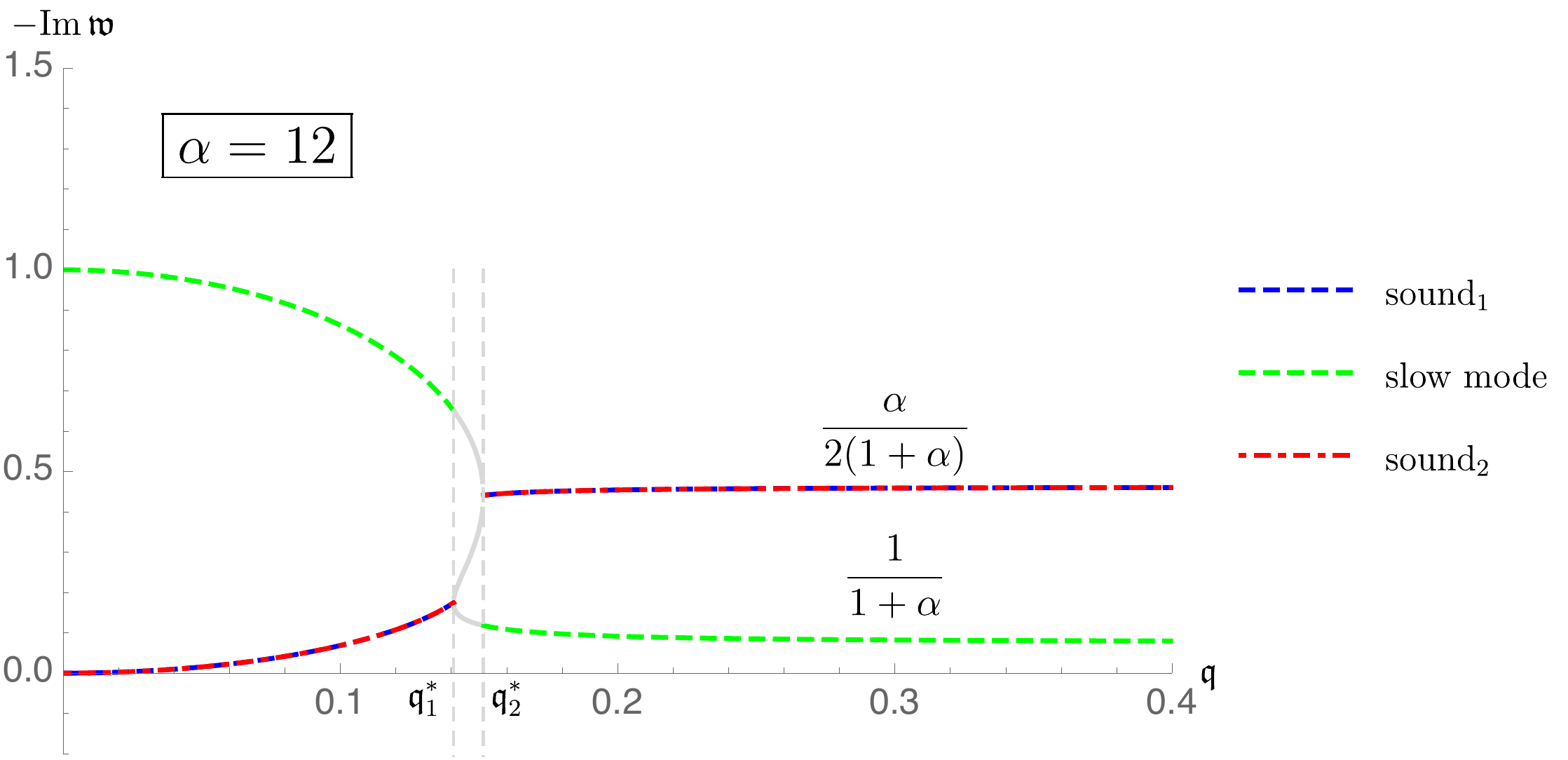}
	\caption{ {\it Dispersion relations for $\alpha > 8 $ (displayed is the example $\alpha=12$).} 
	There is a universal feature in this sector (for all $\alpha>8$): for $\qn > \qn_2^*$ it is the slow mode of Hydro$+$ (green curves) which is the longest-lived one. 
	Within the range $\qn^{*}_1<\qn<\qn^*_2$, all three modes become purely imaginary. 
	For this reason, one cannot distinguish between the type of modes in this range. 
	Hence, within this range, in the right panel we display all three modes as gray curves.
		}
		\label{spectrum_after}
\end{figure}

It turns out that for $\alpha>8$, the dispersion relations \eqref{omega_s} have two unequal real-valued singularities. In other words, $\qn^{*2}_1$ and $\qn^{*2}_2$ in \eqref{singularities} will be two real values in this range of $\alpha$.
In figure~\ref{spectrum_after} we have illustrated the spectrum of modes at $\alpha=12>8$ case.  We find
\begin{equation}\label{branch_alpha_g_8}
\mathcal{Q}_1=\qn^{*2}_1=\,0.03125,\,\,\,\,\mathcal{Q}_1=\qn^{*2}_2=\,0.03200\,.
\end{equation}

\par\bigskip 
\noindent
As it can be seen in the right panel of Figure.\ref{spectrum_after}, the $\phi$ mode is the slowest mode at large momenta, say at $\qn\gtrsim \qn_c$.

\subsection*{Spectrum at $\alpha=8$}
\begin{figure}[tb]
	\centering
	\includegraphics[width=0.55\textwidth]{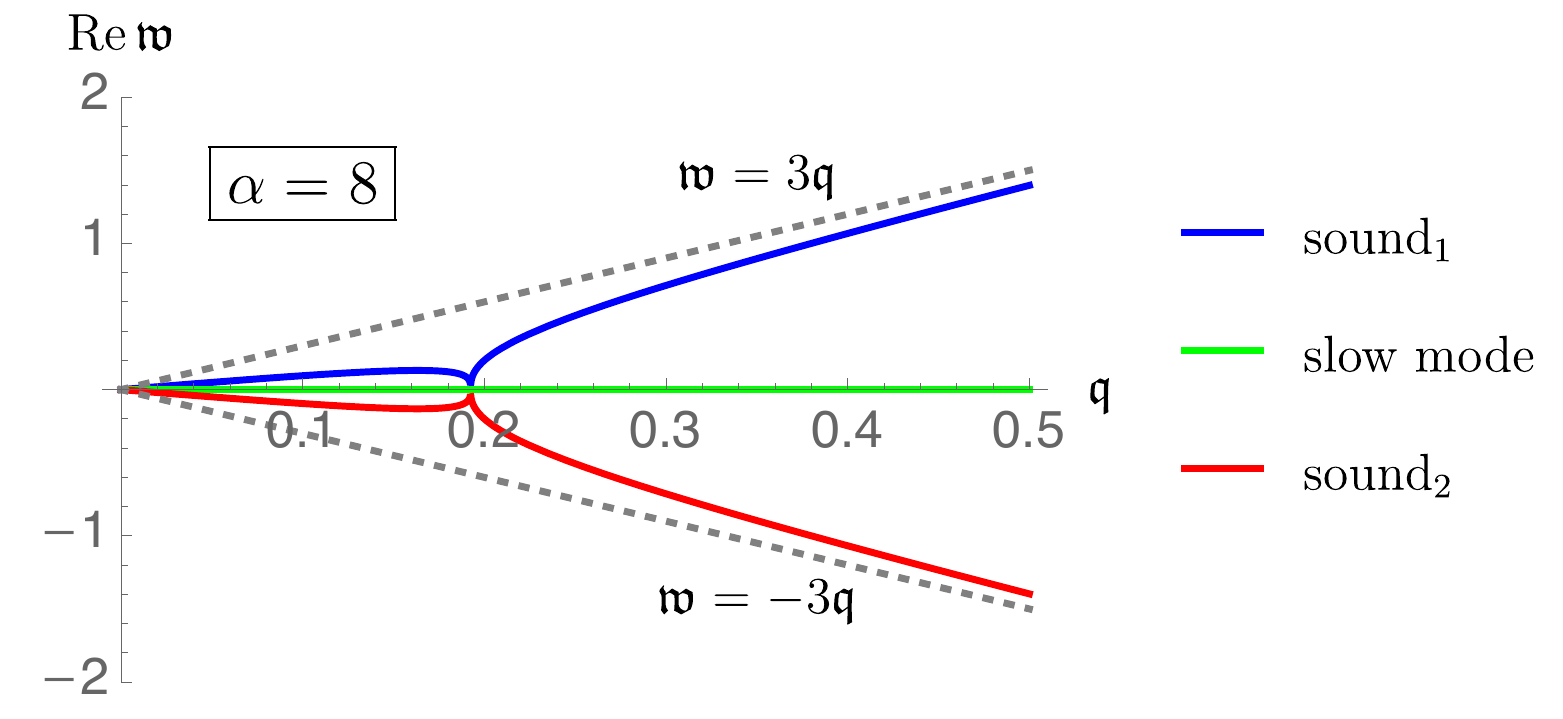}\includegraphics[width=0.55\textwidth]{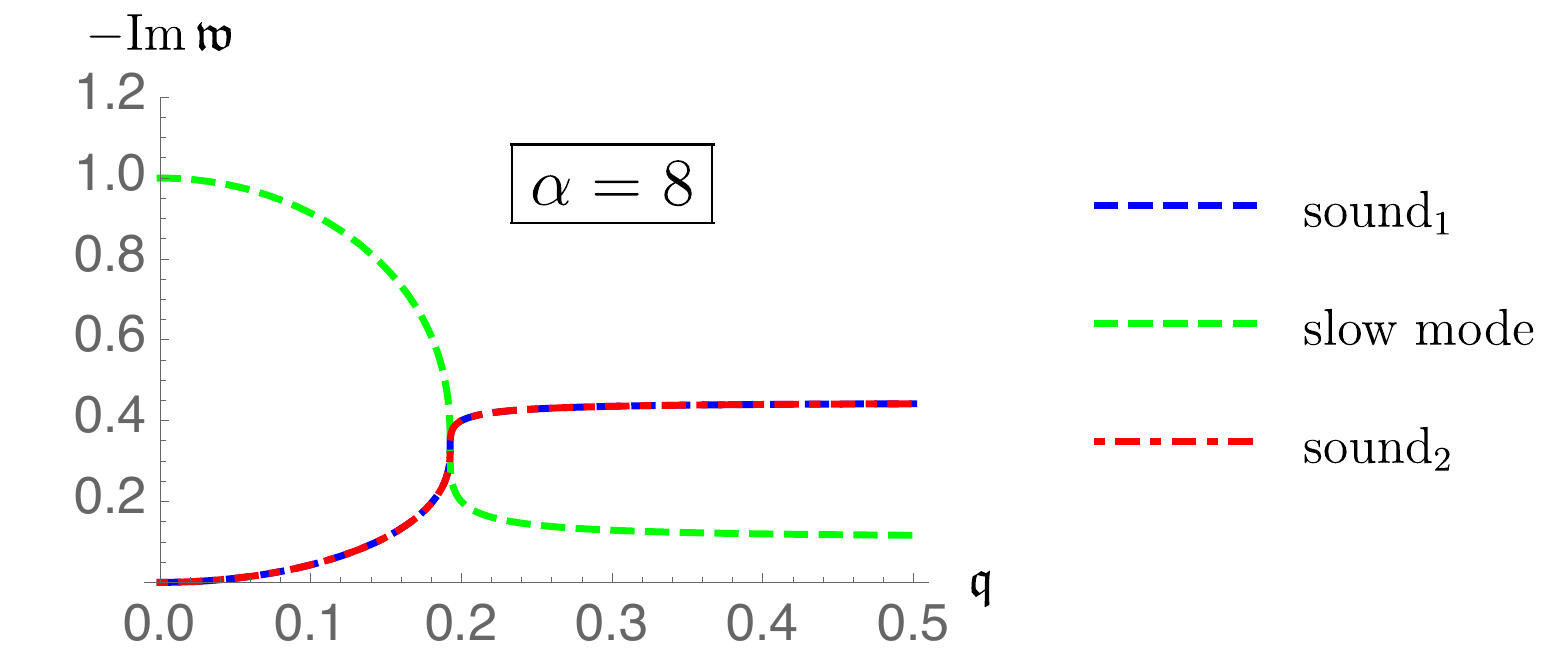}
	\caption{ {\it Dispersion relations for the degenerate case $\alpha = 8 $.}  
	At $\alpha=8$, the four branch point singularities become degenerate, which leads to $|\qn^{*}_1|=|\qn^*_2|$.
	}
	\label{spectrum_degenerate}
\end{figure}

At this special value of $\alpha$, the magnitudes of the square of the critical momenta coincide 
\begin{equation}\label{branch_alpha_eq_8} 
\mathcal{Q}_1=\mathcal{Q}_2=\frac{1}{27}\,.
\end{equation}
All four of the branch points coincide at this value of $\alpha$, see figure~\ref{spectrum_degenerate}. 

\par\bigskip 
\noindent
\section{Hydro+ near the critical point in the QCD phase diagram}
\label{QCD}
The important point about QCD plasma near the critical point in its phase diagram is that its partial-equilibrium states cannot be described only by the conserved densities.  Since the correlation length $\xi$ is much larger than the thermal equilibrium scale $\sim1/T$ there, the two point function of the density or even the higher point functions may be different from the corresponding values in complete equilibrium. Then the dynamics is given by the set of equations governing the evolution of conserved quantities together with those of their two and higher point functions. 

In the simplest setup, one only considers one and two point functions of the conserved densities, $\bar{\Psi}(\boldsymbol{x})$ and $\bar{G}(\boldsymbol{x}_1,\boldsymbol{x}_2)$. 
It is also important to note that in such states, entropy is a functional of  $\bar{\Psi}(\boldsymbol{x})$ and $\bar{G}(\boldsymbol{x}_1,\boldsymbol{x}_2)$:
\begin{equation}\label{entropy_functional}
S\equiv\,S\left[\bar{\Psi}, \bar{G}\right]
\end{equation}

In partial-equilibrium states,  $\bar{\Psi}(\boldsymbol{x})$ is a slowly-varying function of $\boldsymbol{x}$ and similarly $\bar{G}(\boldsymbol{x}_1,\boldsymbol{x}_2)$ is slow on $(\boldsymbol{x}_1+\boldsymbol{x}_2)/2$, associated with scales larger than $\ell$, which is much larger than the scale $\xi$ of $|\boldsymbol{x}_1-\boldsymbol{x}_2|$ dependence.

The above \textbf{separation of scales} is used to perform a Wigner transformation of $G$:
\begin{equation}\label{}
G_{\boldsymbol{Q}}(\boldsymbol{x})=\,\int_{\Delta\boldsymbol{x}} G\left(\boldsymbol{x}+\frac{\Delta \boldsymbol{x}}{2}, \boldsymbol{x}-\frac{\Delta \boldsymbol{x}}{2}\right)e^{i \boldsymbol{Q} \Delta{\boldsymbol{x}}}.
\end{equation}
In fact, $G_{\boldsymbol{Q}}(\boldsymbol{x})$ characterizes states which vary slowly with respect to $\boldsymbol{x}$ compared to $\boldsymbol{Q}$. In other words
\begin{equation}\label{separation_of_scales}
1/\ell\,\ll\,Q \, .
\end{equation}
For such partial-equilibrium states, the problem simplifies in the sense that the entropy functional \eqref{entropy_functional}, \textbf{or equivalently the 2PI effective action}, can be written as a \textbf{local functional} of $G_{\boldsymbol{Q}}(\boldsymbol{x})$ \cite{Stephanov:2017ghc}: 
\begin{equation}\label{S_2}
S_2[\bar{\Psi}, G]\approx S_1[\bar{\Psi}]+\frac{1}{2}\int_{\boldsymbol{x}}\int_{\boldsymbol{Q}}\text{Tr}(1- C_{\boldsymbol{Q}}G_{\boldsymbol{Q}}+\log C_{\boldsymbol{Q}}G_{\boldsymbol{Q}})
\end{equation}
where $S_1$ is the 1PI effective action and  $C_{\boldsymbol{Q}}=\bar{G}_{\boldsymbol{Q}}^{-1}$.

Now we can consider the \textit{slowest} eigenmode of $G_{\boldsymbol{Q}}(\boldsymbol{x})$ locally and call it
\begin{equation}\label{}
\phi_{\boldsymbol{Q}}(\boldsymbol{x})=\,\text{the slowest eigenmode of} \,\,G_{\boldsymbol{Q}}(\boldsymbol{x})
\end{equation}
Substituting it back into \eqref{S_2}, one finds
\begin{equation}\label{}
S_2[\bar{\Psi}, G]\approx S_1[\bar{\Psi}]+\frac{1}{2}\int_{\boldsymbol{x}}\int_{\boldsymbol{Q}}(1-\phi_{\boldsymbol{Q}}/\bar{\phi}_{\boldsymbol{Q}}+\log(\phi_{\boldsymbol{Q}}/\bar{\phi}_{\boldsymbol{Q}}))
\end{equation}
Now, in addition to the conserved densities $\bar{\Psi}$, there is another degree of freedom $\phi_{\boldsymbol{Q}}$. The local equilibrium value of this mode is denoted by $\bar{\phi}_{\boldsymbol{Q}}$ which is related to local value of the hydrodynamic degrees of freedom $\bar{\Psi}$. 

The equation of motion for $\phi_{\boldsymbol{Q}}$ is given by \cite{Stephanov:2017ghc}
\begin{equation}\label{phi_Q_EoM}
D\phi_{\boldsymbol{Q}}=\,-\Gamma_{\boldsymbol{Q}}\left(\phi_{\boldsymbol{Q}}-\bar{\phi}_{\boldsymbol{Q}}\right)
\end{equation}
where\footnote{Note that equation \eqref{phi_Q_EoM} can be also written in the form \eqref{phi_EoM}.} 
$\bar{\phi}_{\boldsymbol{Q}}$ can be approximated by \cite{Rajagopal:2019xwg}
\begin{equation}\label{}
\bar{\phi}_{\boldsymbol{Q}}\approx \frac{c_M\,\xi^2}{1+ (Q\xi)^2}\,.
\end{equation}
In this equation, $c_M$ is a constant and $\xi$ denotes the equilibrium correlation length.
The leading behavior of $Q$-dependent of $\Gamma_{\boldsymbol{Q}}$ near the critical point is given by \cite{Pradeep:2021opj}
\begin{equation}\label{decay_rate}
\Gamma(\boldsymbol{Q})=\frac{2D_0 \xi_0}{\xi^3}K(Q\xi)\,,
\end{equation}
where  $\xi_0$ is the value equilibrium correlation length far from the critical point and $K(x)=\frac{3}{4}[1+x^2+(x^3-x^{-1})\arctan(x)]$ \cite{Stephanov:2017ghc}. 
In our later computations, we will consider two distinct values for $D_0$:
\begin{equation}\label{}
D_0=\, 0.1, \,0.5\, \text{fm}\,. 
\end{equation}
In order to parameterize $\xi$,  we follow  \cite{Rajagopal:2019xwg} and write
\begin{equation}\label{xi_xi_0}
\left(\frac{\xi}{\xi_0}\right)^{-2}=\sqrt{\tanh^2\left(\frac{T-T_c}{\Delta T}\right)\left(1-\left(\frac{\xi_{\text{max}}}{\xi_0}\right)^{-4}\right)+\left(\frac{\xi_{\text{max}}}{\xi_0}\right)^{-4}}\,,
\end{equation}
with  $\xi_0=0.5\,\text{fm}$.
In this ansatz, the equilibrium correlation length, $\xi$, increases to a maximum value $\xi_{\text{max}}$, then decreases to its value at freeze-out. 
The finite value of $\xi_{\text{max}}$ indicates that the trajectory of the QGP droplet is close to the critical point, instead of running directly through it. We will consider two distinct values for $\xi_{\text{max}}$:
	\begin{equation}\label{}
	\frac{\xi_{\text{max}}}{\xi_0}=\,2,6\,.
	\end{equation}
It is clear that by approaching the critical point, $\xi_{\text{max}}$ increases. Very close to the critical point, namely in the scaling limit, $\xi_{\text{max}}$ diverges 
(see \sec{sec_critical_point} for comparison with the case of passing through the critical point).

In summary, we see that near the QCD critical point, a slowly varying variable $\phi_{\boldsymbol{Q}}$ indexed by a continuous index $\boldsymbol{Q}$ does not take its equilibrium value $\bar{\phi}_{\boldsymbol{Q}}(\epsilon, n)$ for given $\epsilon$ and $n$. In  the following subsections we briefly review what exactly this mode is and how it back-reacts on the QGP droplet near the critical point.

\subsection*{What is the lowest mode?}
The eigenmodes of the two point function of conserved densities, namely $G$, can be found by studying the linear perturbations on top of thermal equilibrium.
Considering the linearized hydrodynamic equations as 
\begin{equation}\label{}
D \Psi=\,-L\delta \Psi +\mathcal{O}(\delta \Psi^2),
\end{equation}
the linearized evolution equation for 2-point function is given by the following matrix form equation:
\begin{equation}\label{G_equation}
\partial_t G=\,-L(G-\bar{G})-(G-\bar{G})L^{\dagger}+ \mathcal{O} (G-\bar{G})^2
\end{equation}
As discussed earlier, in the limit $1/\ell\ll Q$, G can be replaced with local $G_{\boldsymbol{Q}}$ modes. Then equation \eqref{G_equation} indicates that the slowest $G_{\boldsymbol{Q}}$ mode, namely $\phi_{\boldsymbol{Q}}$,  corresponds to the smallest eigenvalue of the matrix $L$. It is easy to show that the smallest eigenvalue of $L$ is $0$. Then the projection of an arbitrary perturbation $\delta \Psi$ on the corresponding eigenvector turns out to be proportional to the function $m=s/n$, i.e. the ratio of the entropy density to the baryon density \cite{Stephanov:2017ghc}. Thus one finds that the $\phi_{\boldsymbol{Q}}$ mode is proportional to the two point function of fluctuations of $\delta m$:
\begin{equation}\label{}
\phi_{\boldsymbol{Q}}(\boldsymbol{x})\sim \int_{\Delta \boldsymbol{x}}\bigg\langle\delta m\left(\boldsymbol{x}+\frac{\Delta \boldsymbol{x}}{2}\right)\delta m\left(\boldsymbol{x}-\frac{\Delta \boldsymbol{x}}{2}\right)\bigg\rangle\,.
\end{equation}
See Appendix C of reference~~\cite{Stephanov:2017ghc} for an interesting discussion about the arbitrariness in normalization of $\phi_{\boldsymbol{Q}}$.

\subsection*{Feedback from the slow mode on $c_s^2$}
Following the discussion in \sec{single_section}, we simply find that in the present case, equation \eqref{Delta_cs_2_single} takes the following form~\cite{Rajagopal:2019xwg}
\begin{equation}\label{Deltacs2}
\Delta c_s^2(\omega)\approx\,\frac{c_s^4}{2s}\int\frac{d^3\boldsymbol{Q}}{(2\pi)^3}\,\,[f_2(Q\xi)]^2\,\left(\frac{\xi}{\xi_0}\right)^4\,\left(T\frac{\partial}{\partial T}\left(\frac{\xi}{\xi_0}\right)^{-2}\right)^2\frac{\omega^2}{\omega^2+\Gamma^2_{\boldsymbol{Q}}}\,,
\end{equation}
with the decay rate $\Gamma_{\boldsymbol{Q}}$ given by \eqref{decay_rate}. 

\textit{Ultimately, what  we want to do is to show how the 
characteristic momentum of Hydro+ discussed in \sec{single_section} limits the range of momenta of the critical fluctuations contributing to $\Delta c_s^2$, discussed in the previous paragraph.}

Our strategy is to think of any of the points near the critical point (in the phase diagram) as one distinct equilibrium state. In other words, we neglect the time-evolution of the QGP droplet on a specific trajectory through the phase space. Then for each of these equilibrium states, the calculation of \sec{single_subsection} is applicable. 
One may consider this an {\it adiabatic} approximation. 
Needless to say that in this case, the temperature dependence of single-mode Hydro+ quantities will be through the special parameterization of $\xi$ in \eqref{xi_xi_0} as well as the equation of state we will construct below.    
It should also be emphasized that  \eqref{xi_xi_0} ignores the dependence of the state on the baryon density. We just consider  states near $n_{B}=0$ in the phase diagram. 

\subsection*{Radius of convergence of derivative expansion near the QCD critical point }
Near the QCD critical point, equation \eqref{phi_EoM} is replaced with \eqref{phi_Q_EoM}. Correspondingly, $\alpha$, given by equation \eqref{dimensionless} becomes a function of $\boldsymbol{Q}$ in this case; we will refer to it as $\alpha_{\boldsymbol{Q}}$. This function can be simply extracted from \eqref{Deltacs2} by reading off the individual contribution of each mode to the integral
\begin{equation}\label{Deltacs2Q}\boxed{
\alpha_{\boldsymbol{Q}}	(\omega\gg\Gamma_{\boldsymbol{Q}})=\frac{\Delta c_{s,\boldsymbol{Q}}^2(\infty)}{c_s^2}\approx\,\frac{c_s^2}{2s}\,\frac{Q^2\,\Delta Q}{2\pi^2}\,[f_2(Q\xi)]^2\,\left(\frac{\xi}{\xi_0}\right)^4\left(T\frac{\partial}{\partial T}\left(\frac{\xi}{\xi_0}\right)^{-2}\right)^2} \, , 
\end{equation}
where $\Delta Q$ is the range over which we take $Q$ to be approximately constant. 

In order to use our theoretical results obtained from single-mode Hydro$+$, namely \eqref{singularities}, we need to know whether $\phi_{\boldsymbol{Q}}$ falls into the $\alpha<8$ or into the $\alpha>8$ sub-sector. So our main task is to evaluate $\alpha_{\boldsymbol{Q}}$ near the QCD critical point.
To this end, three things need to be determined: 1) $\Delta Q$, 2) the entropy density $s$ and 3) the sound velocity $c_s$. 
In our numerical computations, we take $\Delta Q=0.002\, \text{fm}^{-1}$.

\subsection*{The entropy density $s$ and the sound velocity $c_s$}
\begin{figure}[tb]
	\centering
	\includegraphics[width=0.48\textwidth]{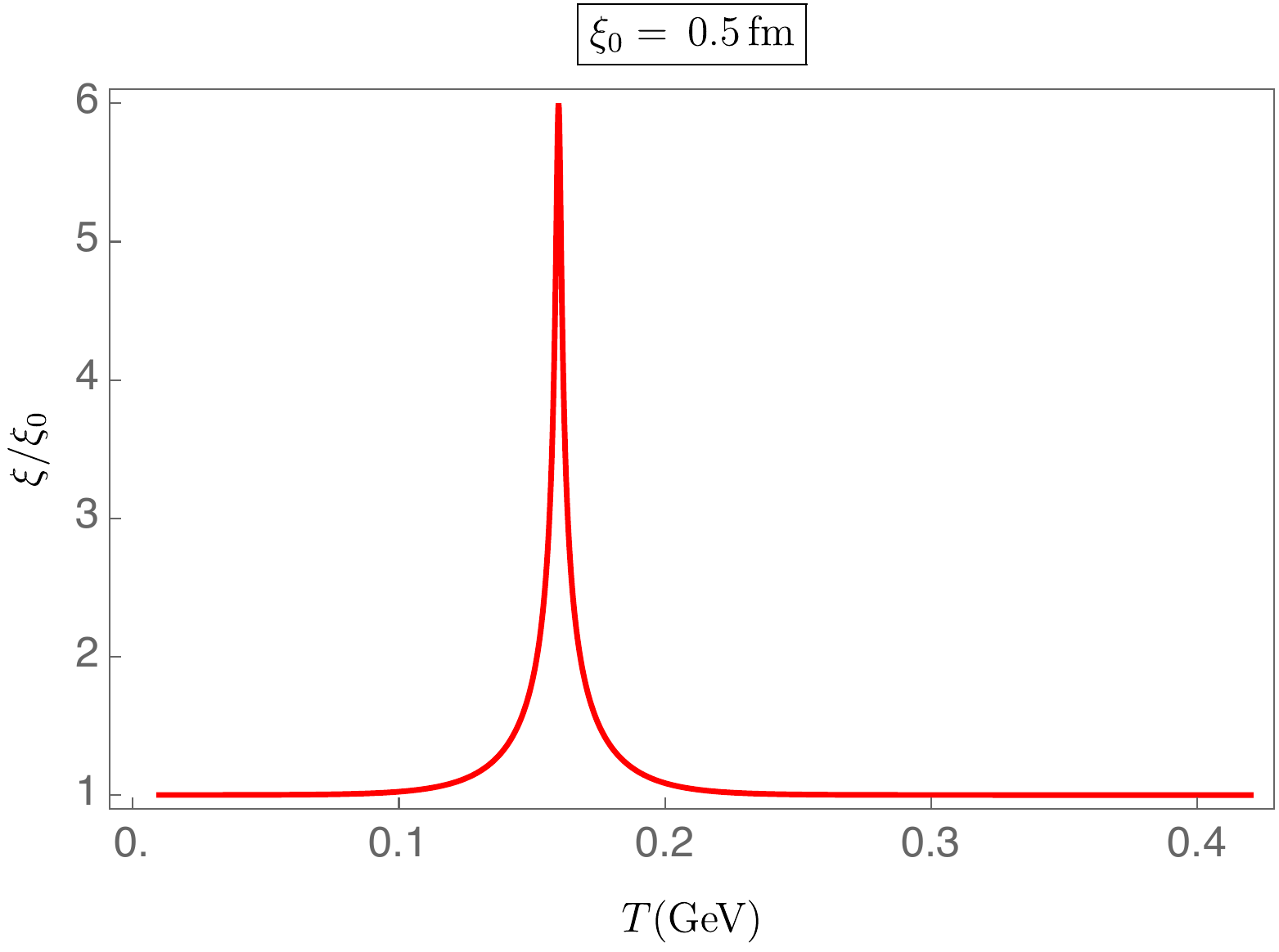}\,\,\includegraphics[width=0.48\textwidth]{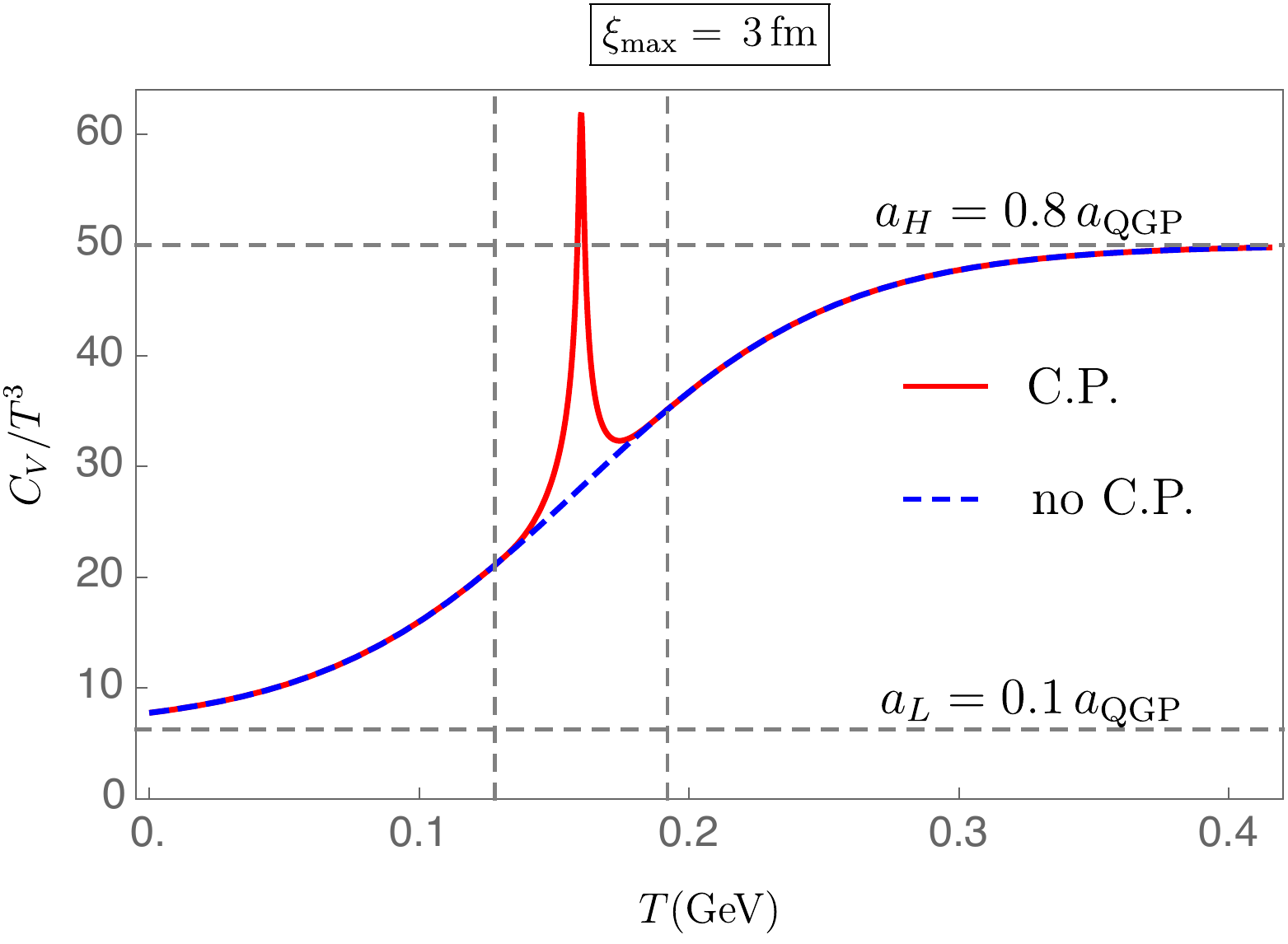}
	\includegraphics[width=0.48\textwidth]{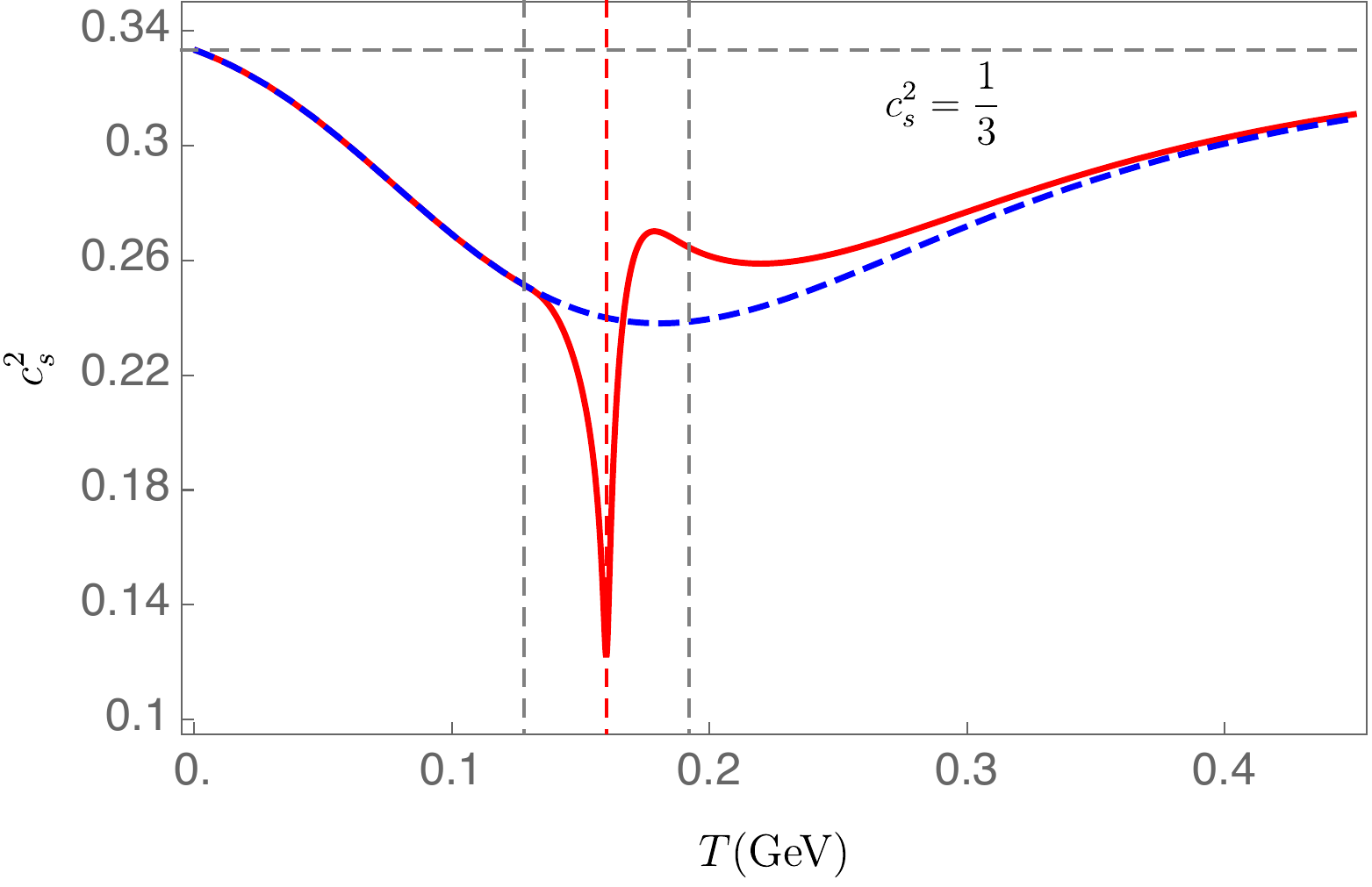}\,\,\includegraphics[width=0.48\textwidth]{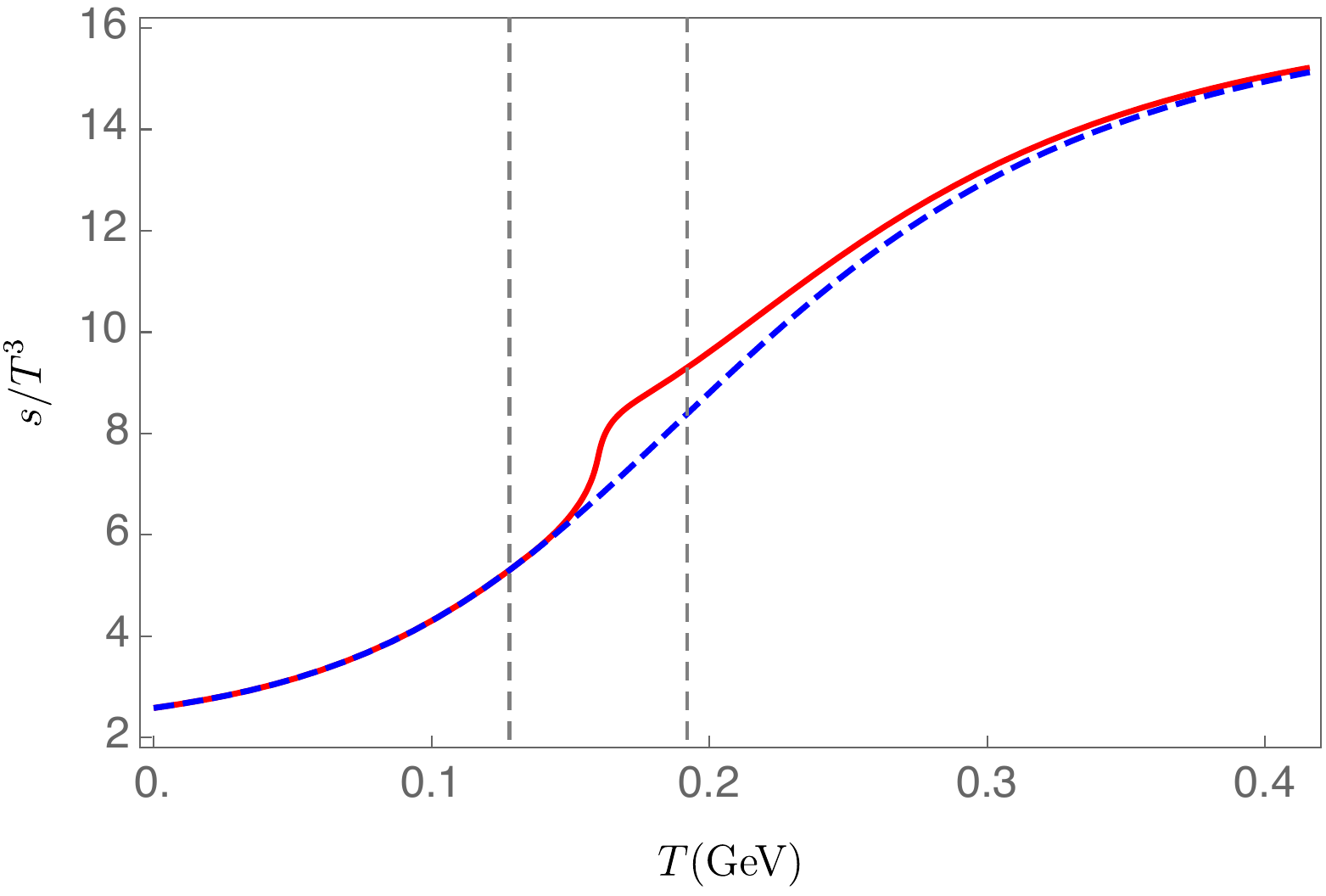}
	\caption{ The left (right) vertical dashed line shows the location of $T_L = T_c-\Delta T\,(T_H = T_c+\Delta T )$. In all panels, solid red curves represent thermodynamic quantities taking the critical contributions into account, 
	dashed blue curves represent results without critical contributions. 
	In the bottom left panel, the red dashed line indicates the critical temperature, $T_c=0.16\, \text{GeV}$.  }
	\label{EoS}
\end{figure}

In order to compute the entropy density, one needs to construct the thermodynamic equation of state. Without taking into account the effect of the critical point, $c_V$ may be expressed by the following ansatz~\cite{Rajagopal:2019xwg}
\begin{equation}\label{c_V_no_C_P}
\frac{c_V^{\text{no\,C.P.}}}{T^3}=\,\left[\left(\frac{a_H+a_L}{2}\right)+\left(\frac{a_H-a_L}{2}\right)\tanh
\left(\frac{T-T_{\text{C.O.}}}{\Delta T_{\text{C.O.}}}\right)\right]\,,
\end{equation}
which interpolates between two temperature independent asymptotic values $a_L$ and $a_H$, corresponding to the value of $c_V/T^3$ at $T_L=T_{c}-\Delta T$ and  $T_H=T_{c}+\Delta T$, respectively. We assume $\Delta T= 0.2 T_c$.
In \eqref{c_V_no_C_P}, the data corresponding to the crossover is given by
\begin{equation}\label{}
T_\text{C.O.}=\,T_c\,\,\,\,\text{and} \,\,\,\,\Delta T_{\text{C.O.}}=\,0.6\, T_c\,.
\end{equation}
Similar to reference~\cite{Rajagopal:2019xwg}, we also choose 
\begin{equation}\label{}
a_L=\,0.1\,a_{\text{QGP}},\,\,\,\,\,a_H=\,0.8\,a_{\text{QGP}}\,,
\end{equation}
where $a_{\text{QGP}}$ is the value of $c_V/T^3$ for the non-interacting ideal gas QGP, given by
\begin{equation}\label{}
a_{\text{QGP}}=\,\frac{4\pi^2(N_c^2-1)+21 \pi^2 N_f}{15}\,,
\end{equation}
with  $N_c=3$ and $N_f=3$ corresponding to the number of colors and flavors, respectively.\
Having specified $c_V$, one can then directly compute $s$ and then $c_s^2$ as the following
\begin{equation}\label{s_cs_2}
\begin{split}
s(T)=&\,\int_{0}^{T}dT'\frac{c_V(T')}{T'}\,,\\
c_s^2=&\,\frac{s}{c_V}\,.
\end{split}
\end{equation}
These functions are displayed as blue-dashed curves in figure~\ref{EoS}. However, we have not considered the effect of the critical point so far. For this reason we refer to these curves as the no-critical-point (no C.P.) result.
%
\par\bigskip 
\noindent

In order to construct $c_V$ near the critical point, we need to include the critical contribution. We follow~\cite{Rajagopal:2019xwg} and take the textbook result $C_V\propto\xi$~\cite{Kardar} to write
\begin{equation}\label{}
c_V^{\text{crit}}(T)=\frac{1}{2}\frac{1}{\xi_0^3}\frac{\xi(T)}{\xi_0}\,.
\end{equation}
with $\xi(T)/\xi_0$ given by \eqref{xi_xi_0} \footnote{See \sec{sec_critical_point} for another construction of the EoS near the critical point.}. 
It is expected that the effect of critical fluctuations on $c_V$ is only important  near the critical point. In addition, the value of $c_V$ should approach that without a critical point far from $T_c$. So, $c_V$ can be constructed as follows~\cite{Rajagopal:2019xwg}
\begin{equation}\label{C_V}
c_V(T)=\begin{cases}
c_V^{\text{no C.P.}}(T)\,, \quad & \, T \le T_L \\
 c_V^{\text{cirt}}(T)+T^3\sum_{n=0}a_n\left(\frac{T-T_c}{\Delta T}\right)^n\,, \quad & \, T_L\le T\le T_H \\
c_V^{\text{no C.P.}}(T)\,, \quad & \, T \ge T_H
\end{cases}\,.
\end{equation}
Demanding $c_V/T^3$ and its derivatives to be continuous at $T=T_{L,H}$, one can specify coefficients $a_n$ to any arbitrary order. We choose to go up to the second order in derivatives. We then find six equations leading to six non-zero coefficients\footnote{For $\xi_{\text{max}}=6\xi_{0}=3\, \text{fm}$, we find
$a_0=17.16,\,\,a_1=13.80,\,\,a_2=1.29\,\,a_3=-1.54,\,\,a_4=-1.16,\,\,a_5=0.64$.}.
Having specified $c_V$ (see the red curve in the top right panel in figure~\ref{EoS}), one can use \eqref{s_cs_2} to find $s(T)$ and $c_s^2$. The result is given by red curves in the two bottom panels of figure~\ref{EoS}.

Note that in our present case the sound velocity does not approach zero as $T\rightarrow T_c$. The reason is that the states we are considering here are near (never at) the critical point in the phase diagram. In appendix \ref{sec_critical_point}, we will show how the behavior of above quantities change when passing through the critical point.

\subsection*{Computing $\alpha_{\boldsymbol{Q}}$}
\begin{figure}
		\centering
	\includegraphics[width=0.45\textwidth]{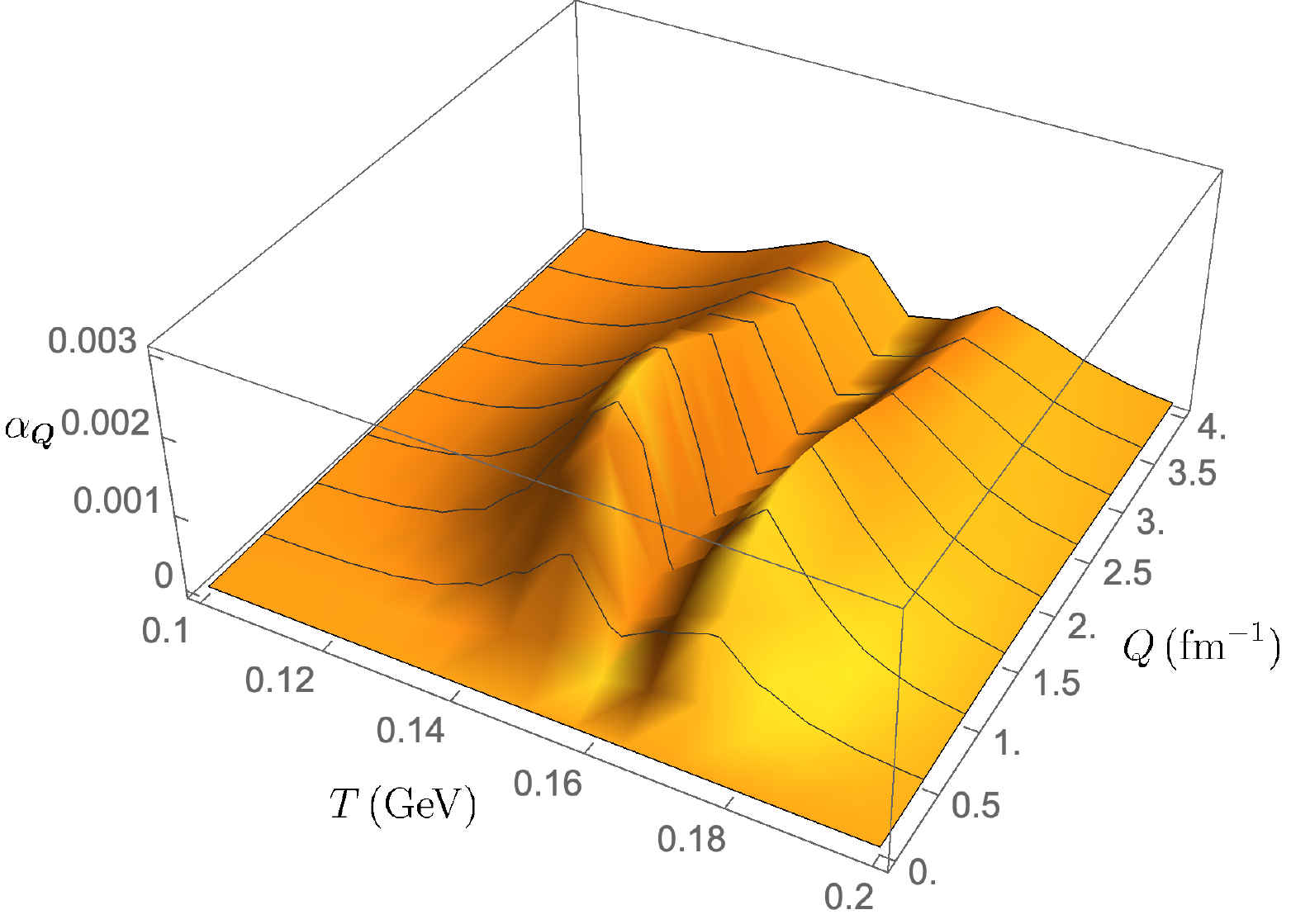}	\includegraphics[width=0.45\textwidth]{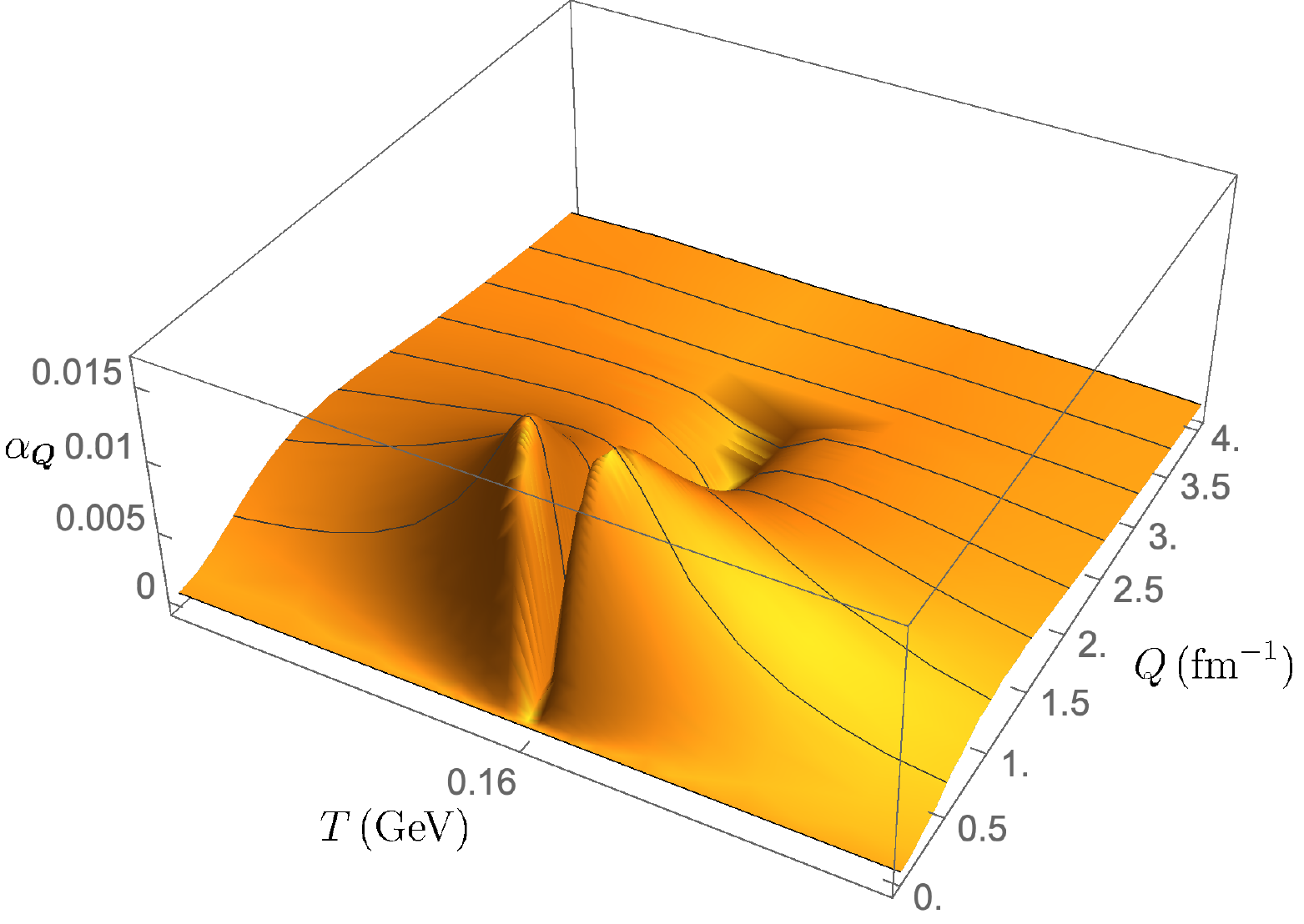}
	\caption{
	The contribution, $\alpha_{\boldsymbol{Q}}$, of each mode $\phi_{\boldsymbol{Q}}$ (labelled by the magnitude $Q$) to the stiffness as a function of temperature.  
	The left panel corresponds to  $\xi_{max}=1 \,\text{fm}$, and the right panel to $\xi_{\text{max}}=3\,\text{fm}$.  
	Values of $\alpha_{\boldsymbol{Q}}$ range from $\alpha_{\boldsymbol{Q}}<8$ to $\alpha_{\boldsymbol{Q}}>8$ with a transition between these two regimes occurring at $\alpha_{\boldsymbol{Q}}=8$, as described in section~\ref{single_subsection}. 
	}
	\label{alpha}
\end{figure}

Considering the  specifications mentioned above, in figure~\ref{alpha} we display~$\alpha$ as a function of $T$ for various $\phi_{\boldsymbol{Q}}$ modes in two cases. The left panel corresponds to $\xi_{\text{max}}=1\,\text{fm}^{-1}$ while the right panel  corresponds to $\xi_{\text{max}}=3\,\text{fm}^{-1}$.

For any mode, each plot features two peaks around the critical temperature. The presence of the two peaks can be understood as follows. Approaching the critical temperature, the correlation length $\xi$ increases. At the same time, the rate of change of $\xi$, namely $T \partial_T (\xi/\xi_0)$ increases, too. However, at some point, the rate of the latter begins to decrease, resulting in $T \partial_T (\xi/\xi_0)=0$ at the critical temperature. This can be obviously seen in the top left panel of figure~\ref{EoS}.  Therefore, this is the competition between the last two factors in eq.~\eqref{Deltacs2Q} that leads to the appearance of the peaks in figure~\ref{alpha}. On the other hand, at larger values of $\xi_{\text{max}}$, the maximum value of $\xi/\xi_0$ in figure~\ref{EoS} will tend to become a singularity. Then the interval between the two peaks becomes smaller and the peaks become sharper.
	
\section{Constraint on the stiffness of EoS from radius of convergence}
\label{stiffness}

Having found the value of $\alpha_{\boldsymbol{Q}}$ as a function of temperature, we are now ready to evaluate \eqref{singularities}. We want to apply these equations to each of the $\phi_{\boldsymbol{Q}}$-modes and find the characteristic momentum $q_c$, associated with that mode, as a function of temperature. 

Because of the transition between the $\alpha<8$ and $\alpha>8$ cases, shown in figure~\ref{alpha}, we should be careful how to use \eqref{singularities}. As discussed earlier, for $\alpha<8$, the convergence radius is set as  $\qn_c=|\qn^*_1|=|\qn^*_2|$.
For $\alpha>8$, however, the correct value of $\qn_c$ at each temperature is determined by the minimum value of $|\qn^*_1|$ and $|\qn^*_2|$ at that temperature: 
\begin{equation}\label{q_c}
\qn_c=\min\{|\qn^*_1|,|\qn^*_2|\}\,.
\end{equation}
Note that in the present case, $|\qn^*_1|$ and $|\qn^*_2|$ are functions of $T$ and $Q$ through the dependence on $\alpha_{\boldsymbol{Q}}(T)$.

\begin{figure}[tb]
	\centering
	\includegraphics[width=0.44\textwidth]{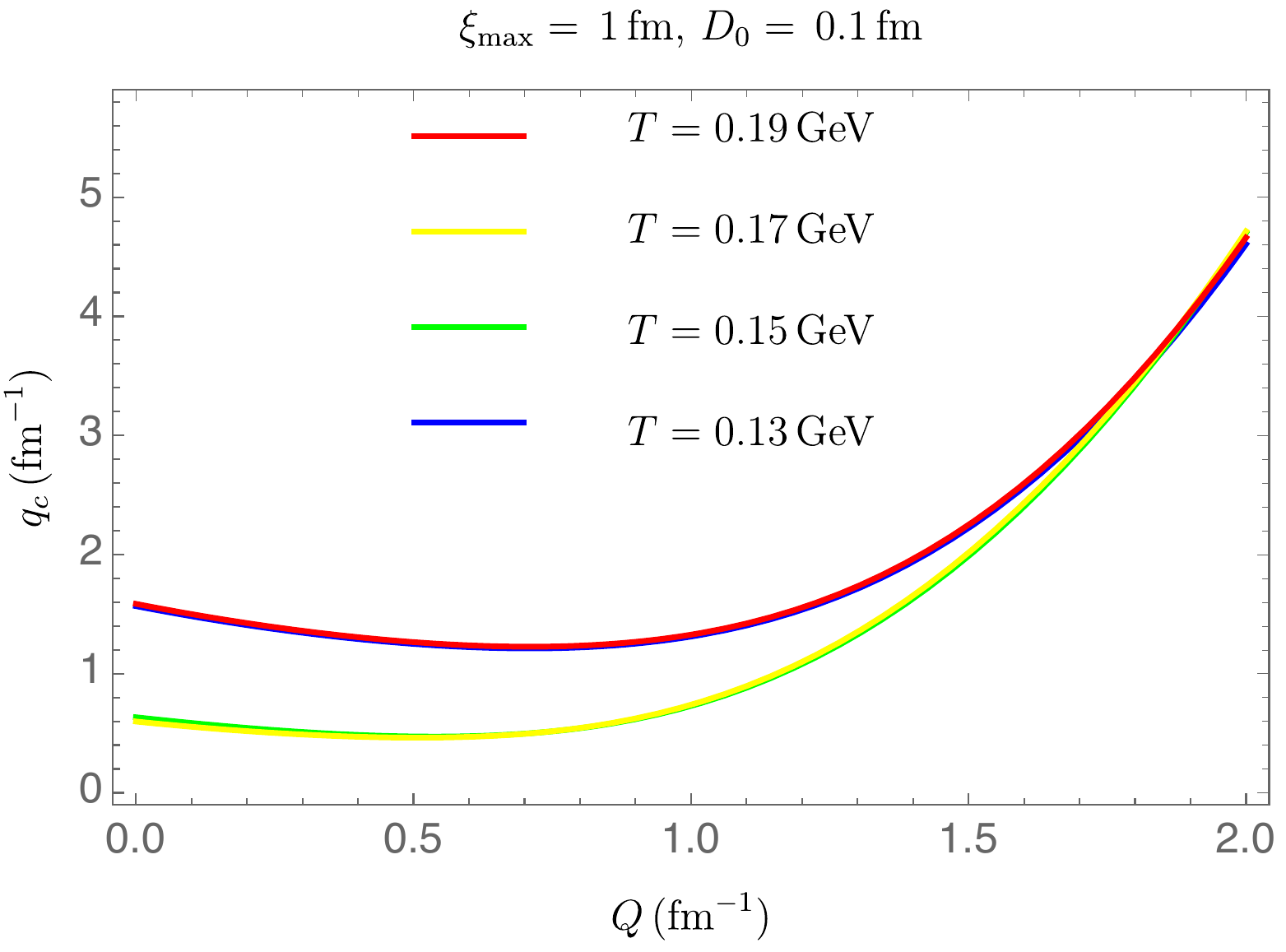}	\includegraphics[width=0.44\textwidth]{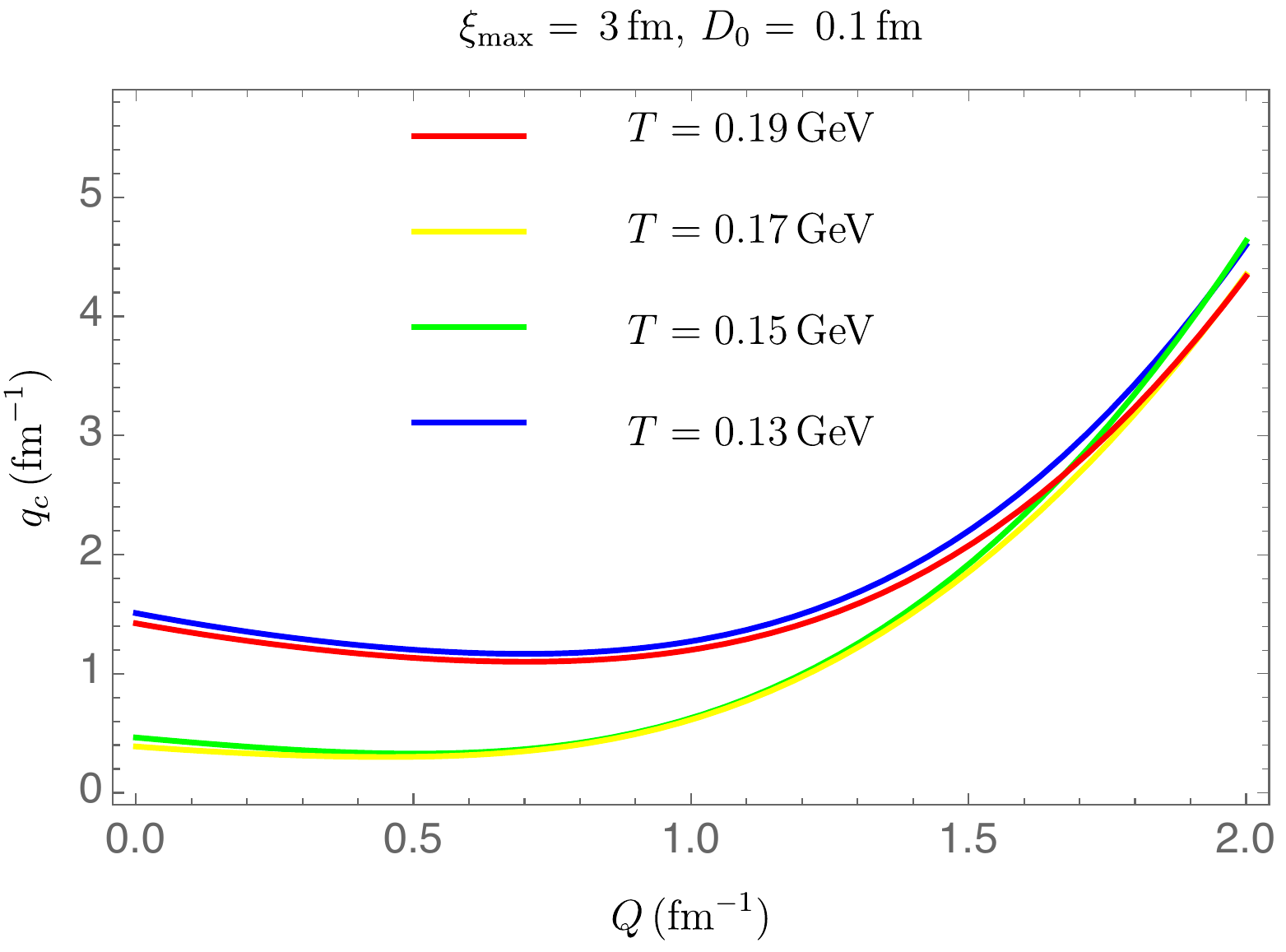}
	\includegraphics[width=0.44\textwidth]{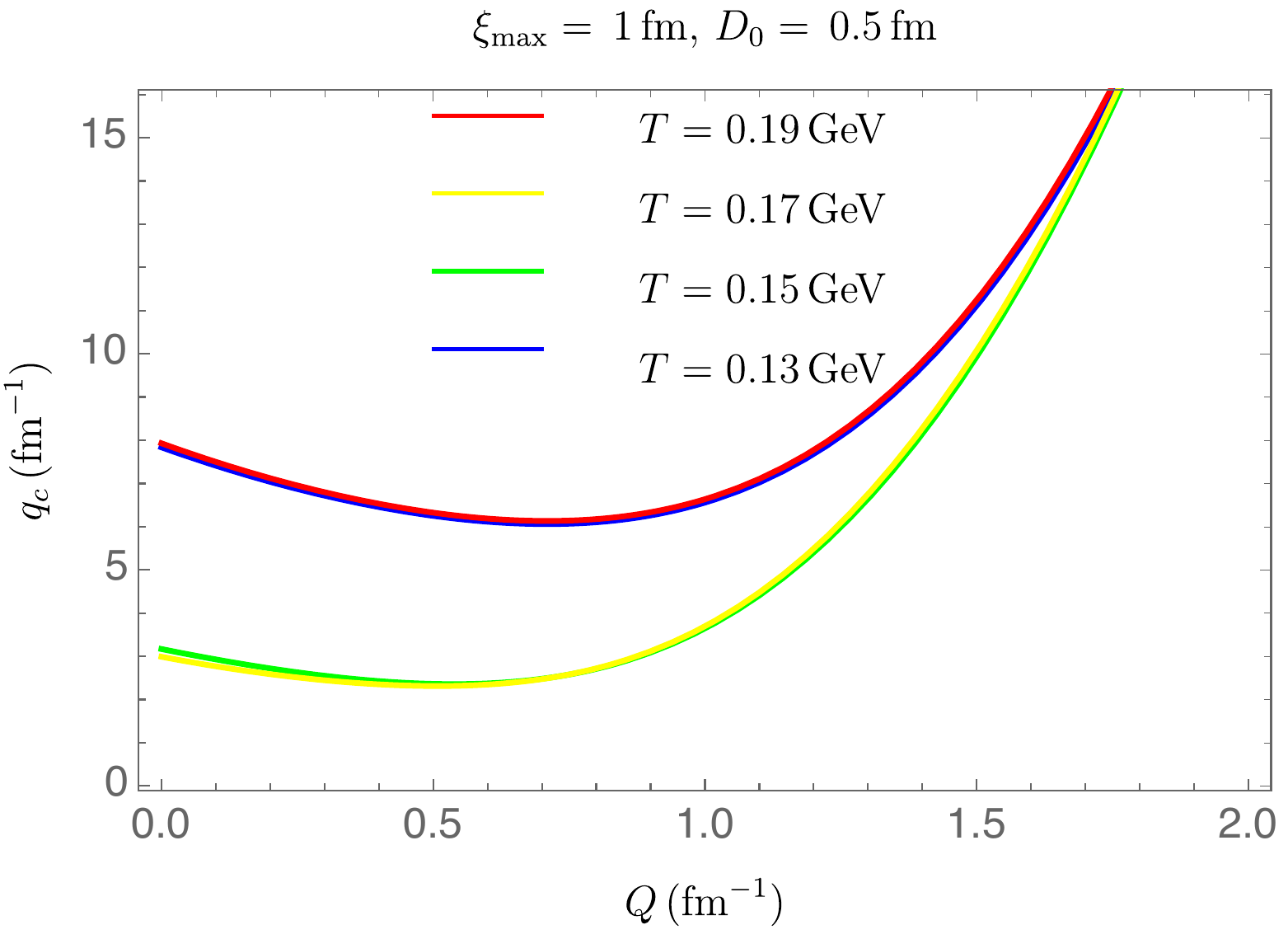}	\includegraphics[width=0.44\textwidth]{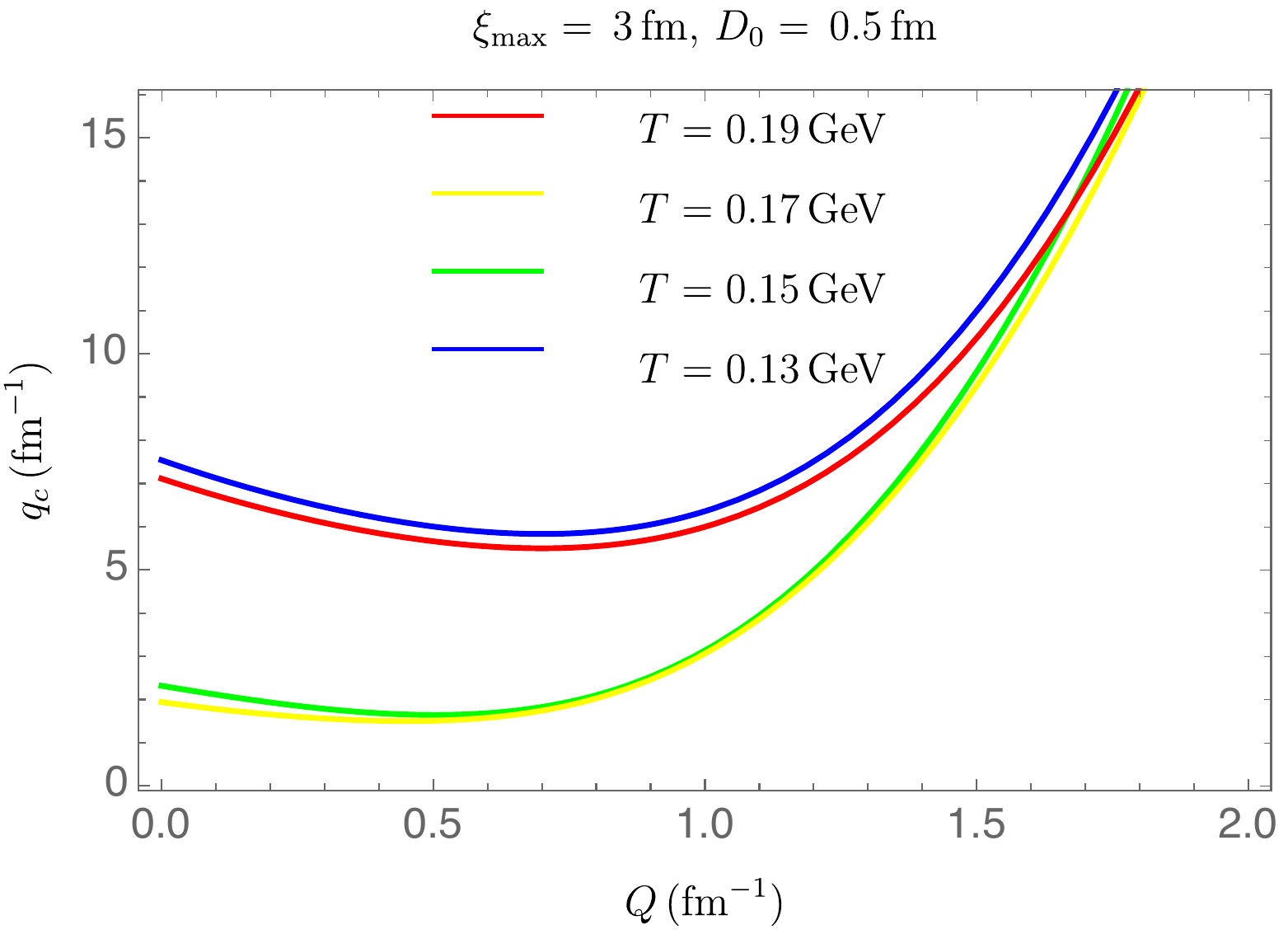}
	\caption{The characteristic momentum $q_c$ at four temperatures near $T_c$ as a function of the momentum $Q$ labeling the critical fluctuation. From the left to the right panel, $\xi_{\text{max}}$ has tripled. As expected, this caused the value of $q_c$ to decrease. 
	This is because the relaxation rate of modes in the right panels is four time larger than that of the corresponding modes in the left panels. }
	\label{q_c_3d}
\end{figure}

\begin{figure}[tb]
	\centering
	\includegraphics[width=0.45\textwidth]{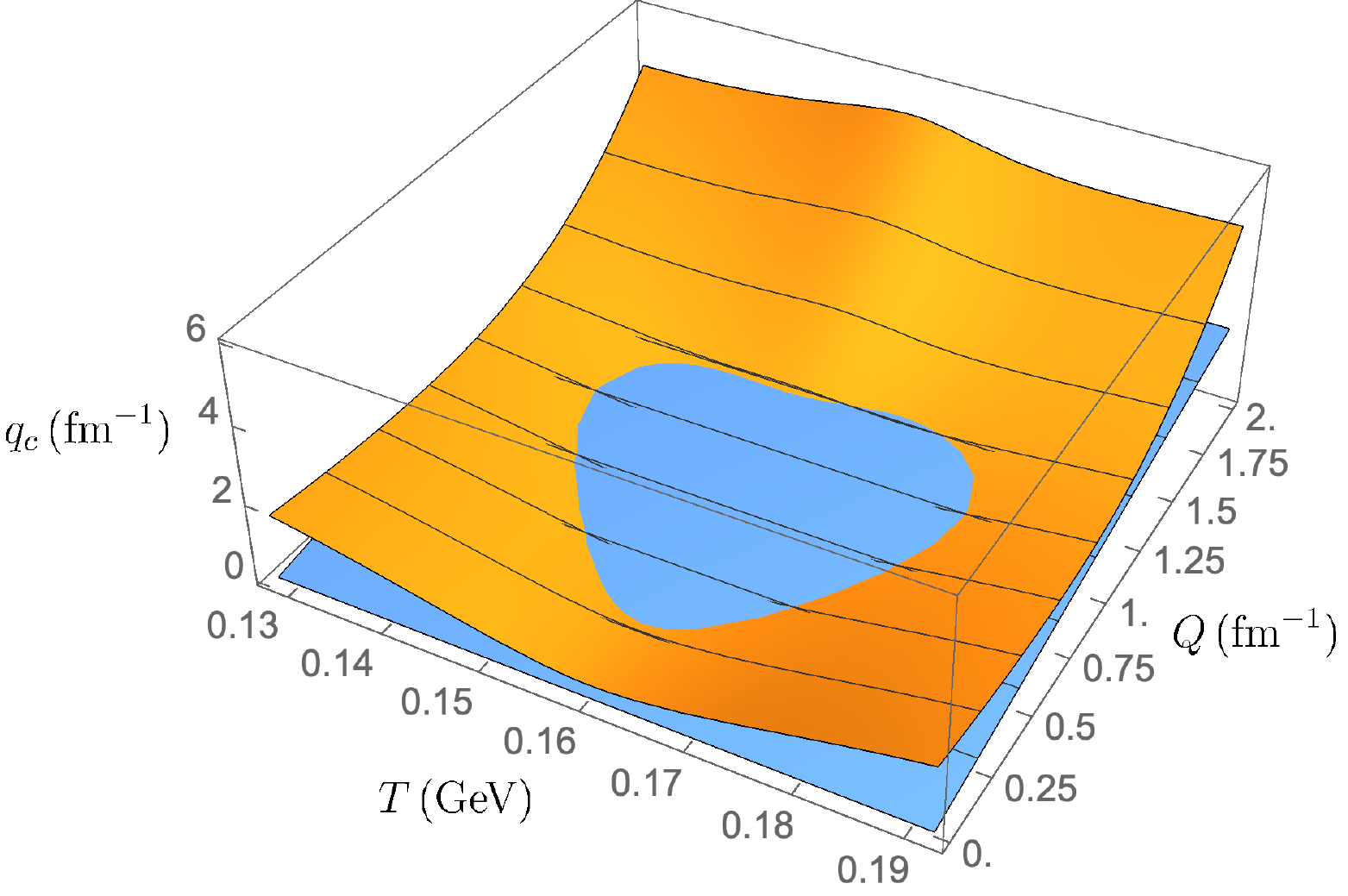}	\includegraphics[width=0.45\textwidth]{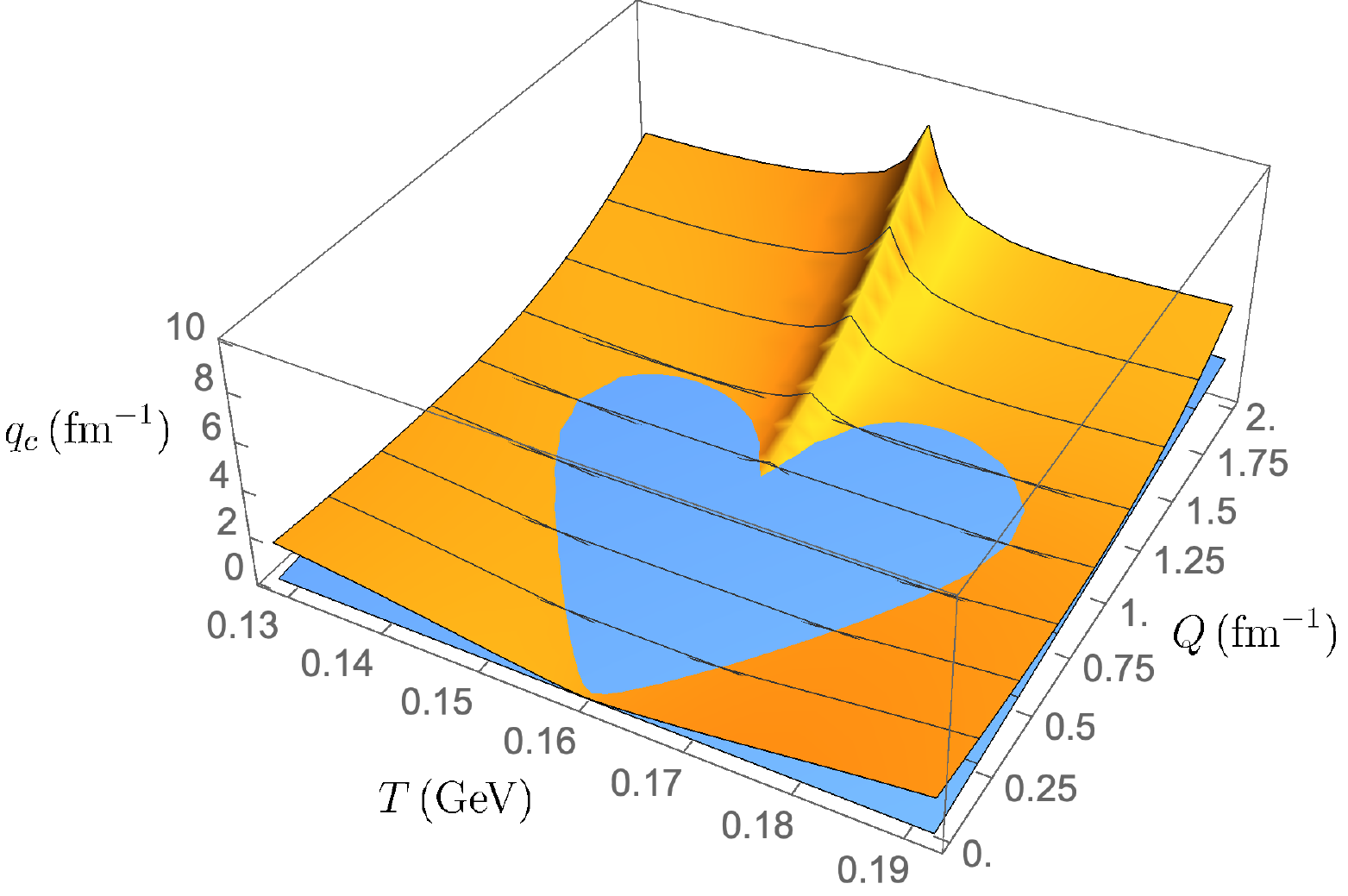}	\includegraphics[width=0.45\textwidth]{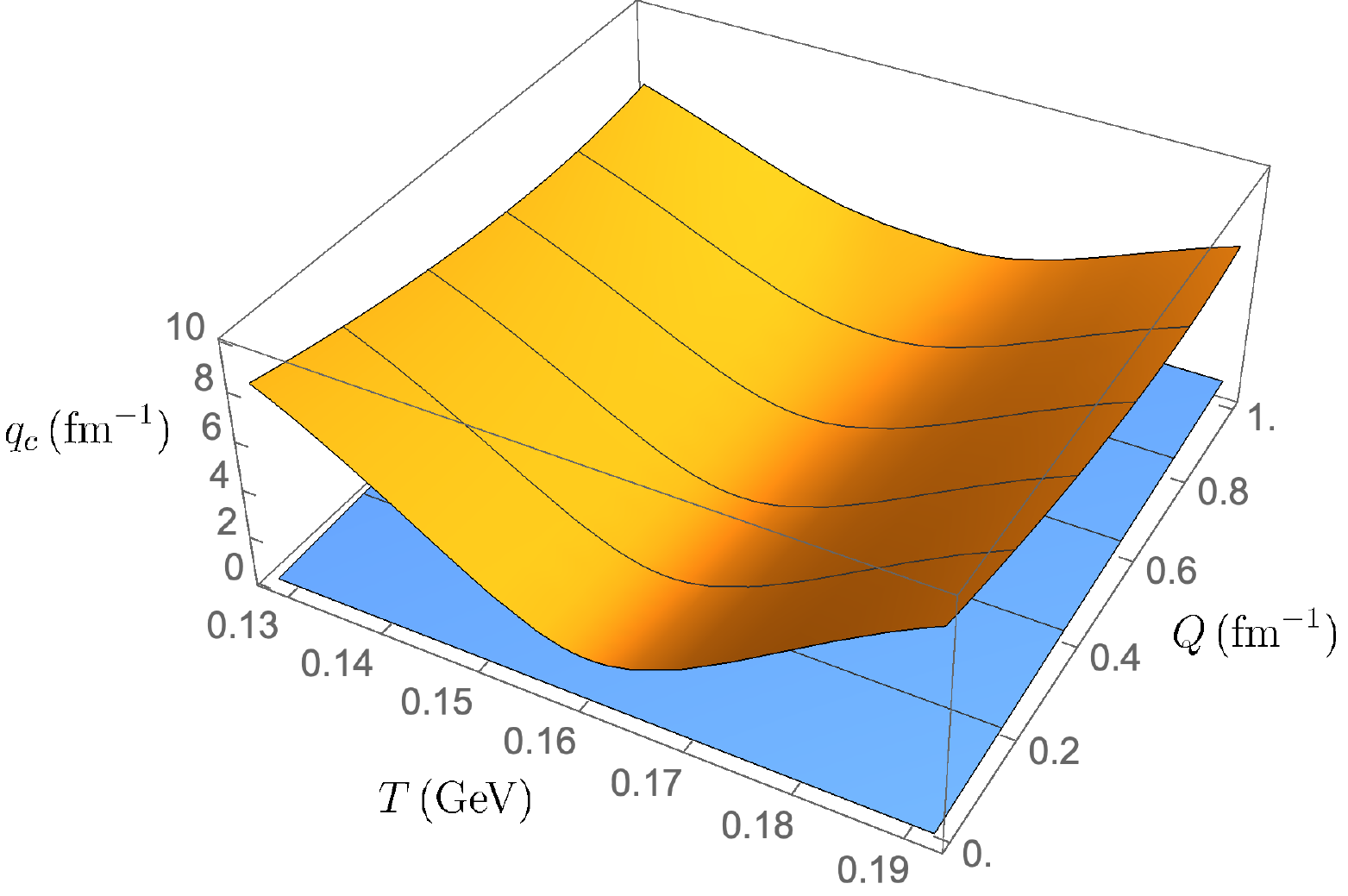}	\includegraphics[width=0.45\textwidth]{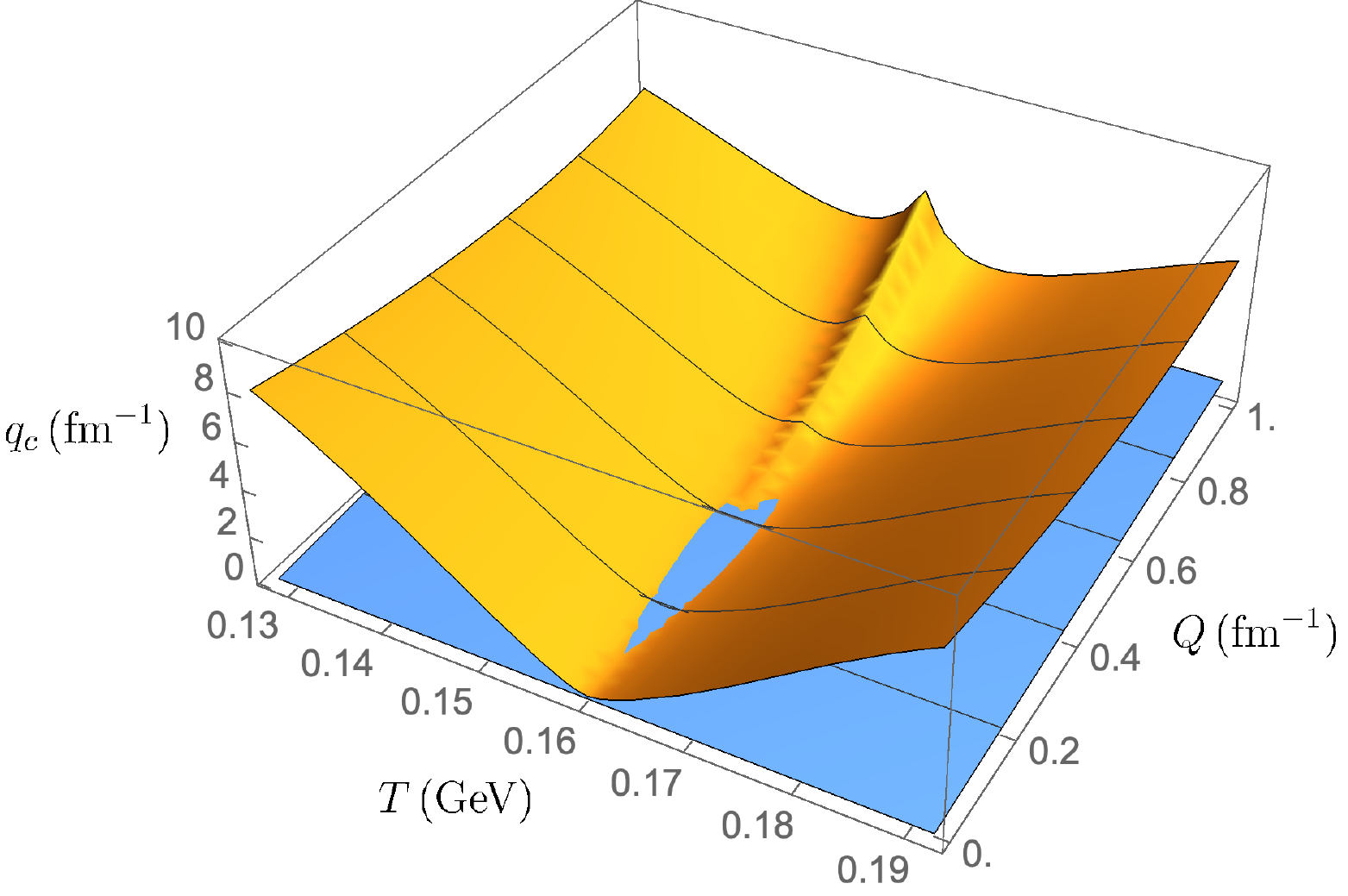}
	\caption{For any mode $\boldsymbol{Q}$, $q_c$ indicates when a single mode Hydro+ with the slow mode $\phi_{\boldsymbol{\phi}}$ must be applied. We have illustarated $q_c$ as a function of $T$ and $Q$ for four cases: \newline Top left panel: $\xi_{\text{max}}=1\,\text{fm}$ and  $D_0=0.1 \,\text{fm}$. Top right panel: $\xi_{\text{max}}=3\,\text{fm}$ and  $D_0=0.1 \,\text{fm}$.
	Bottom left : $\xi_{\text{max}}=1\,\text{fm}$ and  $D_0=0.5 \,\text{fm}$. Bottom right : $\xi_{\text{max}}=3\,\text{fm}$ and  $D_0=0.5 \,\text{fm}$.  The blue plane shows $q_c=Q$. According to \eqref{scales}, regions where this blue plane is higher than the orange surface  correspond to modes that may contribute to the stiffness. The large value of $q_c$ near $T_c$ in the two right panel plots is due to very  small rate of change of the slow mode  with respect to temperature there. On the other hand, the large values of $q_c$ at the two end of the temperate interval  the fact that the corresponding $\xi$ is much smaller than $\xi_{\text{max}}$ there.}
	\label{qstar_Q}
\end{figure}

Let us recall that $\qn_c$ is a dimensionless momentum (see \eqref{dimensionless}); thus having $\alpha_{\boldsymbol{Q}}$ is not enough to find a sensible result for it in QCD. The corresponding dimensionful quantity is given by (see \eqref{dimensionless})
\begin{equation}\label{q_c_dimension_full}
q_c=\frac{\Gamma_{{\pi}}\,\qn_c}{c_s^2}\,.
\end{equation}
In this equation, 
\begin{enumerate}
	\item $\Gamma_{\pi}$ is exactly $\Gamma_{\boldsymbol{Q}}$ defined by \eqref{decay_rate} (see equation (94) in reference~\cite{Stephanov:2017ghc}).
		\item $c_s^2$ is found in the bottom left panel of figure~\ref{EoS}.
		\item $\qn_c$ is found via applying \eqref{singularities} together with \eqref{q_c} to the $\alpha_{\boldsymbol{Q}}(T)$ found in figure~\ref{alpha}.
	\end{enumerate}
In Figure~\ref{q_c_3d}, we have shown $q_c$ for various $\phi_{\boldsymbol{Q}}$ modes at four temperatures near $T_c$.  In each row, we keep the value of $\xi_{\text{max}}$ fixed and consider two different values of $D_0$. From the top panels to the bottom ones, we change the value of $\xi_{\text{max}}$. 
One observes that
\begin{itemize}
	\item The larger the value of $\xi_{\text{max}}$ becomes, the smaller the characteristic momentum gets. 
	It can be intuitively attributed to the fact that a larger $\xi_{\text{max}}$ corresponds to an equilibrium state closer to the critical point and consequently to an earlier breakdown of standard hydrodynamics.  
	\item From these plots it seems that by approaching the critical temperature, either from above or from blow, $q_c$ decreases. As we will show in figure.\ref{qstar_Q}, this behavior continues to persist up to temperatures very close to $T_c$. 
	\end{itemize}
\par\bigskip 
\noindent

As mentioned earlier, specifically in eq.~\eqref{Deltacs2}, all $\phi_{\boldsymbol{Q}}$ modes seem to contribute to the stiffness of the EoS, and consequently to the enhancement in magnitude of $c_s^2$. However, our analysis of the Hydro+ spectrum and its characteristic momentum proves that it is necessary to narrow this range, resulting in a significant decrease in the value of $\Delta c_s^2$, compared to the case in which all modes contribute.
\newline\newline
To investigate this issue in detail, let us first list some of our assumptions and results:
\begin{enumerate}
	\item In all three cases shown in figures~\ref{spectrum_before}, \ref{spectrum_after} and \ref{spectrum_degenerate}, the enhancement in $c_s^2$ becomes remarkable when $\qn$ exceeds $\qn_c$, namely $\qn\gtrsim \qn_c$. Here $\qn$ is the dimensionless momentum of flow in the single-mode Hydro+. It corresponds to the length scale over which sound modes and the slow $\phi$ mode vary. The critical momentum $\qn_c$ is the dimensionless characteristic momentum of the theory. In a single-mode Hydro+, $\qn_c$ depends on $T$: $\qn_c\equiv \qn_c(T)$.
 	\item  In the case of QCD near the critical point, we deal with a spectrum of slow modes, i.e. $\phi_{\boldsymbol{Q}}$.
	In order to apply Hydro+ to this case, it is necessary to explicitly distinguish between $\boldsymbol{Q}$ and $\boldsymbol{q}$. To this end, let us denote that each mode $\phi_{\boldsymbol{Q}}$ can be decomposed into a set of Fourier modes $\hat{\phi}_{\boldsymbol{Q}}$ with momenta $\boldsymbol{q}$:   
\begin{equation}
\phi_{\boldsymbol{Q}}(\boldsymbol{x})=\int_{\boldsymbol{q}}\hat{\phi}_{\boldsymbol{Q}}(\boldsymbol{q})e^{i \boldsymbol{\boldsymbol{q}}\cdot \boldsymbol{x}}
\end{equation}
Here $\boldsymbol{q}$ is the momentum of the flow while $\boldsymbol{Q}$ denotes the momentum of the critical fluctuation.
 Let us recall that the main assumption based on which we include $\phi_{\boldsymbol{Q}}$ as a local slow mode is the separation of scales given by \eqref{separation_of_scales}, or equivalently
	\begin{equation}\label{speration_of_scales_2}
q\ll Q\,.
	\end{equation}
	\item Based on \eqref{speration_of_scales_2}, we can think of each $\phi_{\boldsymbol{Q}}$ as the slow mode of a single-mode Hydro+. 
	Then one naturally expects $q_c$ to become a function of both $Q$ and $T$: $q_c\equiv q_c(Q,T)$. 
	\item[$\Rightarrow$] Considering the three items above, we conclude that the modes contributing to the enhancement of  $c_s^2$ near the critical point are only those satisfying 
\begin{equation}\label{scales}
q_c(Q,T)\ll Q\,.
\end{equation}
In order to understand why this is the case, we define $Q_{j}^*(T)$ as the two roots of the following equation:
\begin{equation}\label{Q_s}
q_c(Q^*_j,T)=Q^*_j,\,\,\,\,\,j=1,2\,, 
\end{equation}
It turns out that at any temperature around $T_c$, only the modes within the interval $[Q^*_1,Q^*_2]$ satisfy \eqref{scales}. 
If \eqref{scales} is not satisfied, namely $Q<q_c(Q,T)$, then \eqref{speration_of_scales_2} requires the flow to vary  slowly, i.e. $q\ll q_c$. This is indeed the standard hydrodynamic range, and there is no need to consider the impact of critical slowing down. Therefore \eqref{scales} is a necessary condition for a mode with magnitude $Q$ to be regarded as a slow mode.

Figure~\ref{qstar_Q}, displays $q_c(Q,T)$ for exactly the four situations discussed in figure~\ref{q_c_3d}. The blue plane shows $q_c=Q$. Then the modes satisfying \eqref{scales} exist in the regions where the blue plane is located above the orange surface. The modes corresponding to the intersection of the blue plane with the orange surface are the $Q^*$ modes defined by eq.~\eqref{Q_s}. 
\end{enumerate}
\par\bigskip 
\noindent
\begin{figure}[tb]
	\centering
	\includegraphics[width=0.45\textwidth]{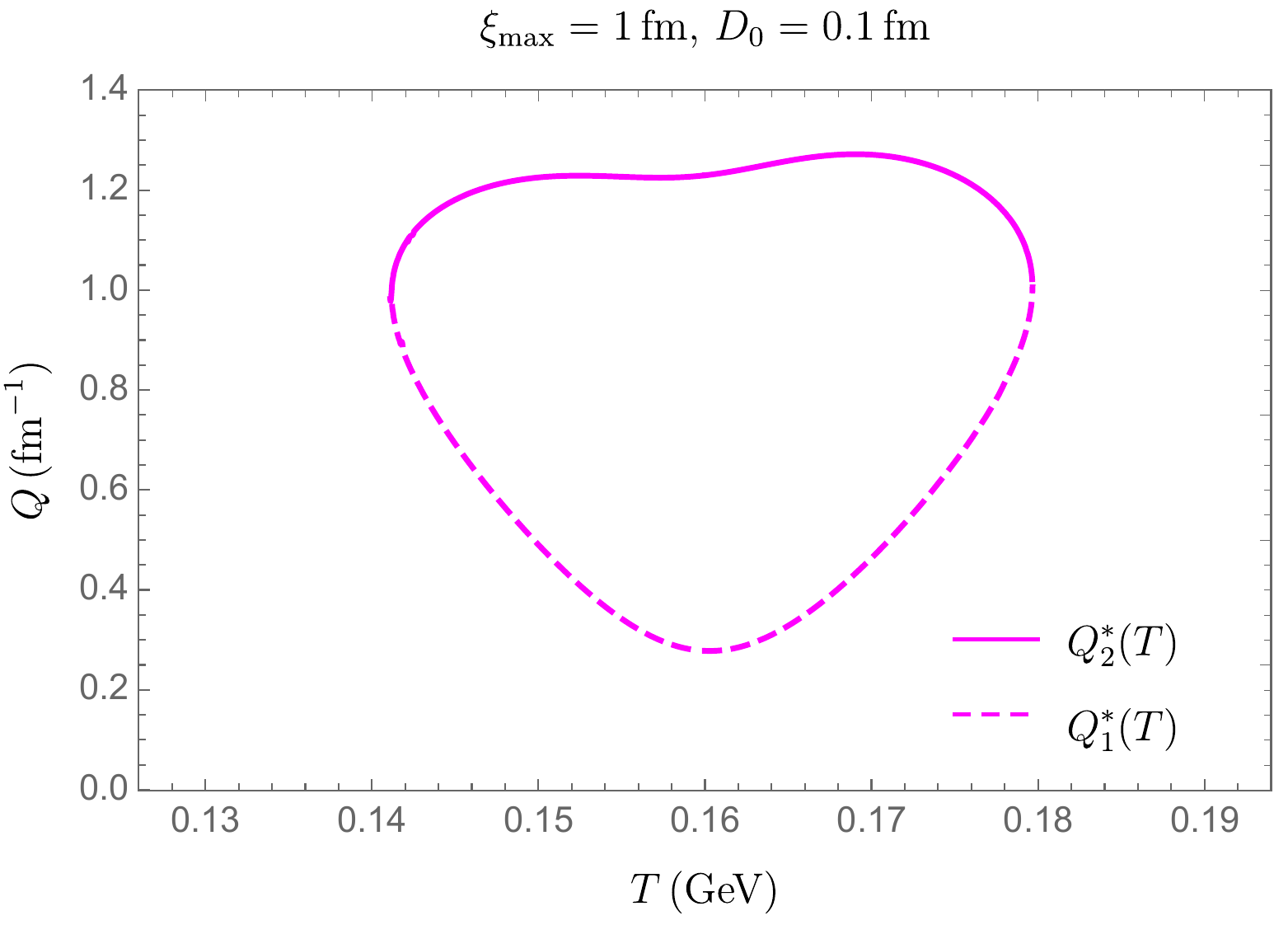}
		\includegraphics[width=0.45\textwidth]{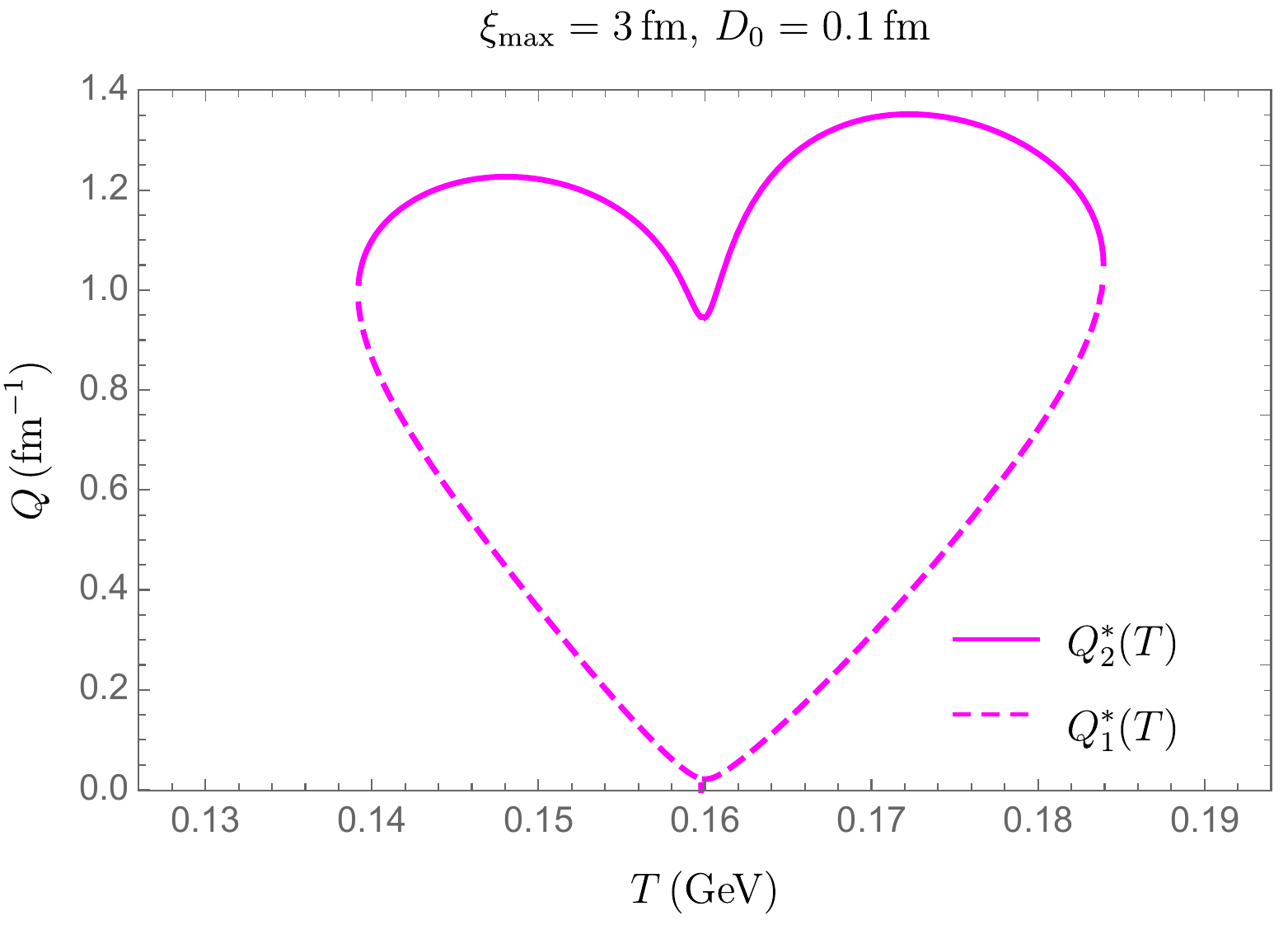}
			\includegraphics[width=0.45\textwidth]{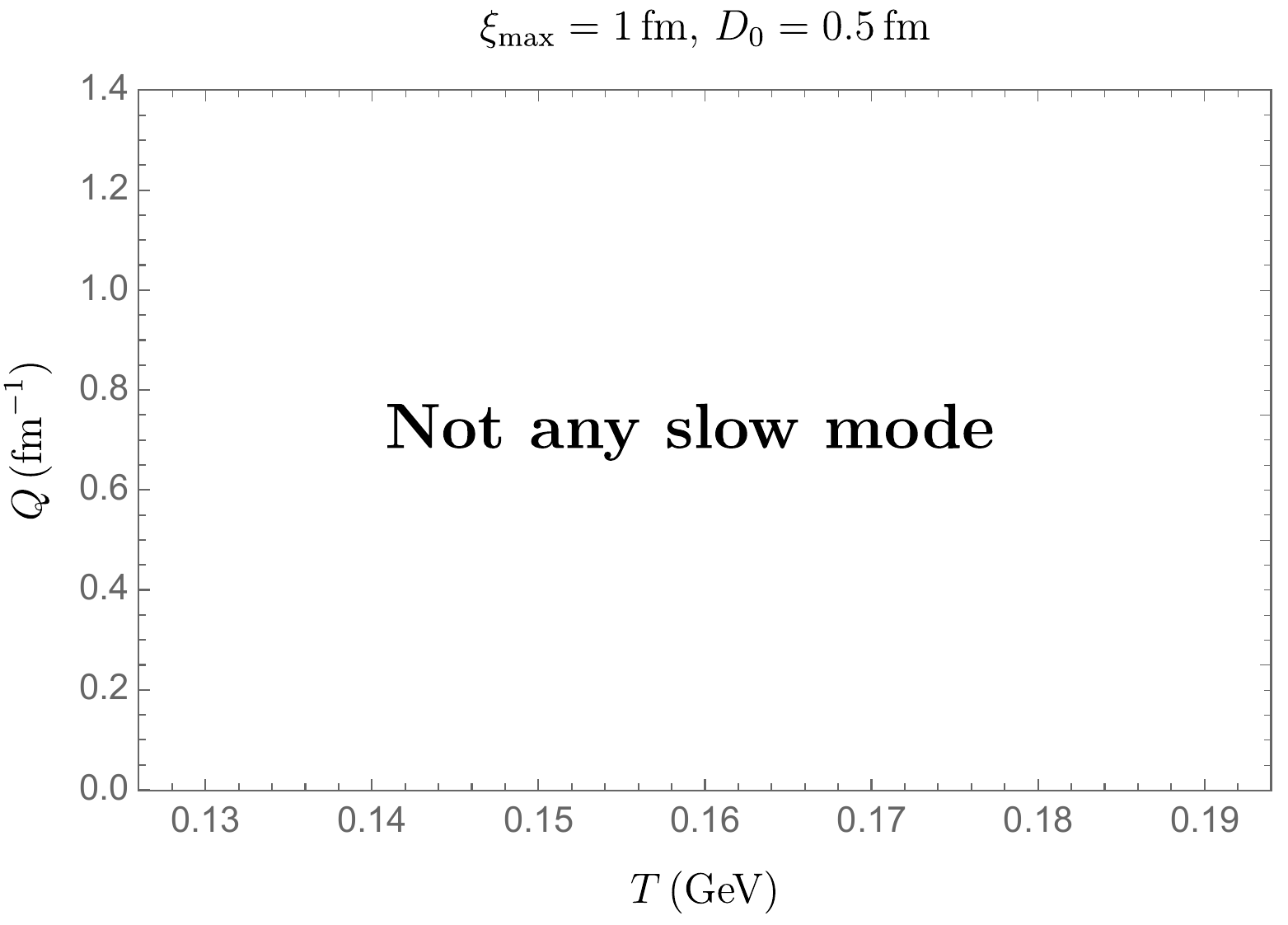}
				\includegraphics[width=0.45\textwidth]{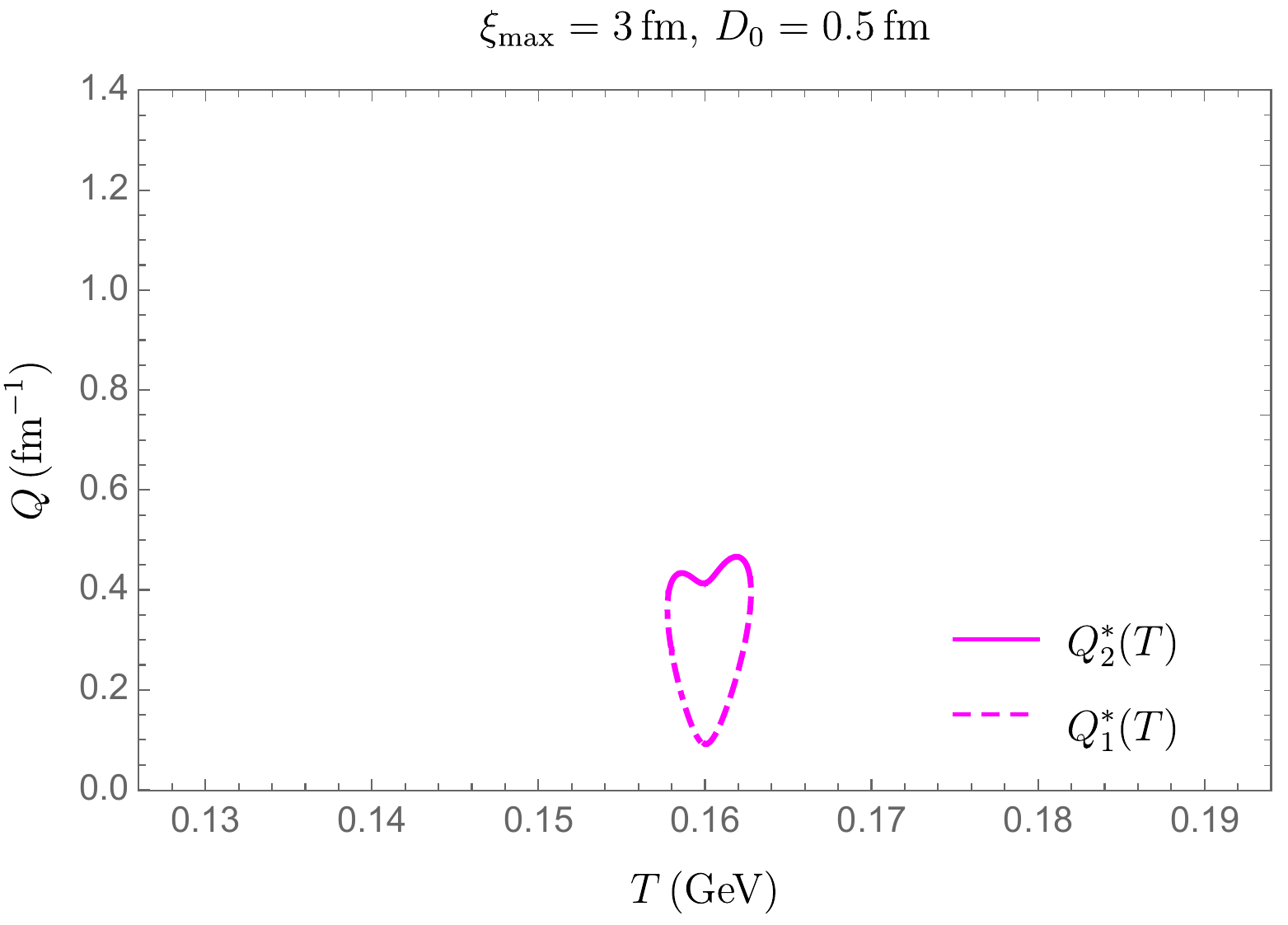}
	\caption{Each panel is related  to the corresponding panel in figure~\ref{qstar_Q}, showing the line of intersection between the blue and orange surfaces.  
	The bottom left figure illustrates that there is no intersection between the blue and orange surfaces in figure~\ref{qstar_Q}, which implies that there is not any slow mode in this case. 
	The dashed (solid) curves indicate the value of $Q_1^*(T)$ (of $Q_2^*(T)$) defined in eq.~\eqref{Q_s}, representing the lower (upper) integral boundary on which eq.~\eqref{Deltacs2} has to be evaluated. 
	}
	\label{roots}
\end{figure}
\par\bigskip 
\noindent
Considering the discussion above, we understand that at any temperature $T$, only fluctuations with $Q^*_1(T)\ll Q \ll Q^*_2(T)$  contribute to $\Delta c_s^2$. Thus the actual value of $\Delta c_s^2$ 
cannot exceed the following upper bound
\begin{equation}\label{range}\boxed{
\Delta c_s^2\,\,\,<\,\,\,\text{eq.}~\eqref{Deltacs2}\,\text{integrated over} \,\big[Q^*_1(T),Q^*_2(T)\big]} \, .
\end{equation}

In order to determine the range of integration on the right side of \eqref{range}, we have found the roots of the equation $q_c(Q,T)=Q$,  shown
in figure~\ref{roots}. The blue curves correspond to the intersection of the magenta plane with the orange surfaces in figure~\ref{qstar_Q}. 
\textit{As a result, the range of integration in the right side of \eqref{range} is determined by the interval between the $Q^*_1$- and $Q^*_2$-curves.}

\begin{figure}[tb]
	\centering
	\includegraphics[width=0.5\textwidth]{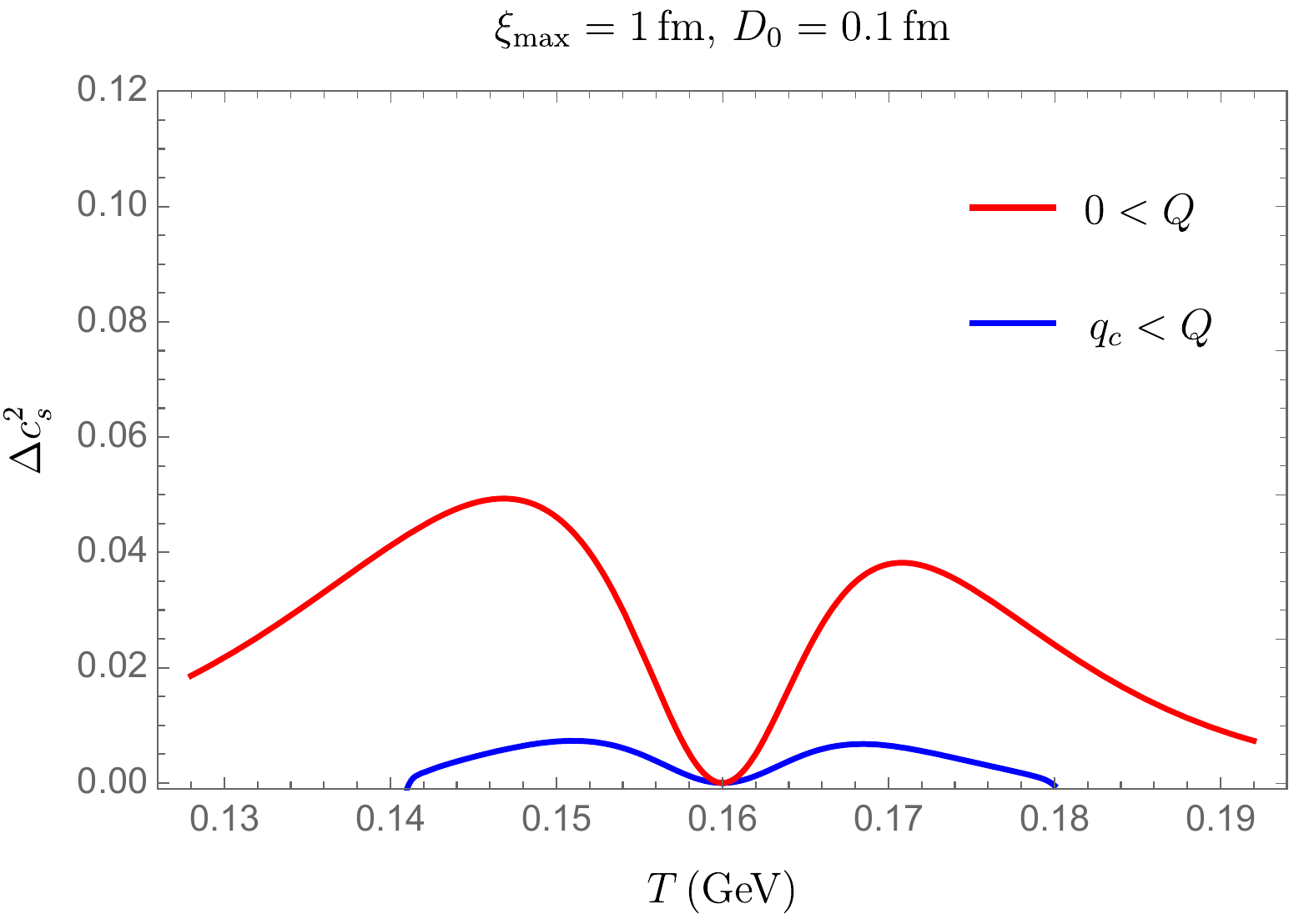}\includegraphics[width=0.5\textwidth]{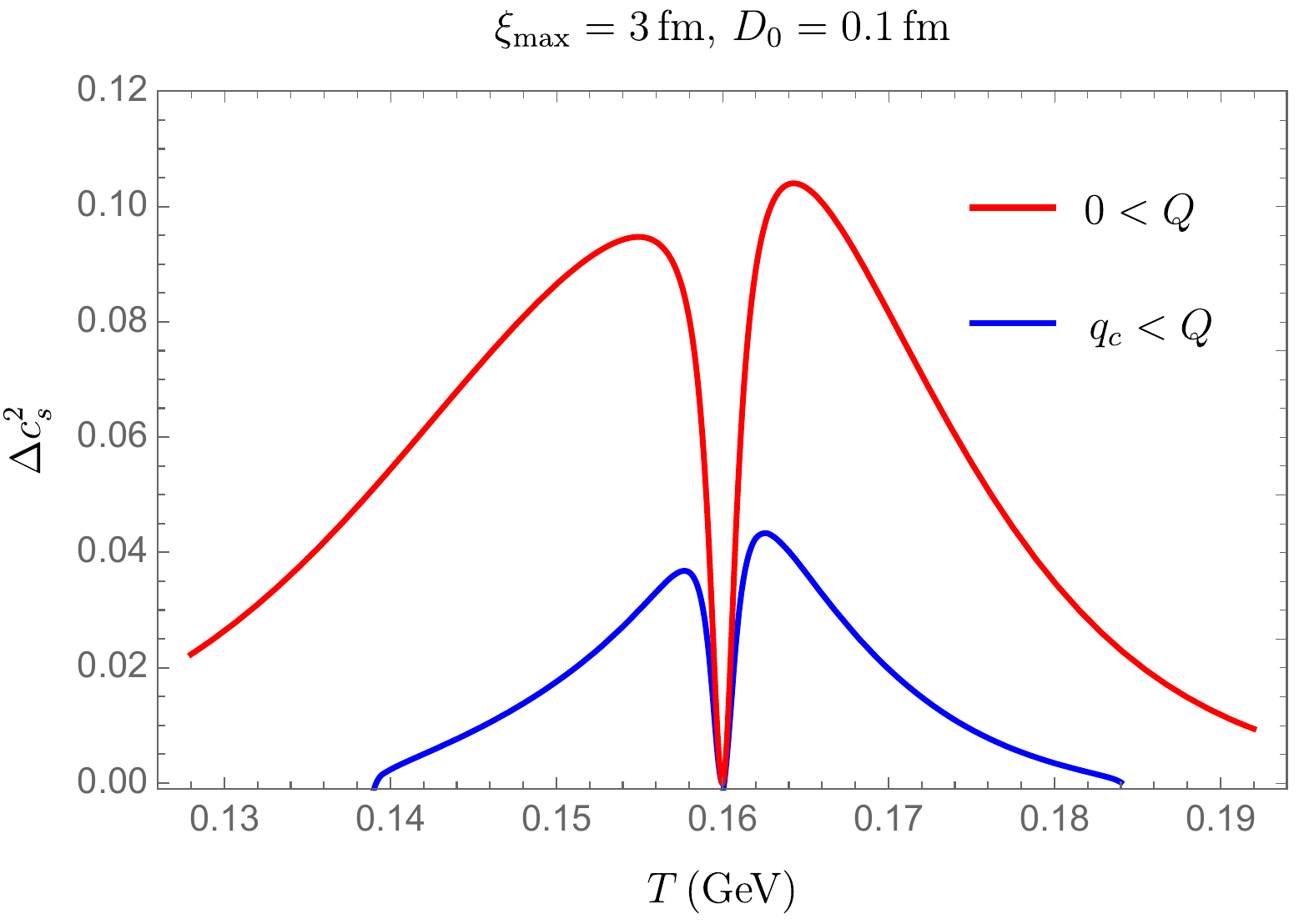}	
		\includegraphics[width=0.5\textwidth]{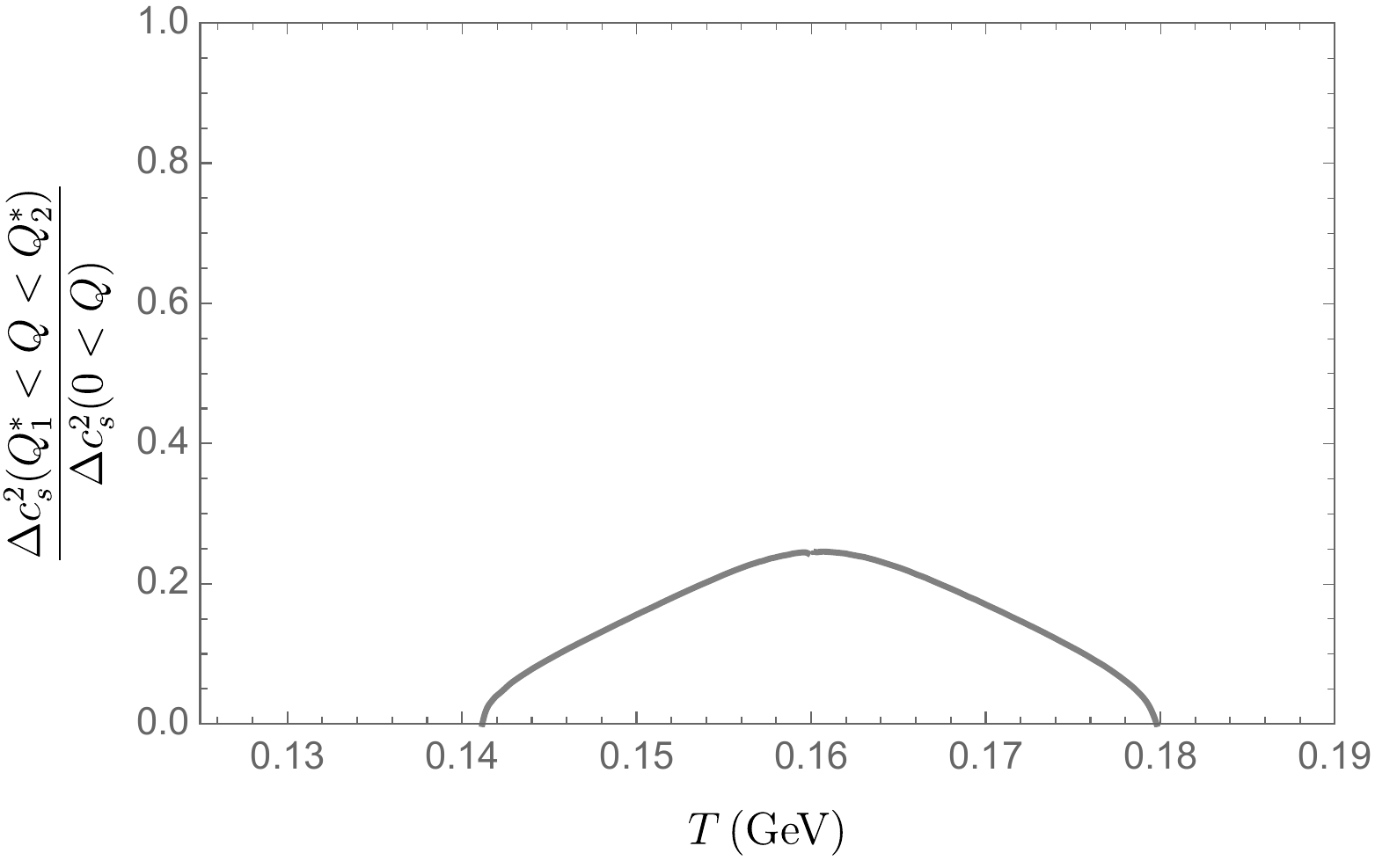}\includegraphics[width=0.5\textwidth]{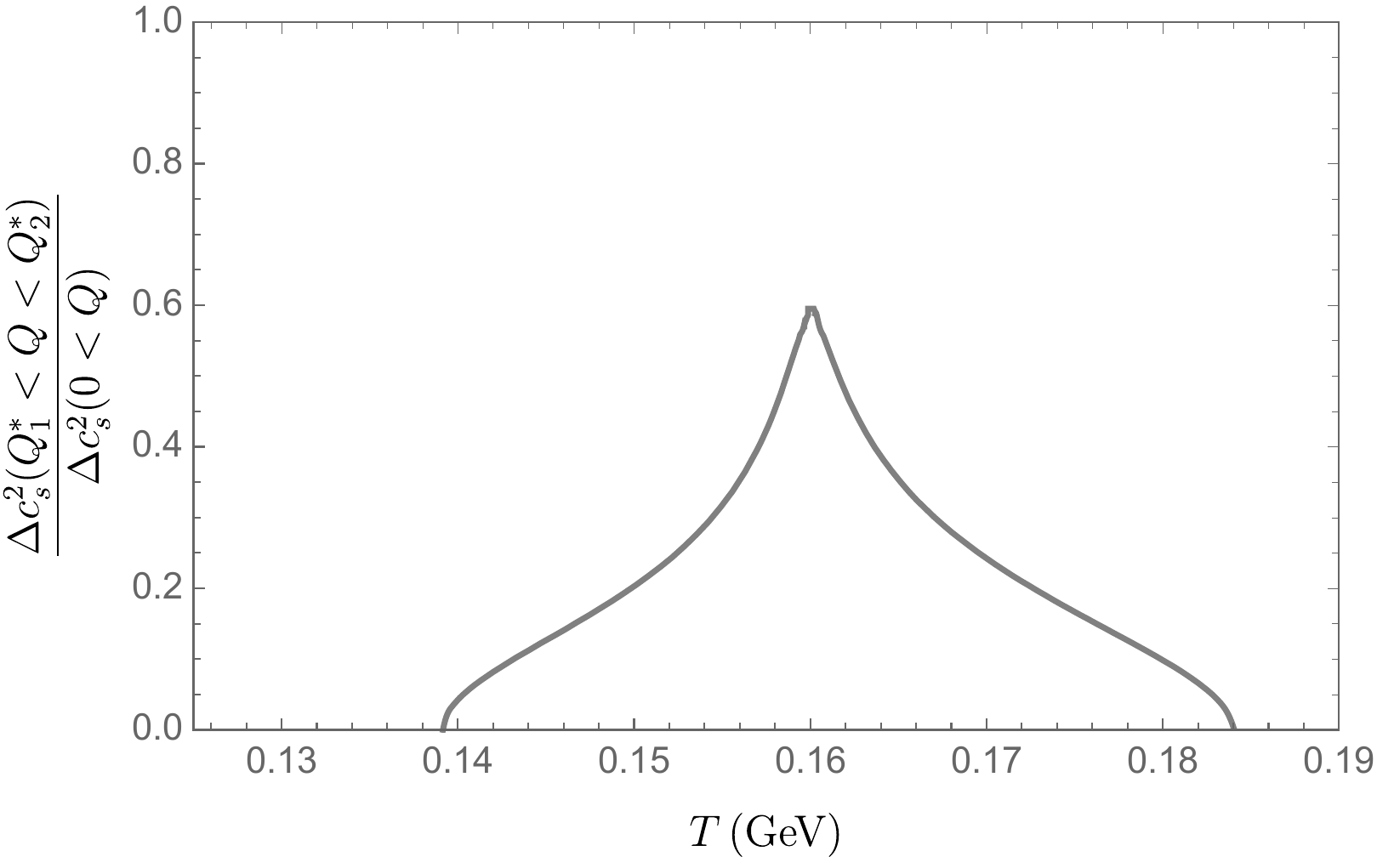}	
	\caption{ Top panel: The blue curve shows the upper bound of the stiffness of EoS near the QCD critical point. The red curve corresponds to the known estimate in which all modes $0<Q$ contribute to the stiffness.  Bottom panel: The ratio of the upper bound to the known  estimate of $\Delta c_s^2$.}
	\label{comparison_Delta_cs2}
\end{figure}

As one would expect, all panels of figure~\ref{qstar_Q} illustrate that we find only modes within an interval near the critical point,  $T=T_c$, may contribute to $\Delta c_s^2$. Why this is the case can be understood as follows. 
At a temperature sufficiently far from $T_c$, there are indeed three length scales in the system, separated as
\begin{equation}\label{cut_off}
\xi(T)\ll\frac{1}{T}\ll \ell\,.
\end{equation}
 Here $1/T$ represents the size of the smallest hydrodynamic cell. The right inequality indicates that hydrodynamics is valid over the scale $\ell$ while the left one tells us that no effect of critical slowing down is seen; all critical fluctuations fall in equilibrium faster than the local equilibration time.  
However, when approaching the critical temperature from above, both $\xi(T)$ and $1/T$ will increase, while the former will change faster. Then at some temperature near $T_c$, $\xi(T)$ will exceed $1/T$. It is clear that by approaching $T_c$ from below, $\xi(T)$ will exceed $1/T$ much earlier. Because in this case $\xi(T)$ is increasing while $1/T$ is decreasing\footnote{This is actually the reason why in all panels of figure.\ref{roots} the bounded region with $T<T_c$ is larger than that with $T>T_c$.}. The competition between $\xi(T)$ and $1/T$  on the two sides of $T_c$ will finally end up in a new ordering of scales
  \begin{equation}\label{switch}
 \frac{1}{T}\ll\xi(T)\lesssim \ell\,.
 \end{equation} 	
The appearance of a bounded region in the panels of  figure~\ref{roots} is related to the switch between the ordering of $\xi(T)$ and $1/T$  from \eqref{cut_off} to \eqref{switch}.
 Inside these regions, standard hydrodynamics breaks down,
  the critical slowing down is inevitable and Hydro+ must be applied. 
\newline\newline
An important observation regarding figure~\ref{roots} is that the critical slowing down is less important for larger values of $D_0$. For this reason, from now on, we only focus on the two top panels of this figure, namely on the case $D_0=0.1 \, \text{fm}$. 
In figure~\ref{comparison_Delta_cs2}, we compare the upper bound found in \eqref{range} with the estimate based on eq.~\eqref{Deltacs2}. We have also plotted the ratio of the latter two, in the right panel. 

\par\bigskip 
\noindent
\textbf{Top panel of figure~\ref{comparison_Delta_cs2}}: 
Our results show that naively treating all modes as valid modes in the stiffness calculation would result in a small enhancement in the speed of sound. When considering the effect of characteristic momentum, discussed in this paper, we will find that the enhancement is even smaller than simply found by including all patterns. We also see that the stiffness only appears in the interval $[T_L, T_H]$, as defined in section~\ref{QCD}. 
\\\\
\textbf{Bottom panel of figure~\ref{comparison_Delta_cs2}}: 
Our analysis reveals that the upper bound of enhancement could be significantly smaller than the known estimates resulting from~\cite{Rajagopal:2019xwg}. 
According to \eqref{scales}, only a limited amount of modes has momenta greater than $q_c$.
As a result, the bound given in \eqref{range} becomes smaller than the value that would be obtained from integration over all modes in eq.~\eqref{Deltacs2}. 
	
\begin{figure}[tb]
	\centering
	\includegraphics[width=0.5\textwidth]{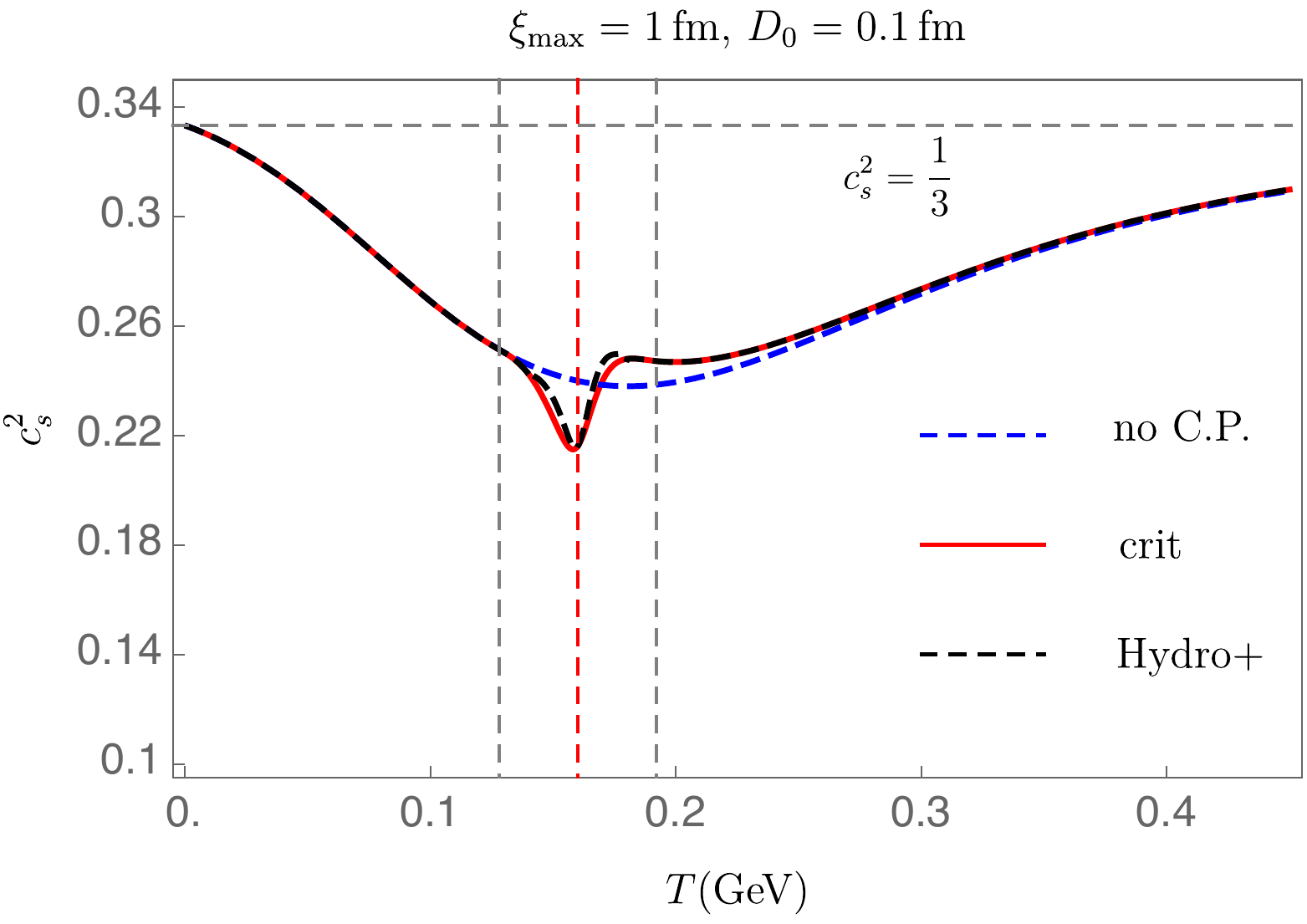}\includegraphics[width=0.5\textwidth]{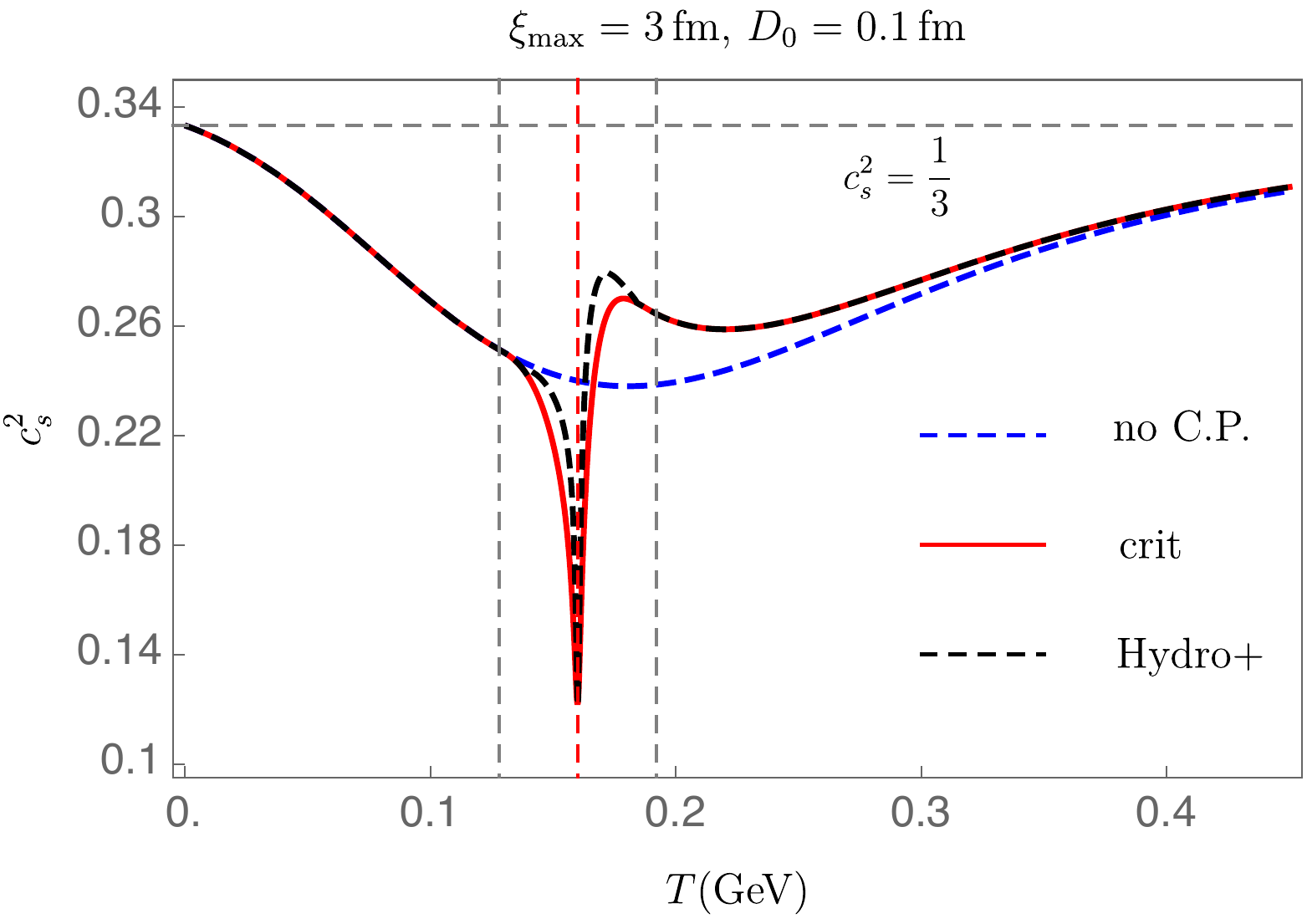}		
	\caption{ 
	{\it Dashed blue curves:} The speed of sound squared when neglecting the influence of the critical point. 
	{\it Red curves:} The speed of sound squared when taking into account the influence of the critical point on the thermodynamic quantities. 
	{\it Dashed black curves:} The speed of sound squared when taking into account the influence of the critical point on both, the thermodynamic and hydrodynamic quantities by extending hydrodynamics to Hydro+. 
	}
	\label{Final_sound}
\end{figure}

In order to show the magnitude of the stiffness, in figure~\ref{Final_sound}, we illustrate the enhanced speed of sound as a function of temperature. 
Comparing the left and the right figure, one observes that increasing the value of $\xi_{\text{max}}$, the enhancement of the speed of sound also increases. 
For larger values of $D_0$, the enhancement will become smaller. 
The enhancement of the speed of sound in any case is small, which is similar to the case of the bulk viscosity enhancement being small~\cite{Martinez:2019bsn}, however, there authors integrated over the whole range of $Q$ considering the Ising equation of state.

\par\bigskip 
\noindent

\section{Range of applicability of our results}
\label{range_of_applicabaility}
There are three important physical scales in our problem that compete with each other \cite{Stephanov:2017ghc}:
\begin{itemize}
	\item The background evolution rate $\omega\sim c_s\, q$.
	\item The slow mode decay rate $\Gamma_{\boldsymbol{Q}}\sim \xi^{-3}$. In fact $\Gamma_{\boldsymbol{Q}}$ corresponds to $\Gamma_{\pi}$ in single-mode Hydro+ (see eq.\eqref{Dpi}).
	\item The shear and diffusion relaxation rate $\Gamma_{\eta}\sim\xi^{-2}$.
	\end{itemize}
\underline{\textbf{Far from any critical point}}, $q \ll \xi^{-1}$ and therefore 
\begin{equation}
\label{}
\omega \ll \Gamma_{\eta} \lesssim \Gamma_{\boldsymbol{Q}}\,.
\end{equation}
Thus standard hydrodynamics is the valid picture at long wave length, and the next major correction comes from statistical fluctuations, i.e. the long time tails \cite{Kovtun:2003vj,Kovtun:2012rj,Akamatsu:2016llw,Chen-Lin:2018kfl,Martinez:2018wia,An:2019osr}. In this case,  there is no any slowing down mode.
\newline
\textbf{\underline{However, near a critical point}}, $\xi^{-1}$ can be so small that  $\Gamma_{\boldsymbol{Q}} \lesssim\Gamma_{\eta}$. This is the well-known critical slowing down limit, discussed in previous sections. 
Now depending on the value of $\omega$, different scenarios can take place.  When 
\begin{equation}
\label{applicablity}
\Gamma_{\boldsymbol{Q}} \lesssim \omega \ll \Gamma_{\eta}\,.
\end{equation}
hydrodynamics breaks down and Hydro+ applies. When  passing by the QCD critical point, depending on the trajectory that QGP droplet follows in the $T-\mu$ phase diagram, it may satisfy \eqref{applicablity}.  \textit{This is exactly the regime of applicability of our result}.
It is worth noting that equations \eqref{Depsilon}-\eqref{Dphi} are consistent with \eqref{applicablity}. Since $\Gamma_{\boldsymbol{Q}} \ll\Gamma_{\eta}$, the shear tensor is completely equilibrated and tracks the hydrodynamic variables and slow mode.

In a more rapidly evolving background, i.e.
\begin{equation}
\label{}
\Gamma_{\boldsymbol{Q}} \ll  \Gamma_{\eta}\lesssim \omega,
\end{equation}
Hydro+ breaks down as well (see \cite{An:2020jjk} for a review). In this regime, it is necessary to take into account the effect of hydrodynamic fluctuations, the long time tails \cite{An:2020vri}. Then the problems translates to studying the radius of convergent of hydrodynamics near the critical point and in the presence of long-time tails \cite{Abbasi:2021fcz}.  It is actually beyond the scope of our analysis in the current study.

\section{Comment on the  gravity dual of Hydro+ }
\label{gravity}
In reference~\cite{Grozdanov:2018fic}, the hydrodynamic theory incorporating  slowly damping gapped modes has been developed. The so-called quasi-hydrodynamics theory of \cite{Grozdanov:2018fic}  describes those systems in which one (or a finite number of gapped) mode(s) is (are) parametrically slow, in the sense that their life-time at $\qn\ll1$ is comparable with that of hydrodynamic modes. The equations of quasi-hydrodynamics then describe the dynamics of hydrodynamic modes together with the slow mode(s) at small momenta.

Authors of  reference~ \cite{Grozdanov:2018fic} introduce several examples for the quasi-hydrodynamics. One of their examples is the Muller-Israel-Stewart theory\footnote{We assume that the decay rate of the shear mode in the MIS theory is parametrically larger then that of other infinitely many non-hydrodynamic modes. This assumption is necessary to extend the range of applicability of MIS theory beyond the standard hydrodynamic approximation~\cite{Stephanov:2017ghc}.}.  They find the following equation for the sound modes in this theory  \footnote{The example illustrated in \cite{Grozdanov:2018fic} is  similar to 
	both our reference\ref{spectrum_before} and reference1 in \cite{Stephanov:2017ghc}. Indeed, in all these three cases, $\alpha=2$. In reference~\cite{Grozdanov:2018fic} the authors choose $\eta/(\epsilon+p)\tau=1/2$ together with $c_s^2=1/3$. Then  by using \eqref{identification} one finds $\alpha=2$.}:
\begin{equation}
\text{MIS}:\,\,\,\,\,\,\omega^3+\frac{i}{\tau}\omega^2-\left(c_s^2+\frac{4}{3}\frac{\eta}{(\epsilon+p)\tau}\right)\omega\,q^2-\frac{i c_s^2}{\tau}q^2=\,0 \, .
\end{equation}
Let us compare it with \eqref{Hydro+} in its simplified form:
\begin{equation}\label{Hydro_simple}
\text{Hydro}+:\,\,\,\,\,\omega^3+i \Gamma\, \omega^2-\big(c_s^2+\,\Delta c_s^2(\infty)\big)\omega\,q^2-\,i\,c_s^2\,\Gamma\,q^2=\,0 \, .
\end{equation}
One immediately notices that \textbf{at the linearized level}, MIS theory is exactly the same as Hydro$+$ by the following identifications 
\begin{equation}\label{identification}
\Gamma\equiv\frac{1}{\tau},\,\,\,\,\,\,\Delta c_s^2(\infty)\equiv\frac{4}{3}\frac{\eta}{(\epsilon+p)\tau} \, .
\end{equation}
As it was expected, we see that Hydro$+$ is one another example of quasi-hydrodynamics. Interestingly,   reference~ \cite{Grozdanov:2018fic}  uncovers the equations of MIS theory from a holographic higher-derivative Einstein-Gauss-Bonnet gravity theory, which was previously shown to be able to exhibit long-lived massive excitations within the low-energy (hydrodynamic) regime \cite{Grozdanov:2016fkt}. \textit{Then we find out that linearized Hydro$+$ can be derived from Einstein-Gauss-Bonnet gravity theory, too}. 

What is the consequence of the above discussion? In last sections we discussed the spectral curve of Hydro$+$ in details. We analytically showed that the \textit{singularities of modes were all square-root type}. When combined with the discussion of the previous paragraph, one concludes that the spectral curve of holographic theories at finite coupling and at small momentum is analytic except for some \textit{square-root singularities}. It is actually in agreement with well-known examples in the literature \cite{Grozdanov:2018gfx}. It may then help us to better understand the full nature of the spectral curve in holographic theories which is still an open question.

\section{Conclusion and Outlook }
\label{conclusion}
Motivated by the recent studies on the convergence of the derivative expansion in holographic theories, in this work we applied the same idea to hydrodynamics near the QCD critical point. For the first time, we computed the radius of convergence, $q_c$, of Hydro+, as well as its effect on the observable $\Delta c_s^2$, the shift in the speed of sound in QCD plasma due to the fluctuations near the critical point. We obtained three distinct results: 
\begin{enumerate}
	\item We computed and analyzed the full spectrum of linear perturbations in single-mode Hydro+.  
	We computed the critical point of the spectral curve and identified the radius of convergence of the derivative expansion, given by the critical momentum  $q_c$, see equations \eqref{singularities} and \eqref{q_c_0}. The key message is that $q_c$ distinguishes between the regime of validity of standard hydrodynamics and the regime where Hydro+ needs to be employed.  
	\item For the hydrodynamic description of QCD near the critical point, $q_c$ is a function of the momentum of critical fluctuations, namely $\boldsymbol{Q}$, as well as the temperature, $T$, as visualized in figure~\ref{qstar_Q}. 
	\item At any temperature, the competition between $Q$ and $q_c(Q)$ determines whether the mode $\phi_{\boldsymbol{Q}}$ actually should be regarded as a slow mode in single-mode Hydro+ or not. If $Q<q_c(Q)$, standard hydrodynamics suffices. By finding $Q_{1,2}^*(T)$ as the roots of the equation $Q^*=q_c(Q^*)$, we discussed that at any temperature only critical modes within the interval $(Q_1^*, Q_2^*)$ may contribute to $\Delta c_s^2$. This leads to an upper-bound for $\Delta c_s^2/c_s^2$ near the critical point, which is calculated numerically and given in Fig~\ref{comparison_Delta_cs2}. 
\end{enumerate}
We think that the upper-bound given in reference\ref{comparison_Delta_cs2} is not saturated. The reason is that our desired separation of scale \eqref{speration_of_scales_2} demands $Q$ to be much larger than $q$. So as we found in \eqref{scales}, $Q$ is much larger than $q_c$. Thus not all the modes within the interval  $(q_{c}(Q^*), \xi^{-1})$ are able to contribute significantly.
In this sense, the bound which is the result of integration over the whole interval, is not saturated. 

In this work we did not consider any non-linear effects in the time-evolution of fluctuations. As discussed in \sec{range_of_applicabaility}, in more rapidly evolving backgrounds than what we study here, the presence of non-linear effects is inevitable. Including such effects in the present study requires extending our knowledge about hydrodynamics in two directions. First, the convergence of the derivative expansion in the presence of long-time tails should be explored \cite{Abbasi:2021fcz}. Second, the non-linear structure of Hydro+ should be understood \cite{An:2019csj}. Then, combing the latter two points will provide us with the background on which we can take into account the effect of non-linearities, which we leave to future work.
It would be also very interesting to pursue this discussion  in the framework of Schwinger-Keldysh effective  field theory .

In a more realistic situation, one should also include the effect of the evolution of the QGP droplet, itself. Following the study of long-time tails  \cite{Akamatsu:2016llw} and the evolution of critical fluctuations \cite{Akamatsu:2018vjr} in a Bjorken flow, it would be very interesting to explore what the effect of the evolution of the background fluid on the results of our paper would be. More realistically, one should also include the fact that the QGP is a highly vortical plasma~\cite{STAR:2017ckg,Cartwright:2021qpp}. Finally, hydrodynamics is the description of deviations from a known state, for example Bjorken flow~\cite{Bjorken:1982qr}, and more generally it is necessary to consider nonlinear flows beyond Bjorken flow as well as higher orders in the hydrodynamic derivative expansion~\cite{Heller:2021oxl,Heller:2021yjh}.

\section*{Acknowledgment}
We would like to thank Misha Stephanov and Yi Yin for valuable discussions, and Sa{\v{s}}o Grozdanov for helpful comments on a draft of this work. 
This work was supported, in part, by the U.S.~Department of Energy grant DE-SC-0012447, and by grant number 561119208 ``Double First Class'' start-up funding of Lanzhou University, China.

\appendix

\section{Mode collision and the radius of convergence}
\label{collision}
According to refs.\cite{Withers:2018srf,Grozdanov:2019kge,Grozdanov:2019uhi,Heller:2020hnq}, in order to find the radius of convergence, one has to find the singularity of dispersion relations, the nearest to the origin. Since in our present case, the spectral curve \eqref{spectral} is analytic, the singular points are those at which $\partial_{\wn}F=\,0$. One can simply check that indeed, singular points given by \eqref{singularities} satisfy this equation.

On the other hand, the points at which $\partial_{\wn}F=\,0$, are the critical points of $F$. In the following, by studying the modes at complex momenta in the single-mode Hydro+, we explicitly demonstrate that complexified modes collide at the critical points.

\begin{figure}[tb]
	\centering
	\includegraphics[width=0.35\textwidth]{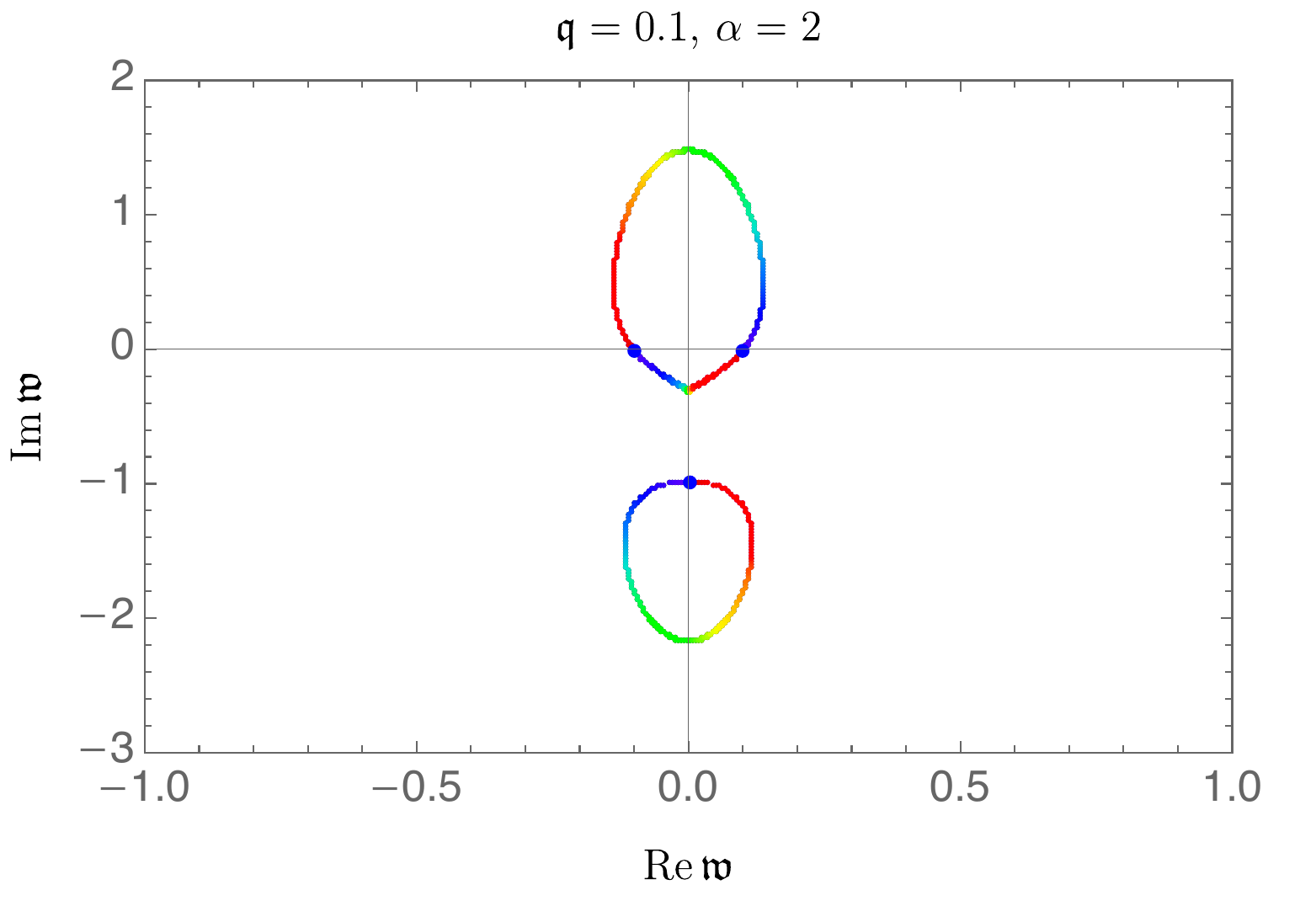}\includegraphics[width=0.35\textwidth]{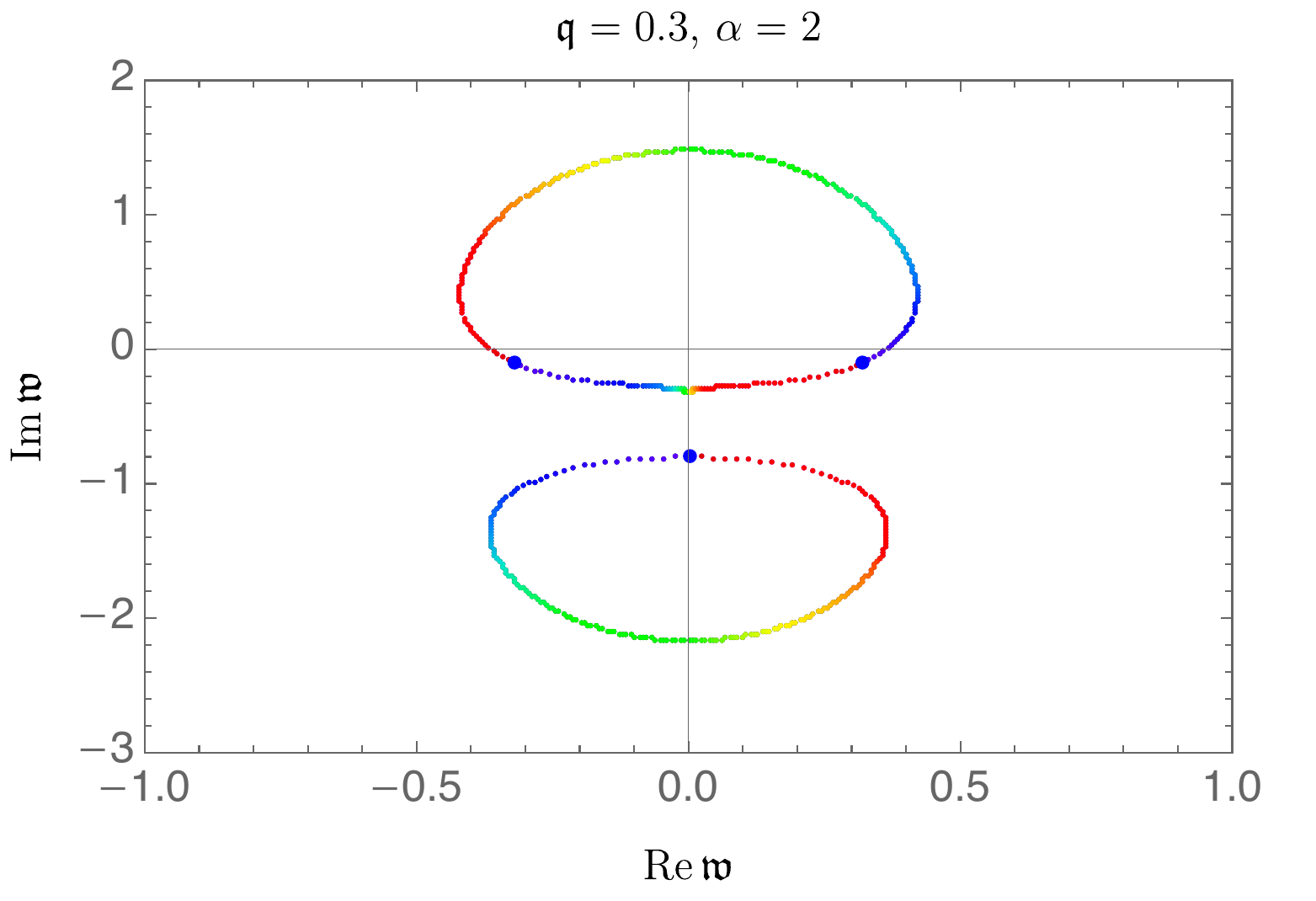}\includegraphics[width=0.35\textwidth]{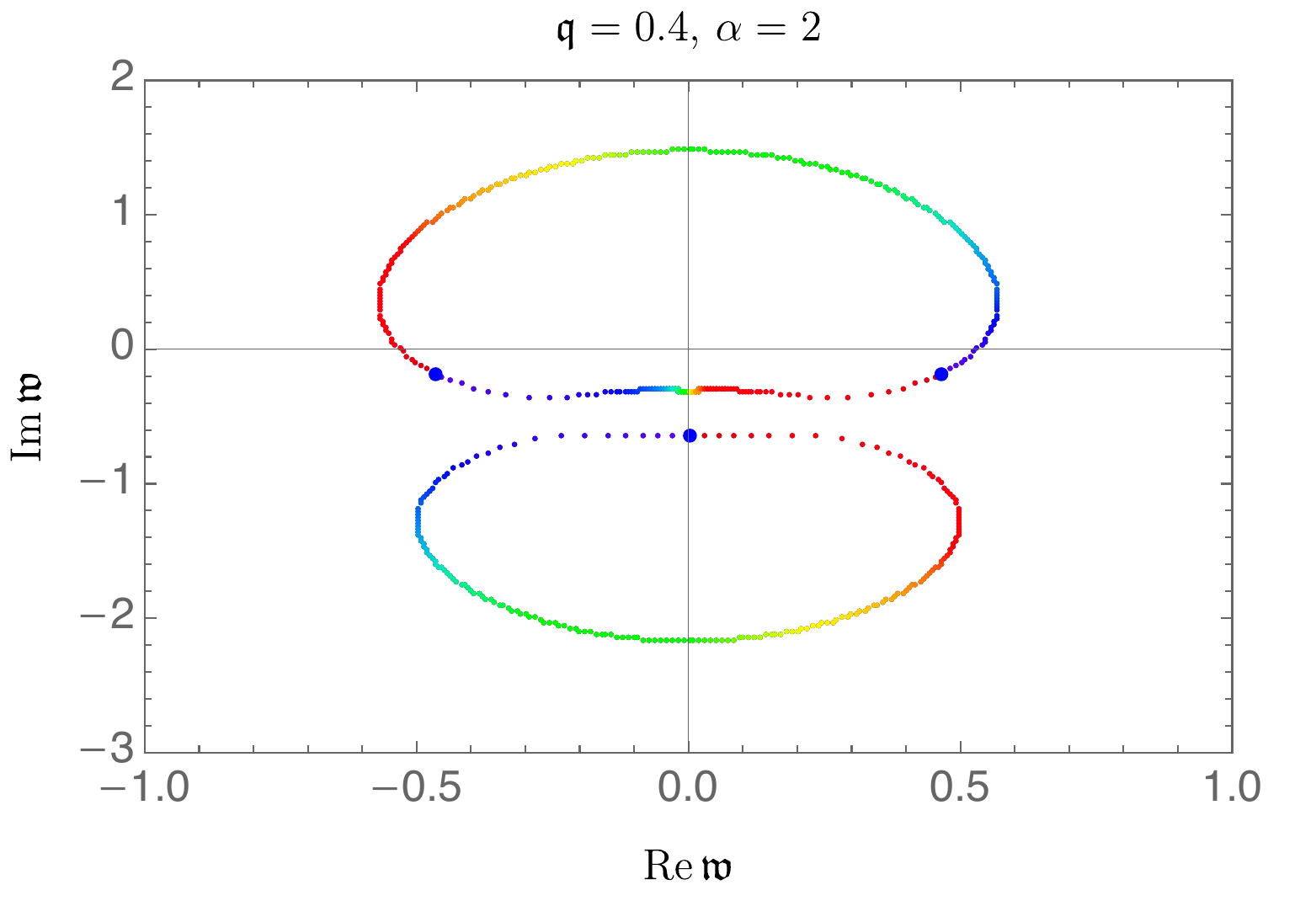}
	\includegraphics[width=0.35\textwidth]{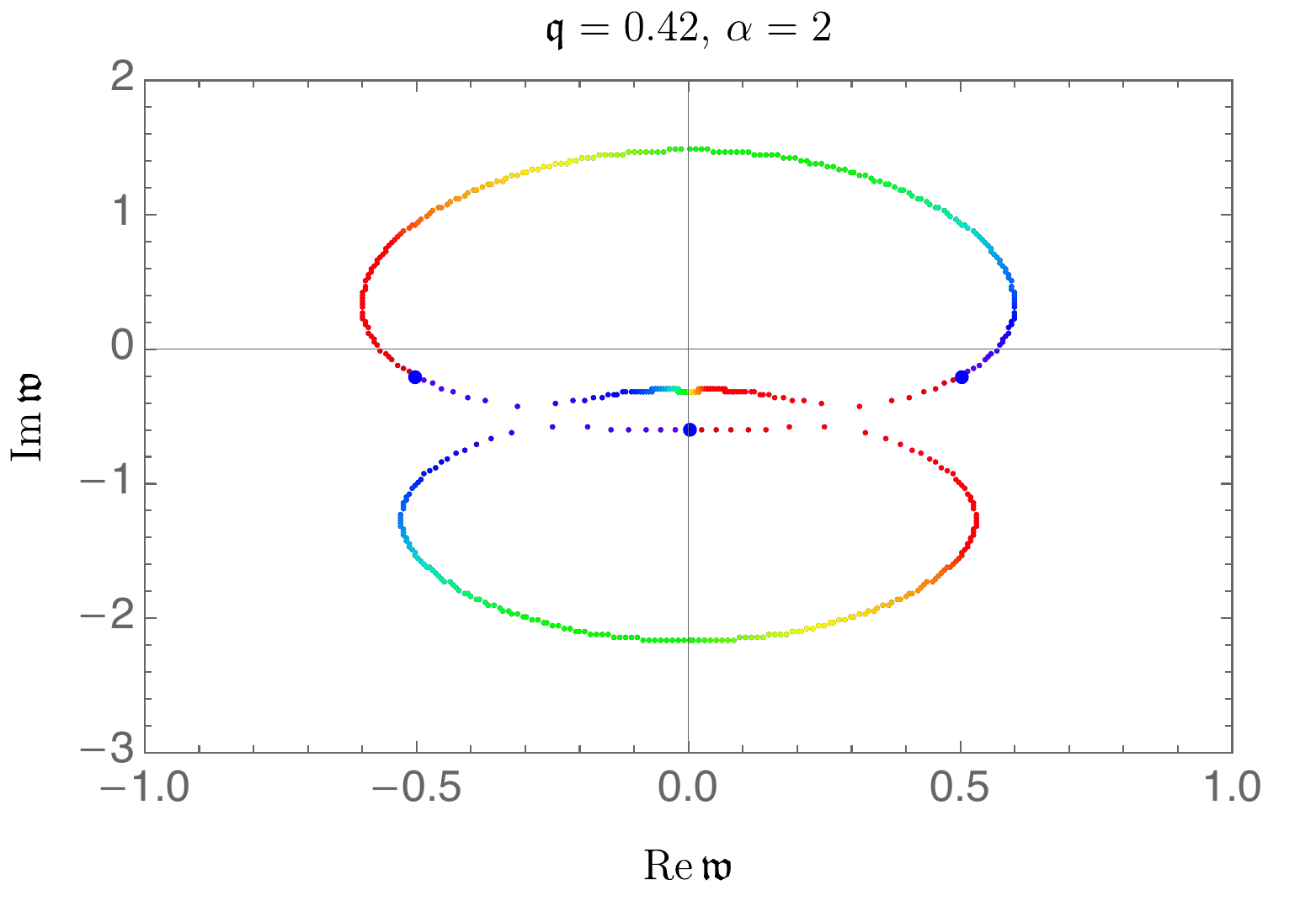}\includegraphics[width=0.35\textwidth]{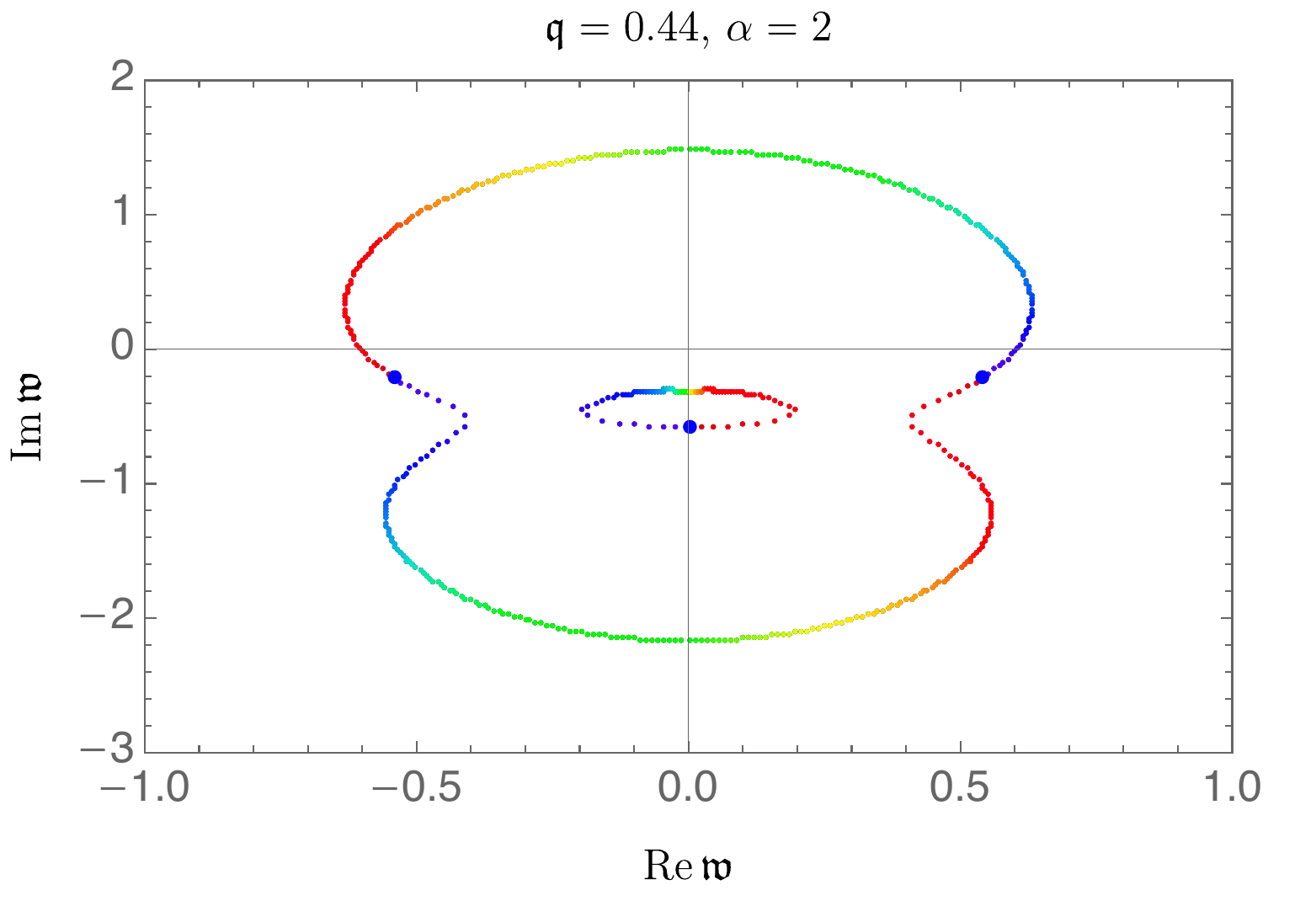}\includegraphics[width=0.35\textwidth]{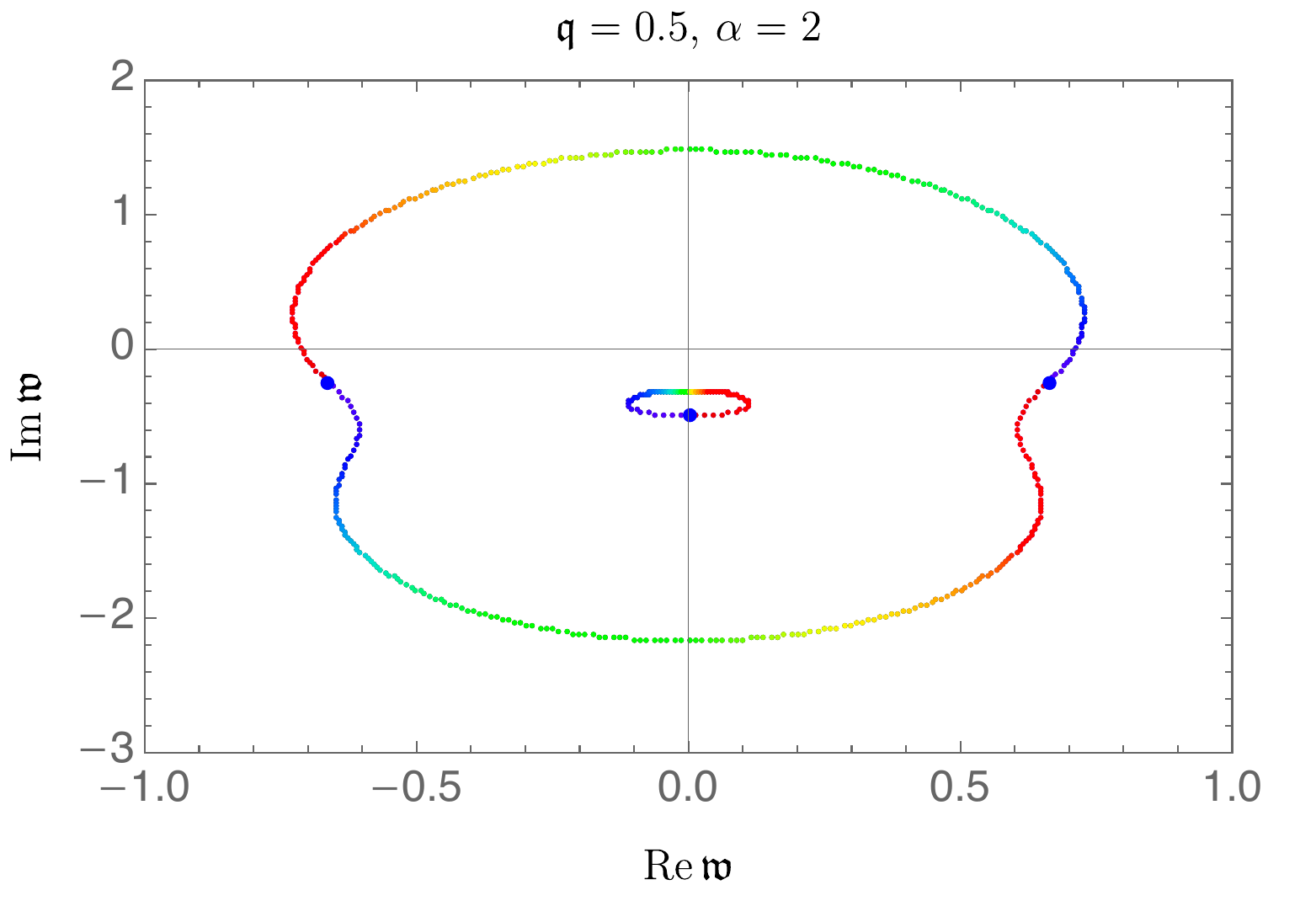}
	\caption{modes for $\alpha=2$ in the complex $\wn-$plane, at various values of
		the complexified momentum $\qn^2=|\qn^2|e^{i \theta}$. Large dots and correspond to the poles with  purely real momentum (i.e. at $\theta=0$). As $\theta$ increases from $0$ to $2\pi$, each pole moves counter-clockwise	following the trajectory whose color changes continuously from blue to red. In all panels, the two blue dots 
		located in the 3rd and 4th quadrants indicate the two sound modes and the single blue dot on the vertical axis illustrates the slow mode, at $\theta=0$. }
	\label{collision_before}
\end{figure}

Figure~\ref{collision_before} is devoted to the case $\alpha<8$. As it is seen, at small values of $|\qn|$, the two sound modes are more slowly decaying than the slow mode. The sound modes can be still described with the standard hydrodynamics. They actually form a low energy decoupled sector in the system. However, by increasing $|\qn|$, their dynamics becomes gradually coupled with the dynamics of the slow mode. Then Hydro+ must be considered as the alternate. As it is seen in the figure, at $\qn_c\approx0.43$, the two trajectories collide. In fact the slow mode collide with the two sound modes at two different values of $\theta$. We see that after the collision, the sound modes cannot be considered as a decoupled low energy sector anymore. We conclude that at $\alpha<8$, the critical momentum $\qn_c$ corresponds to collision of the slow mode with the sound modes.
\par\bigskip 
\noindent
\begin{figure}[tb]
	\centering
	\includegraphics[width=0.35\textwidth]{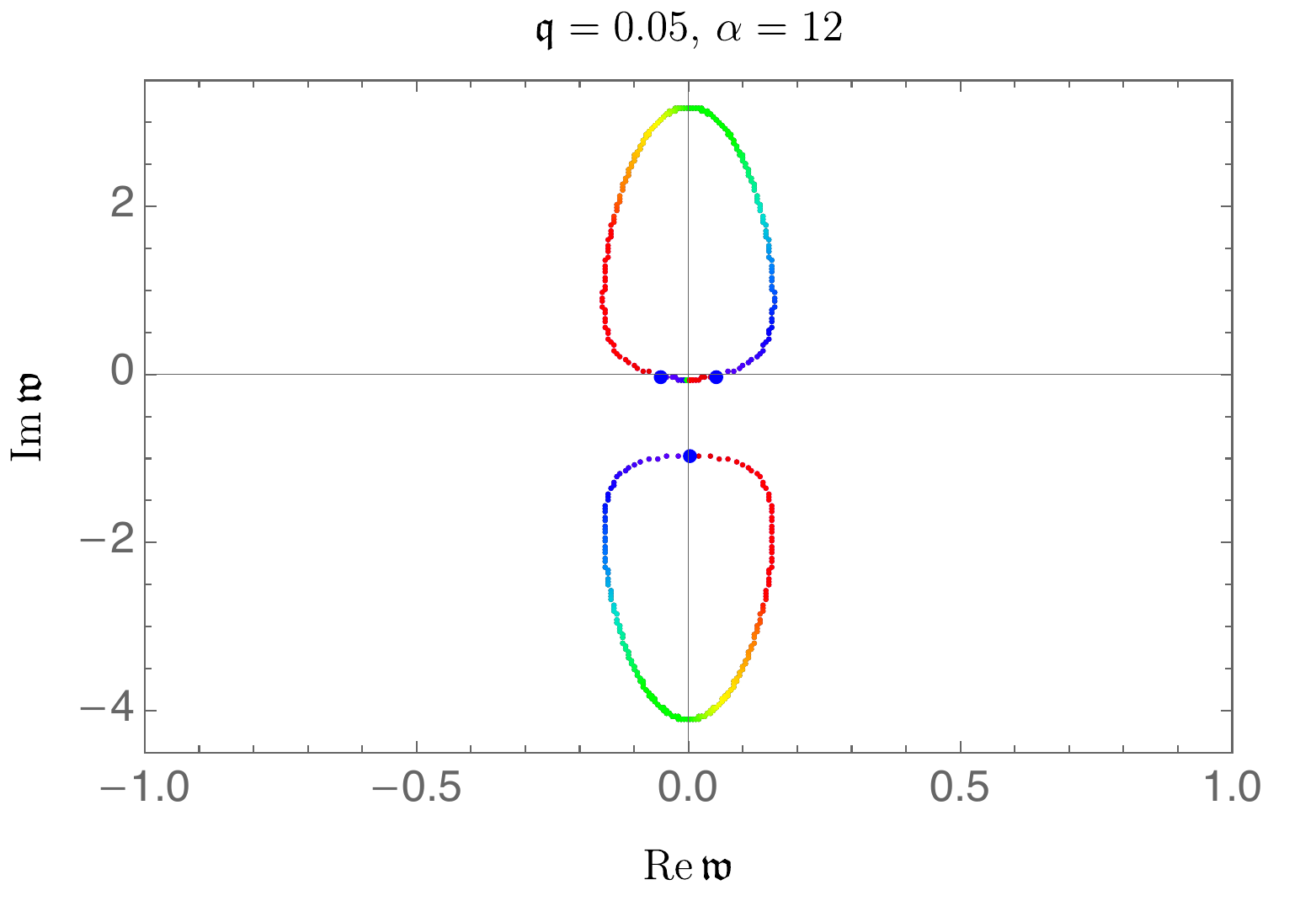}\includegraphics[width=0.35\textwidth]{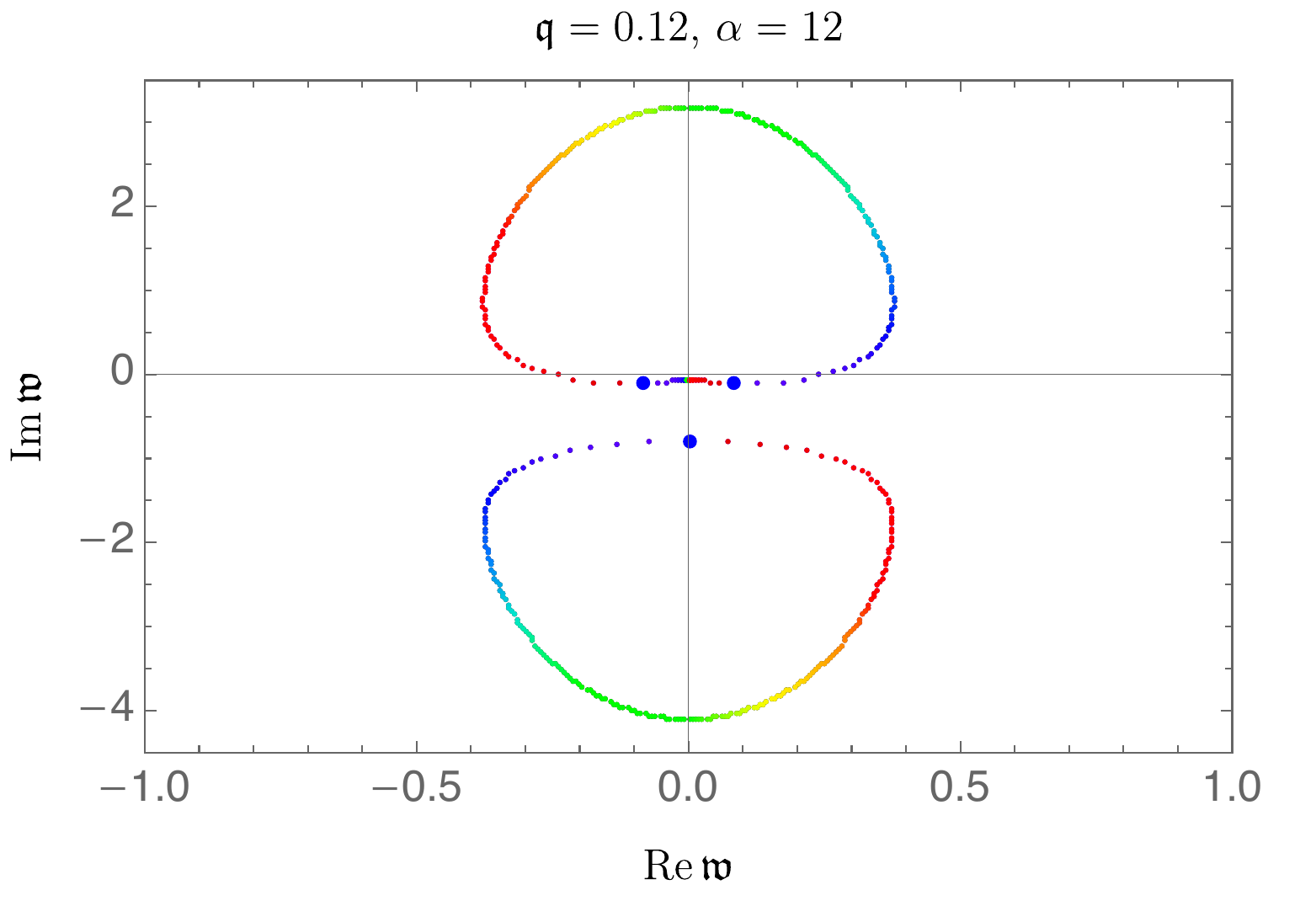}\includegraphics[width=0.35\textwidth]{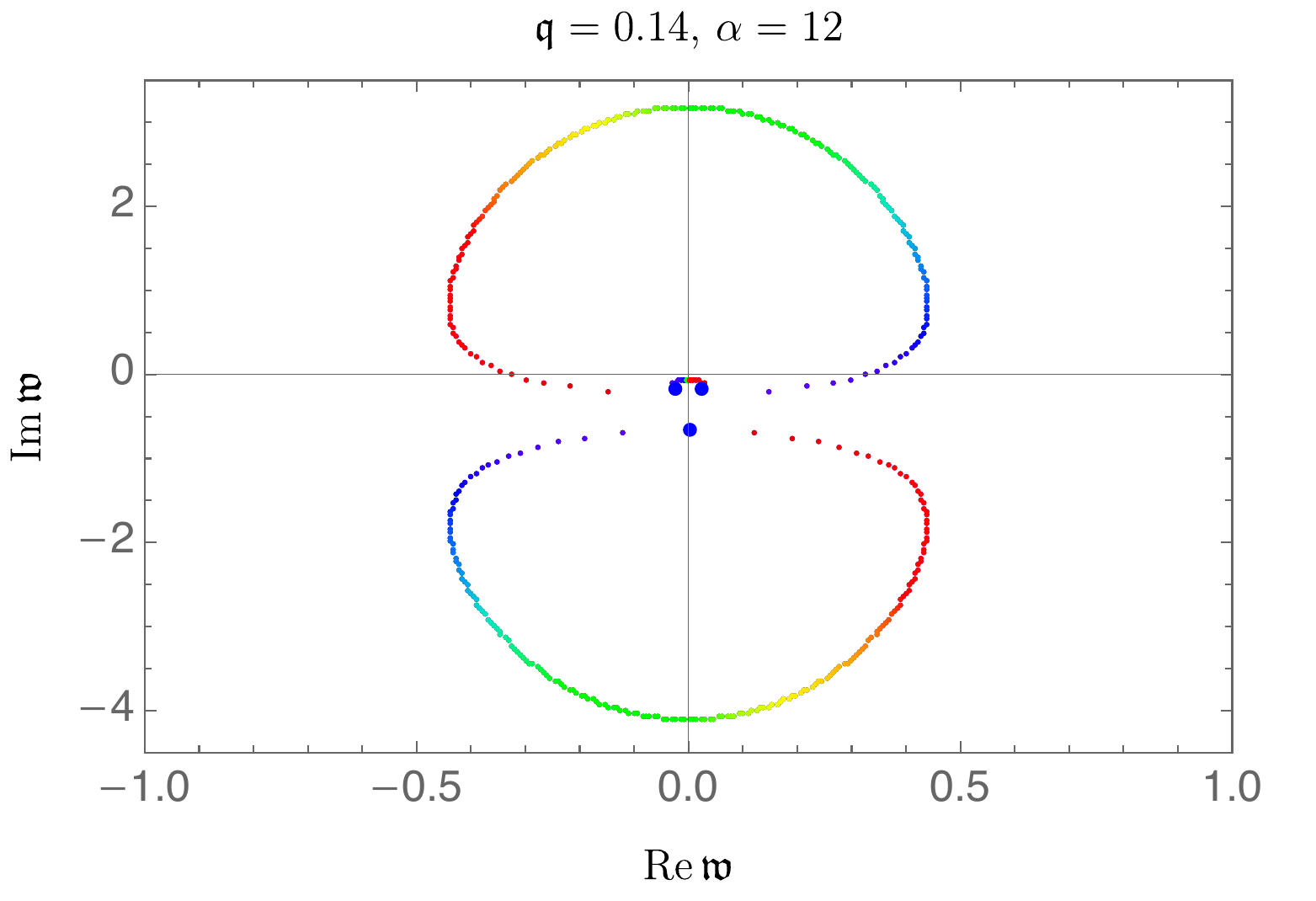}
	\includegraphics[width=0.35\textwidth]{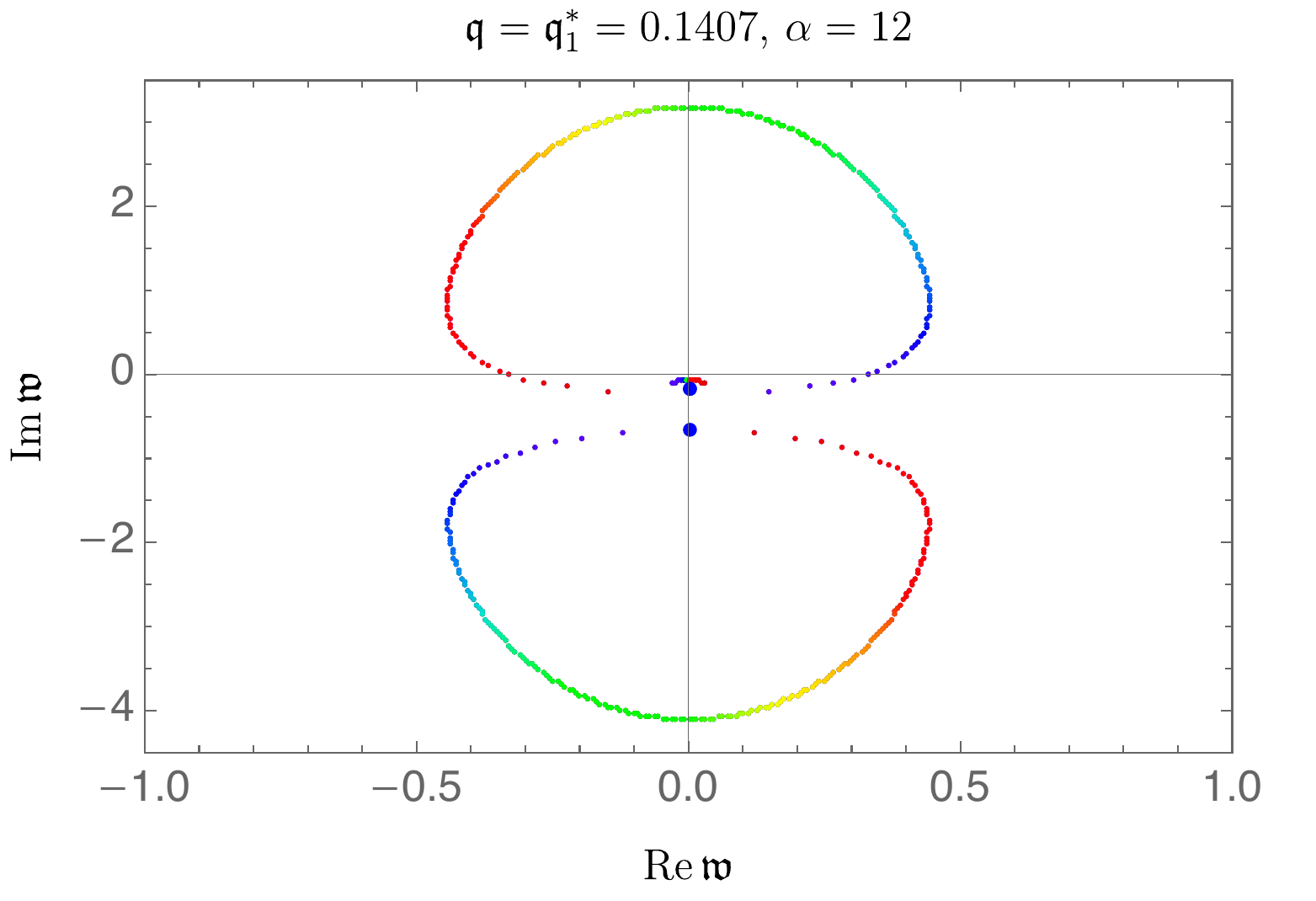}\includegraphics[width=0.35\textwidth]{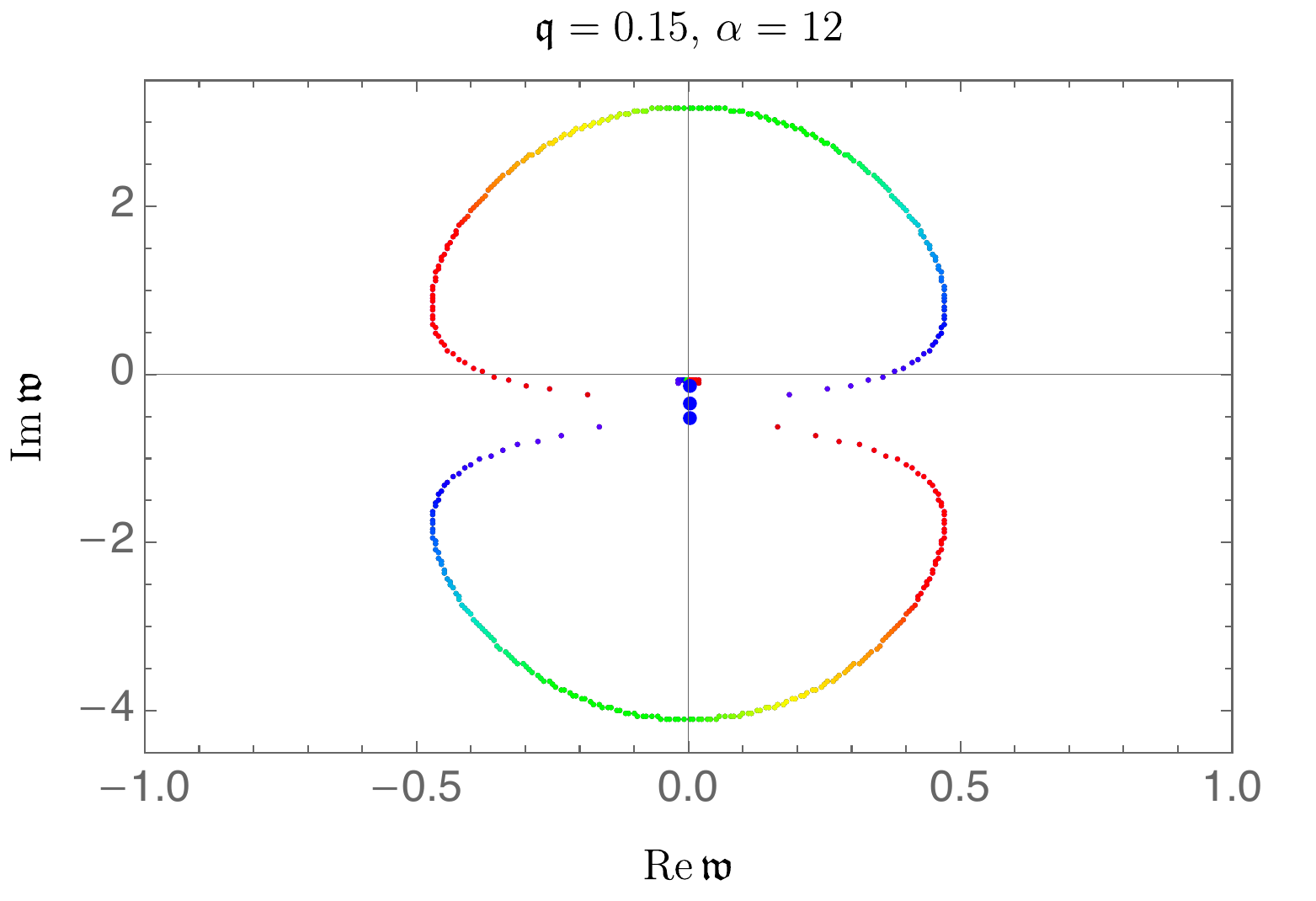}\includegraphics[width=0.35\textwidth]{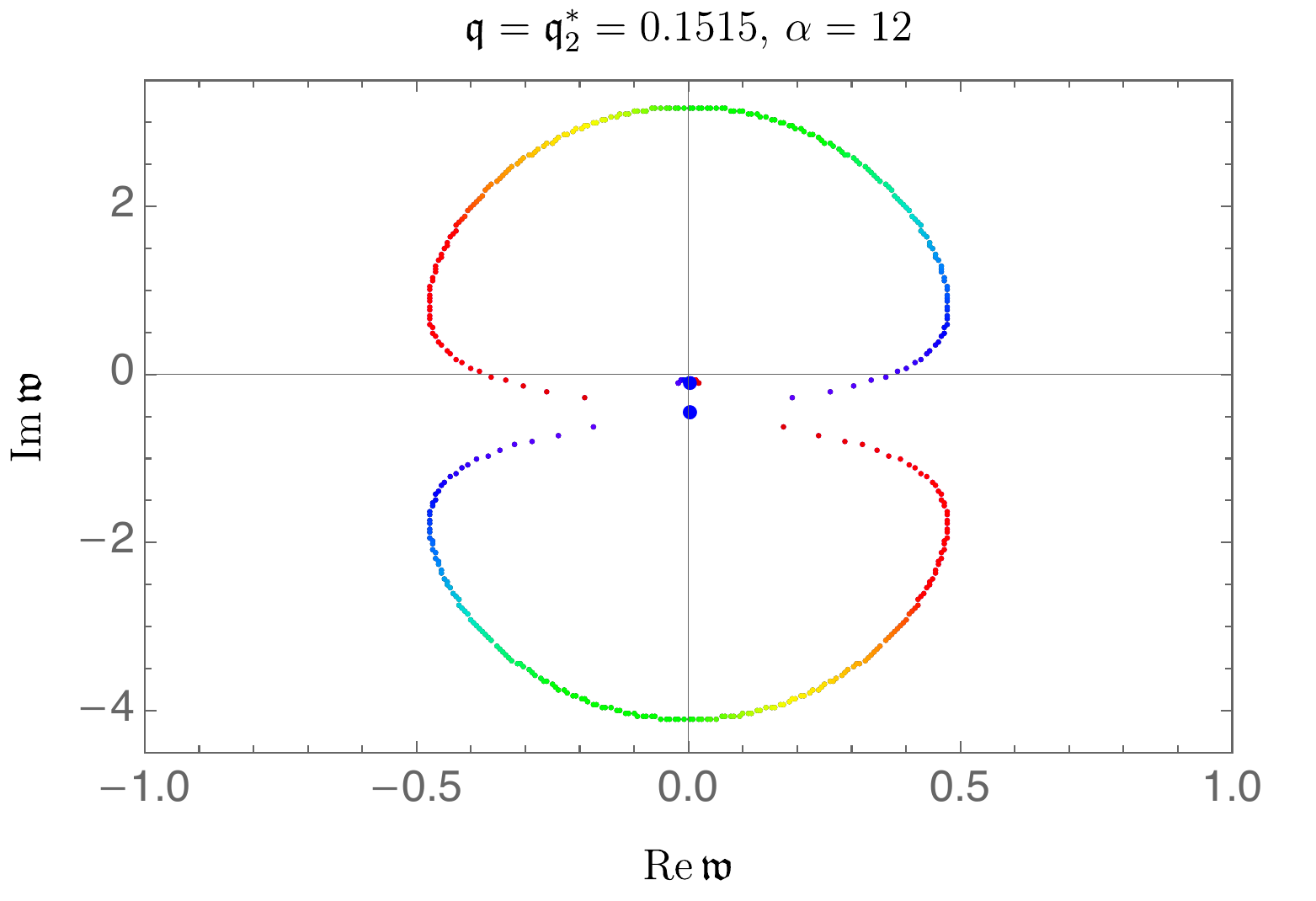}
	\includegraphics[width=0.35\textwidth]{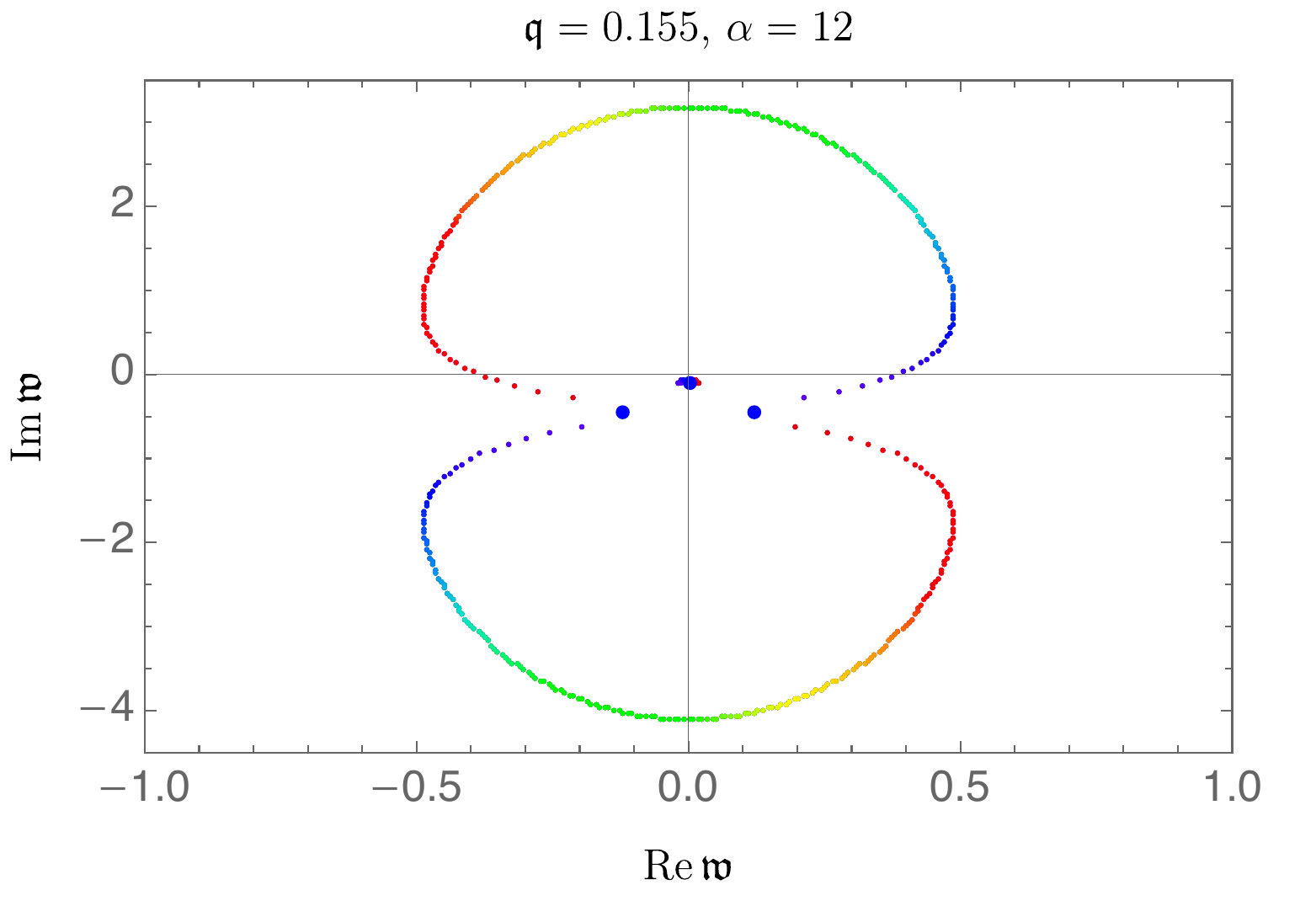}
	\caption{modes for $\alpha=12$ in the complex $\wn-$plane, at various values of
		the complexified momentum $\qn^2=|\qn^2|e^{i \theta}$. Large dots correspond to the poles with  purely real momentum (i.e. at $\theta=0$). As $\theta$ increases from $0$ to $2\pi$, each pole moves counter-clockwise	following the trajectory whose color changes continuously from blue to red. In all panels, the two blue dots 
		located in the 3rd and 4th quadrants indicate the two sound modes and the single blue dot on the vertical axis illustrates the slow mode, at $\theta=0$.  }
	\label{collision_after}
\end{figure}
\par\bigskip 
\noindent

Figure~\ref{collision_after} is devoted to the case $\alpha>8$.  As it is seen, at small values of $|\qn|$, the two sound modes are more slowly decaying than the slow mode. The sound modes can be still described with the standard hydrodynamics.  However, by increasing $|\qn|$, their dynamics becomes gradually coupled with the dynamics of the slow mode. Then Hydro+ must be considered as the alternate. At $\qn_c\approx0.145$, the two sound modes collide, although the trajectory of slow modes does not collide with that of sound modes yet. Thus at $\alpha>8$, the critical momentum $\qn_c$ corresponds to collision of the two sound modes.

As we showed earlier, the QCD near the critical point falls in $\alpha<8$ sector. Thus it is the collision between two sound modes with the slow mode that determine the regime of validity of hydrodynamics, or equivalently the characteristic momentum $q_c$, there. 

\section{Thermodynamics scanning through the critical point}
\label{sec_critical_point}
\begin{figure}[tb]
	\centering
	\includegraphics[width=0.48\textwidth]{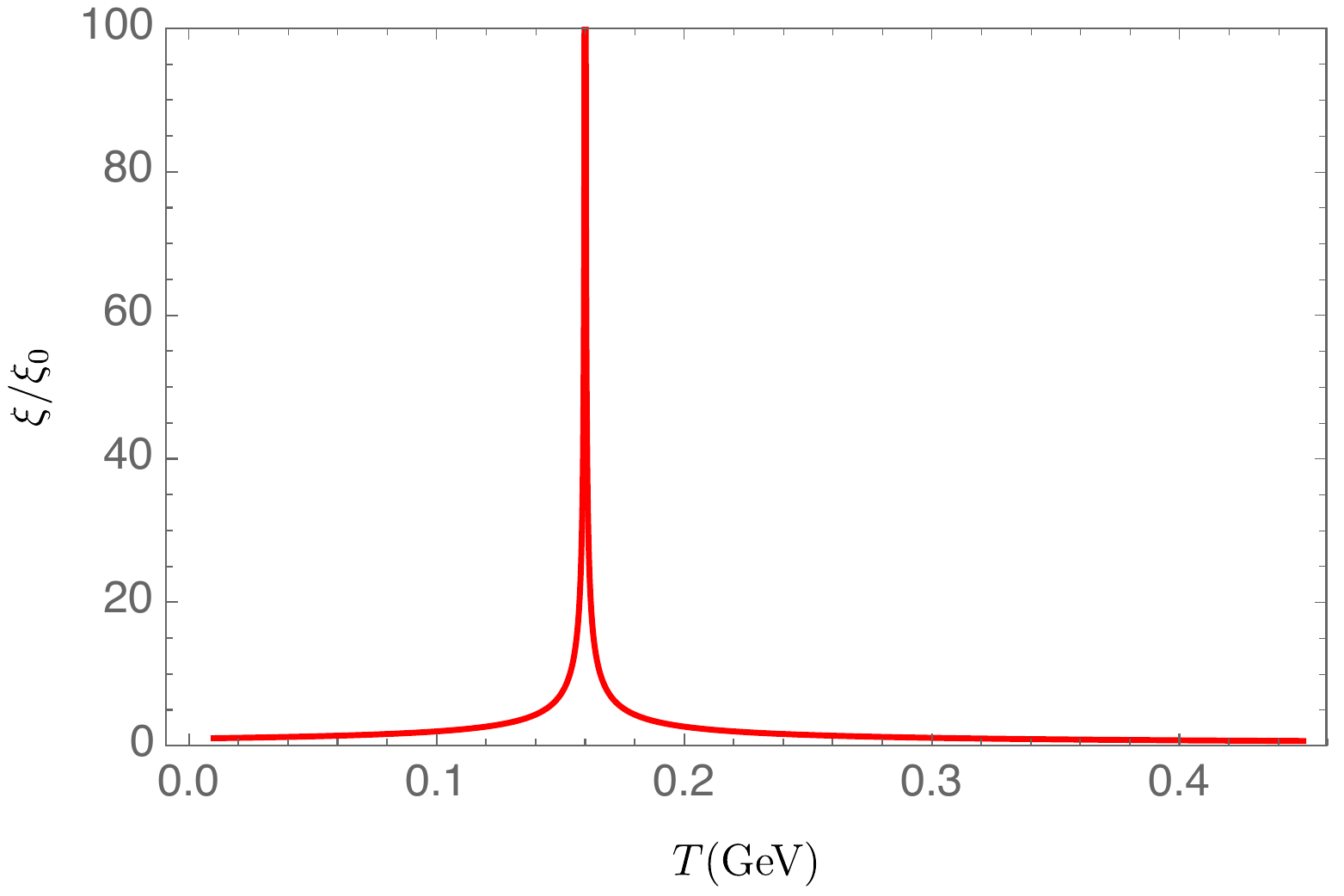}\,\,\includegraphics[width=0.48\textwidth]{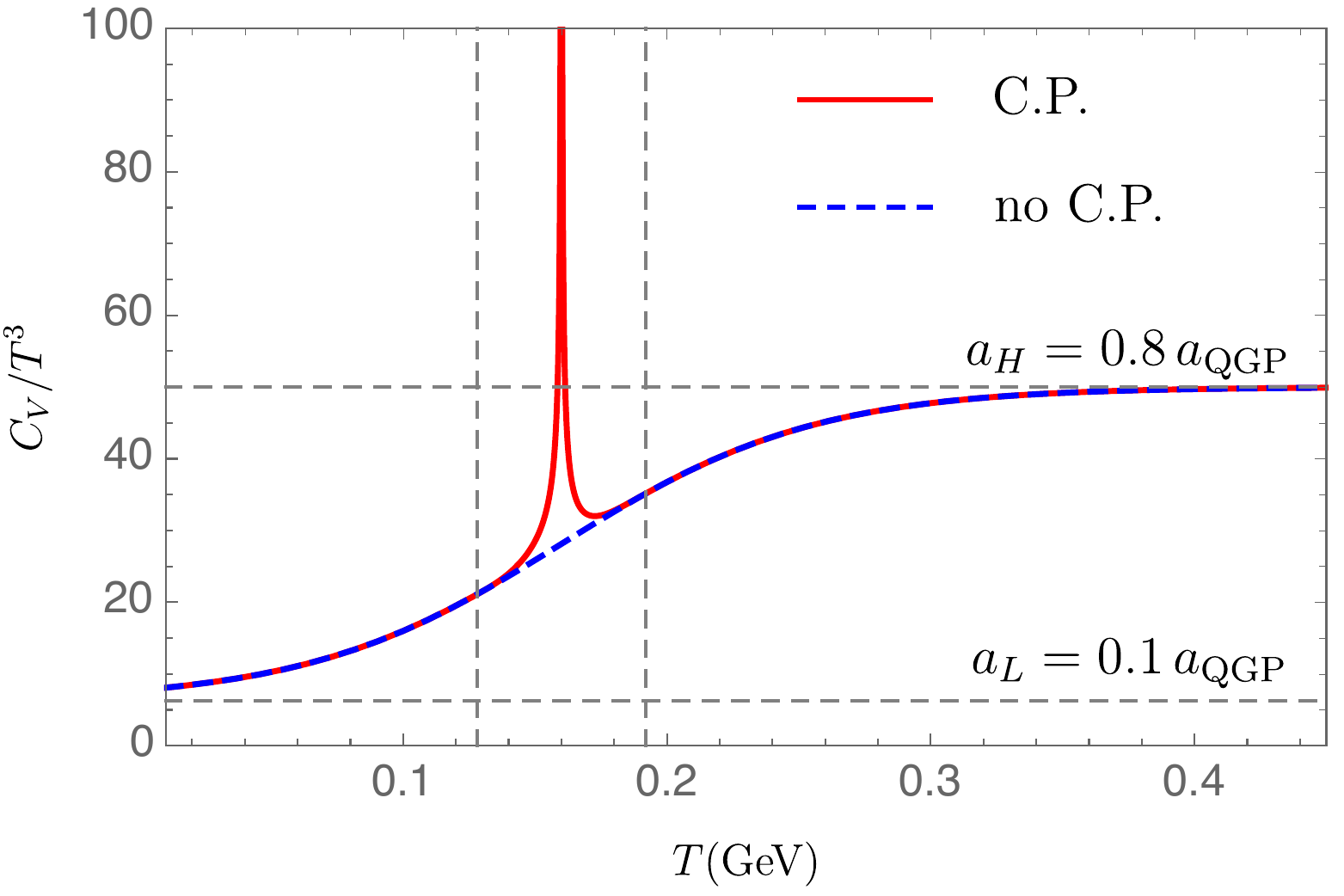}
	\includegraphics[width=0.48\textwidth]{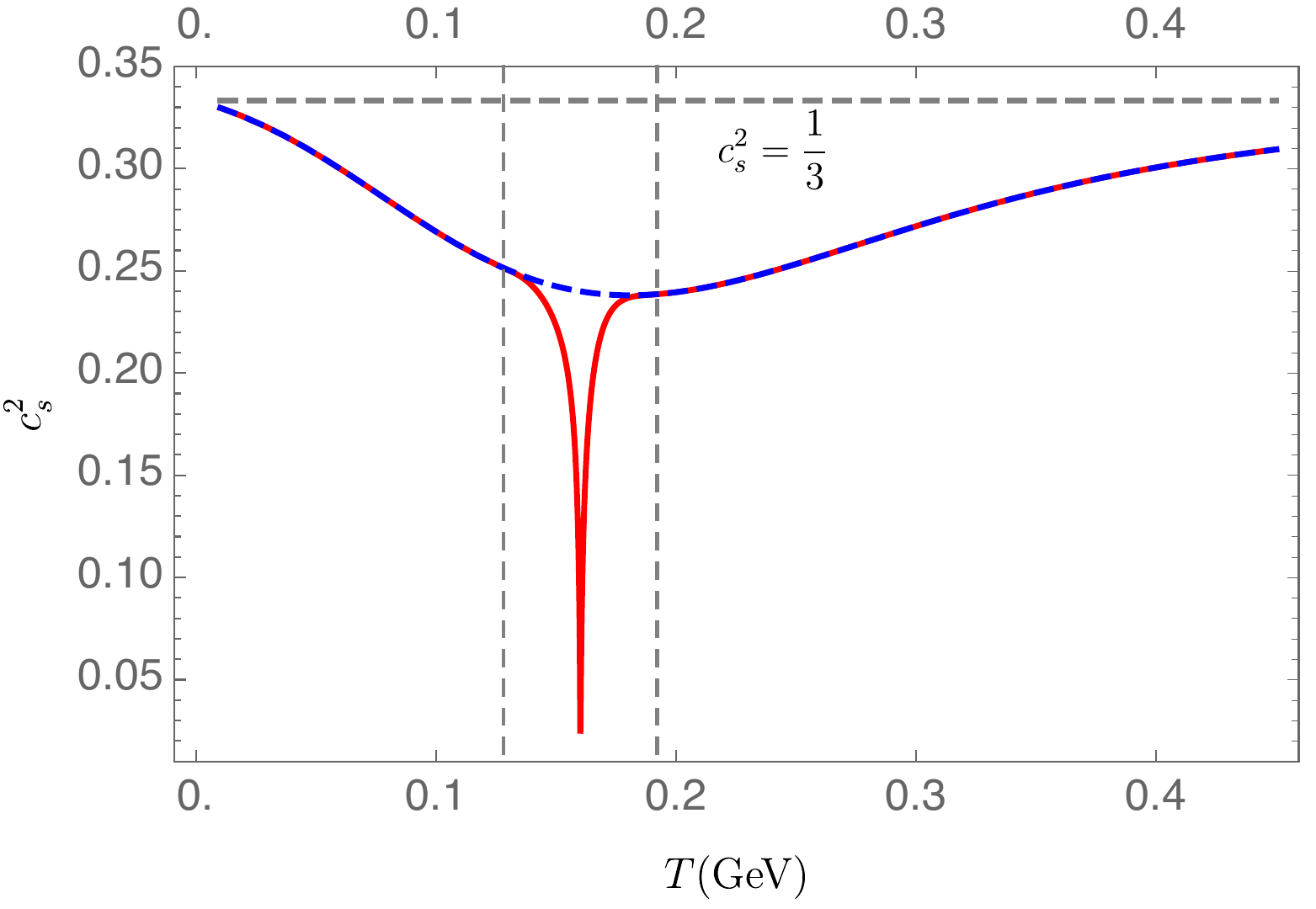}\,\,\includegraphics[width=0.48\textwidth]{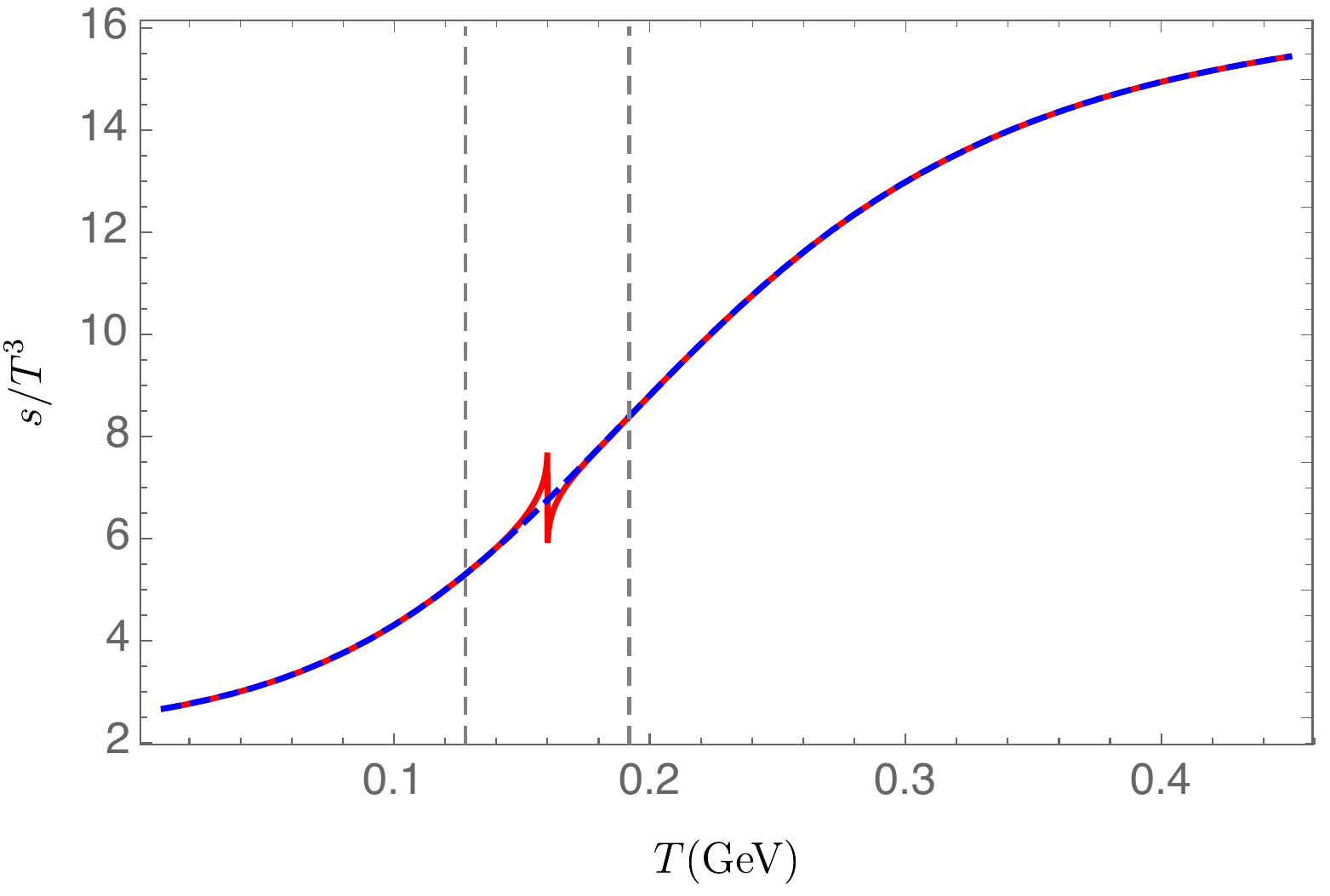}
	\caption{ The left (right) vertical dashed line shows the location of $T = T_c-\Delta T\,(T = T_c+\Delta T )$. In all panels, red curves represent thermodynamic quantities by considering the critical contributions.}
	\label{EoS_2}
\end{figure}

We follow \cite{Parotto:2018pwx,Akamatsu:2018vjr} and map the critical point of QCD to that of the Ising model. Doing so, it turns out that 
\begin{equation}\label{}
\frac{\xi}{\xi_0}=\,\left(\frac{|T-T_c|}{T_c}\right)^{-a \nu},\,\,\,\,a=1.12,\,\,\nu=0.63
\end{equation}
with $T_c=0.16\,\text{GeV}$. Applying this to \eqref{C_V}, we can simply fix the first six non-vanishing coefficients $a_n$ as:
\begin{equation}
a_0=23.12,\,\,a_1=10.25,\,\,a_2=1.85,\,\,a_3=-1.56,\,\,a_4=-0.69,\,\,a_5=0.43\,.
\end{equation}
Having specified $c_V$, one can easily find $c_s^2$ and $s$, as shown in \eqref{EoS_2}.
It can be clearly seen that $c_s^2\rightarrow0$ at the critical point.
\par\bigskip 
\noindent

\bibliographystyle{utphys}

\begin{thebibliography}{10}
	
	\bibitem{Stephanov:1998dy}
	M.~A.~Stephanov, K.~Rajagopal and E.~V.~Shuryak,
	``Signatures of the tricritical point in QCD,''
	Phys. Rev. Lett. \textbf{81} (1998), 4816-4819
	[arXiv:hep-ph/9806219 [hep-ph]].
	
	\bibitem{Stephanov:2004wx}
	M.~A.~Stephanov,
	``QCD phase diagram and the critical point,''
	Prog. Theor. Phys. Suppl. \textbf{153} (2004), 139-156
	[arXiv:hep-ph/0402115 [hep-ph]].
	
	\bibitem{Bzdak:2019pkr}
	A.~Bzdak, S.~Esumi, V.~Koch, J.~Liao, M.~Stephanov and N.~Xu,
	``Mapping the Phases of Quantum Chromodynamics with Beam Energy Scan,''
	Phys. Rept. \textbf{853} (2020), 1-87
	[arXiv:1906.00936 [nucl-th]].
	
	
	\bibitem{Stephanov:2006zvm}
	M.~A.~Stephanov,
	``QCD phase diagram: An Overview,''
	PoS \textbf{LAT2006} (2006), 024
	[arXiv:hep-lat/0701002 [hep-lat]].
	
	\bibitem{Yin:2018ejt}
	Y.~Yin,
	``The QCD critical point hunt: emergent new ideas and new dynamics,''
	[arXiv:1811.06519 [nucl-th]].
	
			\bibitem{Halperin}
	 P. ~C. Hohenberg and B. ~I. ~Halperin,
	 ``Theory of dynamic critical phenomena."
	  Rev. Mod. Phys. 49, 435 (1977).

\bibitem{Berdnikov:1999ph}
B.~Berdnikov and K.~Rajagopal,
``Slowing out-of-equilibrium near the QCD critical point,''
Phys. Rev. D \textbf{61} (2000), 105017
[arXiv:hep-ph/9912274 [hep-ph]].
	
\bibitem{Stephanov:2017ghc}
M.~Stephanov and Y.~Yin,
``Hydrodynamics with parametric slowing down and fluctuations near the critical point,''
Phys. Rev. D \textbf{98} (2018) no.3, 036006
doi:10.1103/PhysRevD.98.036006
[arXiv:1712.10305 [nucl-th]].


\bibitem{Kawasaki}
K. ~Kawasaki,
``Kinetic equations and time correlation functions of critical fluctuations,''
 Annals of Physics 61, 1 (1970).


\bibitem{An:2020vri}
X.~An, G.~Ba\c{s}ar, M.~Stephanov and H.~U.~Yee,
``Evolution of Non-Gaussian Hydrodynamic Fluctuations,''
Phys. Rev. Lett. \textbf{127} (2021) no.7, 072301
[arXiv:2009.10742 [hep-th]].

\bibitem{An:2019csj}
X.~An, G.~Ba\c{s}ar, M.~Stephanov and H.~U.~Yee,
``Fluctuation dynamics in a relativistic fluid with a critical point,''
Phys. Rev. C \textbf{102} (2020) no.3, 034901
[arXiv:1912.13456 [hep-th]].



\bibitem{Withers:2018srf}
B.~Withers,
``Short-lived modes from hydrodynamic dispersion relations,''
JHEP \textbf{06}, 059 (2018)
[arXiv:1803.08058 [hep-th]].



\bibitem{Grozdanov:2019kge} 
S.~Grozdanov, P.~K.~Kovtun, A.~O.~Starinets and P.~Tadić,
``Convergence of the Gradient Expansion in Hydrodynamics,''
Phys.\ Rev.\ Lett.\  {\bf 122}, no. 25, 251601 (2019)
[arXiv:1904.01018 [hep-th]].

\bibitem{Grozdanov:2019uhi} 
S.~Grozdanov, P.~K.~Kovtun, A.~O.~Starinets and P.~Tadić,
``The complex life of hydrodynamic modes,''
JHEP {\bf 1911}, 097 (2019)
[arXiv:1904.12862 [hep-th]].

\bibitem{Abbasi:2020ykq}
N.~Abbasi and S.~Tahery,
``Complexified quasinormal modes and the pole-skipping in a holographic system at finite chemical potential,''
JHEP \textbf{10} (2020), 076
[arXiv:2007.10024 [hep-th]].


\bibitem{Baggioli:2020loj}
M.~Baggioli,
``How small hydrodynamics can go,''
Phys. Rev. D \textbf{103} (2021) no.8, 086001
[arXiv:2010.05916 [hep-th]].

\bibitem{Arean:2020eus}
D.~Arean, R.~A.~Davison, B.~Gout\'eraux and K.~Suzuki,
``Hydrodynamic Diffusion and Its Breakdown near AdS2 Quantum Critical Points,''
Phys. Rev. X \textbf{11} (2021) no.3, 031024
[arXiv:2011.12301 [hep-th]].


\bibitem{Heller:2020hnq}
M.~P.~Heller, A.~Serantes, M.~Spali\'nski, V.~Svensson and B.~Withers,
``Convergence of hydrodynamic modes: insights from kinetic theory and holography,''
[arXiv:2012.15393 [hep-th]].

\bibitem{Heller:2020uuy}
M.~P.~Heller, A.~Serantes, M.~Spali\'nski, V.~Svensson and B.~Withers,
``Hydrodynamic gradient expansion in linear response theory,''
Phys. Rev. D \textbf{104}, no.6, 066002 (2021)
[arXiv:2007.05524 [hep-th]].

\bibitem{Abbasi:2020xli}
N.~Abbasi and M.~Kaminski,
``Constraints on quasinormal modes and bounds for critical points from pole-skipping,''
JHEP \textbf{03} (2021), 265
[arXiv:2012.15820 [hep-th]].


\bibitem{Asadi:2021hds}
M.~Asadi, H.~Soltanpanahi and F.~Taghinavaz,
``Critical behaviour of hydrodynamic series,''
JHEP \textbf{05} (2021), 287
[arXiv:2102.03584 [hep-th]].

\bibitem{Baggioli:2021ujk}
M.~Baggioli, U.~Gran and M.~Torns\"o,
``Collective modes of polarizable holographic media in magnetic fields,''
JHEP \textbf{06} (2021), 014
[arXiv:2102.09969 [hep-th]].




\bibitem{Wu:2021mkk}
N.~Wu, M.~Baggioli and W.~J.~Li,
``On the universality of AdS$_{2}$ diffusion bounds and the breakdown of linearized hydrodynamics,''
JHEP \textbf{05} (2021), 014
[arXiv:2102.05810 [hep-th]].

\bibitem{Jeong:2021zhz}
H.~S.~Jeong, K.~Y.~Kim and Y.~W.~Sun,
``Bound of diffusion constants from pole-skipping points: spontaneous symmetry breaking and magnetic field,''
[arXiv:2104.13084 [hep-th]].


\bibitem{Grozdanov:2021gzh}
S.~Grozdanov, A.~O.~Starinets and P.~Tadi\'c,
``Hydrodynamic dispersion relations at finite coupling,''
JHEP \textbf{06} (2021), 180
[arXiv:2104.11035 [hep-th]].

\bibitem{Jansen:2020hfd}
A.~Jansen and C.~Pantelidou,
``Quasinormal modes in charged fluids at complex momentum,''
JHEP \textbf{10} (2020), 121
[arXiv:2007.14418 [hep-th]].

\bibitem{Heller:2021oxl}
M.~P.~Heller, A.~Serantes, M.~Spali\'nski, V.~Svensson and B.~Withers,
``The hydrodynamic gradient expansion diverges beyond Bjorken flow,''
[arXiv:2110.07621 [hep-th]].

\bibitem{Jeong:2021zsv}
H.~S.~Jeong, K.~Y.~Kim and Y.~W.~Sun,
``The breakdown of magneto-hydrodynamics near AdS$_2$ fixed point and energy diffusion bound,''
[arXiv:2105.03882 [hep-th]].

\bibitem{Huh:2021ppg}
K.~B.~Huh, H.~S.~Jeong, K.~Y.~Kim and Y.~W.~Sun,
``Upper bound of the charge diffusion constant in holography,''
[arXiv:2111.07515 [hep-th]].


\bibitem{Liu:2021qmt}
Y.~Liu and X.~M.~Wu,
``Breakdown of hydrodynamics from holographic pole collision,''
[arXiv:2111.07770 [hep-th]].

\bibitem{Grozdanov:2018fic}
S.~Grozdanov, A.~Lucas and N.~Poovuttikul,
``Holography and hydrodynamics with weakly broken symmetries,''
Phys. Rev. D \textbf{99} (2019) no.8, 086012
doi:10.1103/PhysRevD.99.086012
[arXiv:1810.10016 [hep-th]].

\bibitem{Grozdanov:2016fkt}
S.~Grozdanov and A.~O.~Starinets,
``Second-order transport, quasinormal modes and zero-viscosity limit in the Gauss-Bonnet holographic fluid,''
JHEP \textbf{03} (2017), 166
[arXiv:1611.07053 [hep-th]].

\bibitem{Grozdanov:2018gfx}
S.~Grozdanov and A.~O.~Starinets,
``Adding new branches to the \textquotedblleft{}Christmas tree\textquotedblright{} of the quasinormal spectrum of black branes,''
JHEP \textbf{04} (2019), 080
doi:10.1007/JHEP04(2019)080
[arXiv:1812.09288 [hep-th]].




\bibitem{Rajagopal:2019xwg}
K.~Rajagopal, G.~Ridgway, R.~Weller and Y.~Yin,
Phys. Rev. D \textbf{102} (2020) no.9, 094025
doi:10.1103/PhysRevD.102.094025
[arXiv:1908.08539 [hep-ph]].


\bibitem{Pradeep:2021opj}
M.~Pradeep, K.~Rajagopal, M.~Stephanov and Y.~Yin,
``Freezing out critical fluctuations,''
[arXiv:2109.13188 [hep-ph]].


\bibitem{Parotto:2018pwx}
P.~Parotto, M.~Bluhm, D.~Mroczek, M.~Nahrgang, J.~Noronha-Hostler, K.~Rajagopal, C.~Ratti, T.~Sch\"afer and M.~Stephanov,
``QCD equation of state matched to lattice data and exhibiting a critical point singularity,''
Phys. Rev. C \textbf{101} (2020) no.3, 034901
[arXiv:1805.05249 [hep-ph]].

\bibitem{Akamatsu:2018vjr}
Y.~Akamatsu, D.~Teaney, F.~Yan and Y.~Yin,
``Transits of the QCD critical point,''
Phys. Rev. C \textbf{100} (2019) no.4, 044901
[arXiv:1811.05081 [nucl-th]].

\bibitem{Kardar}
M. ~Kardar, 
``Statistical Physics of Particles,"
 Cambridge University Press, 2007.


\bibitem{An:2020jjk}
X.~An,
``Relativistic Dynamics of Fluctuations and QCD Critical Point,''
Nucl. Phys. A \textbf{1005} (2021), 121957
[arXiv:2003.02828 [hep-th]].


\bibitem{Kovtun:2003vj}
P.~Kovtun and L.~G.~Yaffe,
``Hydrodynamic fluctuations, long time tails, and supersymmetry,''
Phys. Rev. D \textbf{68} (2003), 025007
[arXiv:hep-th/0303010 [hep-th]].

\bibitem{Kovtun:2012rj}
P.~Kovtun,
``Lectures on hydrodynamic fluctuations in relativistic theories,''
J. Phys. A \textbf{45} (2012), 473001
[arXiv:1205.5040 [hep-th]].

\bibitem{Akamatsu:2016llw}
Y.~Akamatsu, A.~Mazeliauskas and D.~Teaney,
``A kinetic regime of hydrodynamic fluctuations and long time tails for a Bjorken expansion,''
Phys. Rev. C \textbf{95}, no.1, 014909 (2017)
doi:10.1103/PhysRevC.95.014909
[arXiv:1606.07742 [nucl-th]].

\bibitem{Chen-Lin:2018kfl}
X.~Chen-Lin, L.~V.~Delacr\'etaz and S.~A.~Hartnoll,
``Theory of diffusive fluctuations,''
Phys. Rev. Lett. \textbf{122} (2019) no.9, 091602
[arXiv:1811.12540 [hep-th]].

\bibitem{Martinez:2018wia}
M.~Martinez and T.~Sch\"afer,
``Stochastic hydrodynamics and long time tails of an expanding conformal charged fluid,''
Phys. Rev. C \textbf{99} (2019) no.5, 054902
[arXiv:1812.05279 [hep-th]].


\bibitem{An:2019osr}
X.~An, G.~Basar, M.~Stephanov and H.~U.~Yee,
``Relativistic Hydrodynamic Fluctuations,''
Phys. Rev. C \textbf{100} (2019) no.2, 024910
[arXiv:1902.09517 [hep-th]].



\bibitem{Abbasi:2021fcz}
N.~Abbasi,
``Long-time tails in the SYK chain from the effective field theory with a large number of derivatives,''
[arXiv:2112.12751 [hep-th]].

\bibitem{STAR:2017ckg}
L.~Adamczyk \textit{et al.} [STAR],
``Global $\Lambda$ hyperon polarization in nuclear collisions: evidence for the most vortical fluid,''
Nature \textbf{548}, 62-65 (2017)
[arXiv:1701.06657 [nucl-ex]].

\bibitem{Cartwright:2021qpp}
C.~Cartwright, M.~G.~Amano, M.~Kaminski, J.~Noronha and E.~Speranza,
``Convergence of hydrodynamics in rapidly spinning strongly coupled plasma,''
[arXiv:2112.10781 [hep-th]].

\bibitem{Bjorken:1982qr}
J.~D.~Bjorken,
``Highly Relativistic Nucleus-Nucleus Collisions: The Central Rapidity Region,''
Phys. Rev. D \textbf{27}, 140-151 (1983)
doi:10.1103/PhysRevD.27.140

\bibitem{Heller:2021oxl}
M.~P.~Heller, A.~Serantes, M.~Spali\'nski, V.~Svensson and B.~Withers,
``The hydrodynamic gradient expansion diverges beyond Bjorken flow,''
[arXiv:2110.07621 [hep-th]].

\bibitem{Heller:2021yjh}
M.~P.~Heller, A.~Serantes, M.~Spali\'nski, V.~Svensson and B.~Withers,
``Relativistic hydrodynamics: a singulant perspective,''
[arXiv:2112.12794 [hep-th]].

\bibitem{Martinez:2019bsn}
M.~Martinez, T.~Sch\"afer and V.~Skokov,
``Critical behavior of the bulk viscosity in QCD,''
Phys. Rev. D \textbf{100} (2019) no.7, 074017
[arXiv:1906.11306 [hep-ph]].


\end{thebibliography}
\providecommand{\href}[2]{#2}\begingroup\raggedright\endgroup

\end{document}